\documentclass[12pt]{article}
\usepackage{epsfig}
\usepackage{amssymb}
\usepackage{amsmath}
\usepackage{amsfonts}
\usepackage{graphicx}
\usepackage{mathrsfs}
\usepackage[dvips]{color}
\usepackage{multirow}
\usepackage{latexsym}

\newcommand{\R}{\mathbb{R}}
\newcommand{\C}{\mathbb{C}}
\newcommand{\Z}{\mathbb{Z}}

\newcommand{\fc}{\mathfrak{c}}

\newcommand{\fn}{{\mathfrak{n}}}

\newcommand{\fs}{\mathfrak{s}}

\newcommand{\fz}{\mathfrak{z}}

\newcommand{\fK}{\mathfrak{K}}

\newcommand{\fM}{\mathfrak{M}}
\newcommand{\bfM}{\boldsymbol{\fM}}

\newcommand{\by}{\mathbf{y}}
\newcommand{\bz}{\mathbf{z}}
\newcommand{\bA}{\mathbf{A}}
\newcommand{\bB}{\mathbf{B}}

\newcommand{\bE}{\mathbf{E}}

\newcommand{\bcF}{\boldsymbol{\cF}}

\newcommand{\bcG}{\boldsymbol{\cG}}
\newcommand{\bH}{\mathbf{H}}
\newcommand{\bcH}{{\boldsymbol{\cH}}}
\newcommand{\bI}{\mathbf{I}}
\newcommand{\bJ}{\mathbf{J}}
\newcommand{\bL}{\mathbf{L}}

\newcommand{\bM}{\mathbf{M}}

\newcommand{\bT}{\mathbf{T}}
\newcommand{\bU}{\mathbf{U}}
\newcommand{\bcU}{{\boldsymbol{\cU}}}
\newcommand{\bV}{\mathbf{V}}
\newcommand{\bcZ}{\boldsymbol{\cZ}}

\newcommand{\bsigma}{{\boldsymbol{\sigma}}}

\newcommand{\cC}{\mathcal{C}}

\newcommand{\cH}{\mathcal{H}}

\newcommand{\cF}{\mathcal{F}}
\newcommand{\cG}{\mathcal{G}}

\newcommand{\cK}{\mathcal{K}}

\newcommand{\cP}{\mathcal{P}}

\newcommand{\cT}{\mathcal{T}}
\newcommand{\cU}{\mathcal{U}}

\newcommand{\cZ}{\mathcal{Z}}
\newcommand{\be}{\begin{equation}}
\newcommand{\ee}{\end{equation}}
\newcommand{\bea}{\begin{eqnarray}}
\newcommand{\eea}{\end{eqnarray}}
\newcommand{\nn}{\nonumber}
\newcommand{\kt}{\rangle}
\newcommand{\br}{\langle}

\newcommand{\ed}{\end{document}}

\newcommand{\rv}{{\rm v}}

\newcommand{\np}{\newpage}

\newcommand{\bi}{\begin{itemize}}
\newcommand{\ei}{\end{itemize}}

\newcommand{\bce}{\begin{center}}
\newcommand{\ece}{\end{center}}

\newcommand{\sE}{\mathscr{E}}
\newcommand{\sF}{\mathscr{F}}

\newcommand{\sH}{\mathscr{H}}

\newcommand{\sS}{\mathscr{S}}
\newcommand{\sT}{\mathscr{T}}
\newcommand{\sU}{\mathscr{U}}
\newcommand{\sV}{\mathscr{V}}
\newcommand{\RE}{{\rm Re}}
\newcommand{\IM}{{\rm Im}}

\newcommand{\bzero}{\mathbf{0}}

\newcommand{\bPsi}{{\boldsymbol{\Psi}}}
\newcommand{\bPhi}{{\boldsymbol{\Phi}}}

\newcommand{\bcK}{{\boldsymbol{\cK}}}

\newcommand{\for}{{\rm for}}

\makeatletter
\newcommand\xleftrightarrow[2][]{%
  \ext@arrow 9999{\longleftrightarrowfill@}{#1}{#2}}
\newcommand\longleftrightarrowfill@{%
  \arrowfill@\leftarrow\relbar\rightarrow}
\makeatother

\newcommand{\ktl}{{k_{_\triangleleft}}}
\newcommand{\ktr}{{k_{_\triangleright}}}

\begin{document}

%\title{Dynamical Formulation of Time-Independent Scattering Theory and Unidirectional Invisibility}

\title{Transfer matrix in scattering theory: A survey of basic properties and recent developments}

\author{Ali~Mostafazadeh\thanks{E-mail address: amostafazadeh@ku.edu.tr}~
%and Sasan Haji-Zadeh\thanks{E-mail address: shacizadeh@ku.edu.tr}\\[6pt]
\\ Departments of Mathematics %$^*$
and Physics, %$^{*,\dagger}$
Ko\c{c} University,\\ 34450 Sar{\i}yer,
Istanbul, Turkey}

\date{ }
\maketitle

\begin{abstract}
We give a pedagogical introduction to time-independent scattering theory in one dimension focusing on the basic properties and recent applications of transfer matrices. In particular, we begin surveying some basic notions of potential scattering such as transfer matrix and its analyticity, multi-delta-function and locally periodic potentials, Jost solutions, spectral singularities and their time-reversal, and unidirectional reflectionlessness and invisibility. We then offer a simple derivation of the Lippmann-Schwinger equation and Born series, and discuss the Born approximation. Next, we outline a recently developed dynamical formulation of time-independent scattering theory in one dimension. This formulation relates the transfer matrix and therefore the solution of the scattering problem for a given potential to the solution of the time-dependent Schr\"odinger equation for an effective non-unitary two-level quantum system. We provide a self-contained treatment of this formulation and some of its most important applications. Specifically, we use it to devise a powerful alternative to the Born series and Born approximation, derive dynamical equations for the reflection and transmission amplitudes, discuss their application in constructing exact tunable unidirectionally invisible potentials, and use them to provide an exact solution for single-mode inverse scattering problems. The latter, which has important applications in designing optical devices with a variety of functionalities, amounts to providing an explicit construction for a finite-range complex potential whose reflection and transmission amplitudes take arbitrary prescribed values at any given wavenumber.  
\vspace{2mm}

%\noindent PACS numbers: 03.65.Nk, 42.25.Bs \vspace{2mm}

\end{abstract}

\np

\tableofcontents

\section{Introduction}

Scattering of waves by obstacles or interactions is a natural phenomenon that we witness in our everyday lives. It is also a primary tool for acquiring information about uncharted territories of physical reality. The indisputable importance of this phenomenon has led to a systematic study of its basic principles in the 19th century, particularly in the realm of optics \cite{Logan-1965} and acoustics \cite{Lord-Rayleigh}. The resulting body of knowledge played a significant role in the formulation of quantum scattering theory \cite{Born-1926}, a monumental achievement realized almost concurrently with the advent of quantum mechanics. The later discoveries of the notion of the S-matrix in 1937 \cite{wheeler-1937} and the Lippmann-Schwinger equation in 1950 \cite{Lippmann-Schwinger} are among the major developments of the 20th century theoretical physics. These immediately followed by important contributions of mathematicians in addressing the inverse problem of determining the properties of the scatterer using the scattering data  \cite{Gelfand-1951,Marchenko-1955,Faddeev-1959,Agranovich-1963} and developing a rigorous theory of scattering \cite{kato,Reed-Simon4}.

Mathematical theories of scattering and inverse scattering have been active areas of research since their inception \cite{Newton-ST,Chadan-IS,yafaev-2010}, but their achievements could not find their way into the standard physics textbooks. This is mainly because they employ mathematical tools that are beyond the reach of most physics students. The present article is not as ambitious as to try making these tools accessible for an average physicist. It rather aims at providing the background required for following the recent progress made in connection with the study of some of the remarkable properties of complex scattering potentials \cite{prl-2009,chong-2010,lin-2011} and a curious dynamical formulation of time-independent scattering theory. The only prerequisite for an effective use of this article is a basic knowledge of linear algebra, linear differential equations, and quantum mechanics.

The organization of this article is as follows. In Sec.~2, we give some basic definitions and facts about time-independent scattering theory in one dimension. Here we discuss the transfer matrix, reflection and transmission amplitudes, Jost solutions, spectral singularities and their time-reversal, unidirectionally reflectionless and invisible potentials, Lippmann-Schwinger equation, Born series, and Born approximation. In Sec.~3, we outline the dynamical formulation of time-independent scattering theory in one dimension. This section starts with a general review of quantum dynamics of a two-level system with a time-dependent Hamiltonian. It then introduces a two-level system whose evolution operator determines the transfer matrix of a given short-range potential and provides a survey of the theoretical implications and applications of this formulation of scattering theory in one dimension. In particular, it reports a powerful alternative to Born series and Born approximation, derives dynamical equations for the scattering data, and uses them to construct the first examples of exact tunable unidirectionally invisible potentials. This section ends with a detailed discussion of the role of these potentials in offering an exact solution for the single-mode inverse scattering problems. In Sec.~4, we give our concluding remarks and comment on  some of the surprising achievements of higher-dimensional generalizations of the dynamical formulation of time-independent scattering theory.

\section{Time-independent scattering theory in one dimension}
\label{Sec2}

\subsection{Left-going and right-going waves}
\label{Sec2-1}

Consider the scalar wave equation in 1+1 dimensions,
    \be
    (-\partial_t^2+\rv^2\partial_x^2)\phi(x,t)=0,
    \label{w-eq}
    \ee
where  $(x,t)\in\R^2$ and $\rv$ is a positive real parameter. The general solution of (\ref{w-eq}), which was obtained by D'Alembert in the 18th century, reads
    \be
    \phi(x,t)=f_+(x-\rv t)+f_-(x+\rv t),
    \label{sol-w-eq}
    \ee
where $f_\pm:\R\to\C$ are twice differentiable functions. Let $\phi_\pm(x,t):=f_\pm(x\mp vt)$ whose sum gives the D'Alembert's solution (\ref{sol-w-eq}). Suppose that $|f_\pm(x)|$ has a peak at $x=a_\pm\in\R$. Then it is not difficult to see that $|\phi_\pm(x,t)|$ will have a peak at $x=a_\pm\pm \rv\,t$. This shows that, for $t>0$, the position of the peak of $|\phi_\pm(x,t)|$ moves towards $x=\pm\infty$ with a speed $\rv$. For this reason, we refer to $\phi_+(x,t)$ and $\phi_-(x,t)$ as the ``right-going'' and ``left-going'' waves, respectively.\footnote{We use this terminology regardless of whether $|\phi_\pm(x,t)|$ has a peak or not.} The basic examples of $\phi_\pm(x,t)$ are the right- and left-going plane-wave solutions, $e^{i(\pm kx-\omega t)}$, with $k\in\R^+$ and $\omega:=k\,\rv$. We can superpose these solutions to construct right- and left-going wave packets.

Now, consider the following generalization of (\ref{w-eq}).
    \be
    \left[-\widehat\varepsilon(x)\partial_t^2+ \rv^2 \partial_x^2\right]\phi(x,t)=0,
    \label{w-eq-2}
    \ee
where $\widehat\varepsilon:\R\to\C$ is a nowhere-vanishing function such that $\widehat\varepsilon(x)=1$ when $x$ lies outside a close interval in $\R$, say $[a_-,a_+]$. This condition implies that, in the intervals $]\!-\infty,a_-[$ and $]a_+,+\infty[$, every solutions of (\ref{w-eq-2}) coincides with a solution of (\ref{w-eq}). Therefore, we  can express it as the sum of a right-going and a left-going solution of (\ref{w-eq}) for $x<a_-$ and $x>a_+$. We use the term ``right-going'' and ``left-going'' for the asymptotic solutions of (\ref{w-eq-2}) at $-\infty$ (respectively $+\infty$), if they are ``right-going'' and ``left-going'' solutions of (\ref{w-eq}) for $x<a_-$ (respectively $x>a_+$.)

Eq.~(\ref{w-eq-2}) admits a class of solutions of the form,
    \be
    \phi(x,t)=e^{-i\omega t}\psi(x),
    \label{time-harmonic}
    \ee
where $\omega\in\R^+$, $\psi:\R\to\C$ is a twice-differentiable function satisfying the Helmholtz equation,
    \be
    \psi''(x)+k^2\widehat\varepsilon(x)\psi(x)=0,
    \label{helmholtz-eq}
    \ee
and $k:=\omega/\rv$. For  $x<a_-$ and $x>a_+$, every solution of (\ref{helmholtz-eq}) is a linear combination of $e^{ikx}$ and $e^{-ikx}$, i.e., there are complex coefficients $A_\pm$ and $B_\pm$ such that
    \be
    \psi(x)=\left\{\begin{array}{ccc}
    A_- e^{ikx}+B_- e^{-ikx} &{\rm for} & x<a_-,\\
    A_+ e^{ikx}+B_+ e^{-ikx} &{\rm for} & x>a_+.
    \end{array}\right.
    \label{e2}
    \ee
In view of (\ref{time-harmonic}) and (\ref{e2}), we can identify $A_\pm$ and $B_\pm$ with the complex amplitudes of the right- and left-going components of $\psi(x)$ as $x\to\pm\infty$.

\begin{figure}
    \begin{center}
    \includegraphics[scale=0.25]{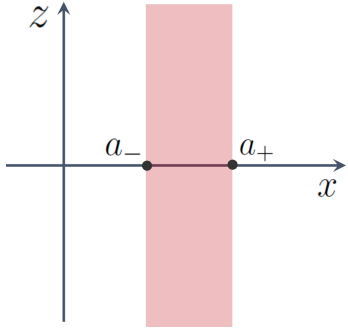}
    \end{center}
    \vspace{-12pt}
    \caption{Schematic view of the cross section of a planer slab of dielectric material (colored in pink) occupying the space between the planes $x=a_\pm$.}
    \label{fig1}
    \end{figure}

The Helmholtz equation (\ref{helmholtz-eq}) arises naturally in the study of electromagnetic waves propagating in a dielectric medium. A typical example is an optical slab made out of possibly lossy or active optical material whose properties are invariant under translations along the $y$- and $z$-axes. Suppose that the slab is immersed in an infinite homogeneous medium with a real and constant permittivity $\varepsilon_\infty$ and occupies the space bounded by the planes $x=a_-$ and $x=a_+$ as depicted in Fig.~\ref{fig1}.
Then the electromagnetic properties of this system is determined by the permittivity profile $\varepsilon(x)$ which satisfies $\varepsilon(x)=\varepsilon_\infty$ for $x\notin [a_-,a_+]$. The Maxwell's equations \cite{Griffiths} for this setup admit monochromatic solutions of the form,
    \begin{align*}
    &\bE(x,y,z,t)=E_0\, e^{-i\omega t}\psi(x)\,\hat\by,
    &&\bB(x,y,z,t)=-\frac{i E_0}{\omega}\, e^{-i\omega t}\psi'(x)\,\hat\bz,
    \end{align*}
where $\bE$ and $\bB$ are respectively the electric and magnetic fields, $E_0$ is a constant amplitude, $\hat\by$ and $\hat\bz$ are unit vectors pointing along the positive $y$- and $z$- axes, $\psi$ satisfies the Helmholtz equation (\ref{helmholtz-eq}) with
    \begin{align}
    &\widehat\varepsilon(x):=\frac{\varepsilon(x)}{\varepsilon_\infty},
    &&\rv:=c\,\sqrt{\frac{\varepsilon_0}{\varepsilon_\infty}},
    \label{e3}
    \end{align}
and $c$ and $\varepsilon_0$ label the speed of light in vacuum and the permittivity of vacuum, respectively. 

Another useful information about the Helmholtz equation (\ref{helmholtz-eq}) is that we can identify it with the time-independent Schr\"odinger equation,
    \be
    -\psi''(x)+v(x)\psi(x)=k^2\psi(x),
    \label{sch-eq}
    \ee
for the potential,
    \be
    v(x):=k^2[1-\widehat\varepsilon(x)].
    \label{optical}
    \ee
Because $k^2$ plays the role of the ``energy'' in (\ref{sch-eq}),  $v$ is a manifestly ``energy-dependent'' potential. We also recall that if the slab is made of lossy or active material, $\widehat\varepsilon(x)$ takes complex values. This shows that $v$ is in general a complex potential. Furthermore, because $\widehat\varepsilon(x)=1$ for $x<a_-$ and $x>a_+$, $v(x)$ vanishes for $x\notin[a_-,a_+]$. Hence, it is a finite-range potential.

\subsection{Short-range potentials and the transfer matrix}
\label{Sec2-2}

The finiteness of the range of the potential (\ref{optical}) stems from the fact that the permittivity of the system we consider is homogeneous outside the slab. We can relax this condition by considering the situation that $\varepsilon$ is a function of $x$ such that $\varepsilon(x)\to\varepsilon_\infty$ for $x\to\pm\infty$. In view of (\ref{e3}) and (\ref{optical}) this is equivalent to the requirement that $v(x)\to 0$ for $x\to\pm\infty$. We use the term ``scattering potential'' to refer to potentials satisfying this condition.

Because for a scattering potential the Schr\"odinger equation (\ref{sch-eq}) tends to that of a free particle, it is tempting to conclude that every solution of  (\ref{sch-eq}) satisfies the asymptotic boundary condition,
    \be
    \psi(x)\to\left\{\begin{array}{ccc}
    A_- e^{ikx}+B_- e^{-ikx} &{\rm for} & x\to-\infty,\\
    A_+ e^{ikx}+B_+ e^{-ikx} &{\rm for} & x\to+\infty,
    \end{array}\right.
    \label{asym1}
    \ee
for some $A_\pm,B_\pm\in\C$. This expectation turns out to fail in general. It is true provided that $v$ is a ``short-range potential.'' This means that there are positive real numbers $C$, $M$, and $\alpha$ such that $\alpha>1$ and
    \be
    |v(x)|\leq \frac{C}{|x|^\alpha}~~~\for~~~|x|\geq M.
    \label{short-range}
    \ee
This conditions means that, as $|x|$ tends to $\infty$, the potential decays to zero more rapidly than the Coulomb potential in one dimension. In this article, we will only be concerned with short-range potentials. A similar treatment of the long-range scattering potentials that satisfy (\ref{short-range}) for some $\alpha>1/2$ is provided in \cite{p159}.

A finite-range potential, $v:\R\to\C$, which vanishes outside some interval $[a_-,a_+]$, is a short-range potential; it certainly satisfies (\ref{short-range}) for $M=|a_+|+|a_-|$ and any positive real numbers $C$ and $\alpha$. Solutions $\psi$ of the Schr\"odinger equation (\ref{sch-eq}) for this potential fulfill (\ref{e2}) which implies (\ref{asym1}). To determine these solutions we need to solve (\ref{sch-eq}) in the interval $[a_-,a_+]$ and then match the result and its derivative to the ones given by (\ref{e2}) at $x=\pm a$.

Because (\ref{sch-eq}) is a second-order homogeneous linear differential equation, its general solution $\psi$ in $[a_-,a_+]$ is a linear combination of a pair of linearly-independent solutions, $\psi_1$ and $\psi_2$; there are $c_1,c_2\in\C$ such that $\psi(x)=c_1\psi_1(x)+c_2\psi_2(x)$. This in turn implies,
    \be
    \left[\begin{array}{c}
    \psi(x)\\
    \psi'(x)\end{array}\right]=\bcF(x)\left[\begin{array}{c}
    c_1\\
    c_2\end{array}\right]
    \label{e4}
    \ee
where $\bcF(x)$ is the Fundamental matrix \cite{boyce-diprima} given by
    \be
    \bcF(x):=\left[\begin{array}{cc}
    \psi_1(x) & \psi_2(x)\\
    \psi'_1(x) & \psi'_2(x)\end{array}\right].
    \label{e5}
    \ee
Notice that the determinant of $\bcF(x)$ is the Wronskian of $\psi_1$ and $\psi_2$, 	
	\be
	W[\psi_1(x),\psi_2(x)]:=\psi_1(x)\psi_2'(x)-\psi_2(x)\psi_1'(x).
	\label{Wronskian-def}
	\ee
Because $\psi_1$ and $\psi_2$ solve the Schr\"odinger equation, $\partial_xW[\psi_1(x),\psi_2(x)]=0$, i.e.,
$W[\psi_1(x),\psi_2(x)]$ does not depend on $x$. Because $\psi_1$ and $\psi_2$ are linearly independence, $W[\psi_1(x),\psi_2(x)]$ is a nonzero constant \cite{boyce-diprima}. These considerations show that $\bcF(x)$ is an invertible matrix for every $x\in[a_-,a_+]$.

It is easy to see that in view of (\ref{e4}),
    \be
    \left[\begin{array}{c}
    \psi(a_+)\\
    \psi'(a_+)\end{array}\right]=\bcF(a_+)\bcF(a_-)^{-1}\left[\begin{array}{c}
    \psi(a_-)\\
    \psi'(a_-)\end{array}\right].
    \label{e6}
    \ee
We can also use (\ref{e2}) to show that
    \be
    \left[\begin{array}{c}
    \psi(a_\pm)\\
    \psi'(a_\pm)\end{array}\right]=\bcF_0(a_\pm)
    \left[\begin{array}{c}
    A_\pm\\
    B_\pm\end{array}\right],
    \label{e7}
    \ee
where
    \be
    \bcF_0(x):=\left[\begin{array}{cc}
    e^{ikx} & e^{-ikx}\\
    ik e^{ikx} & -ik e^{-ikx}\end{array}\right].
    \label{F-zero}
    \ee
Substituting (\ref{e7}) in (\ref{e6}), we can relate the coefficients $A_\pm$ and $B_\pm$ appearing in (\ref{e2}) according to
    \be
    \left[\begin{array}{c}
    A_+\\
    B_+\end{array}\right]=\bM
    \left[\begin{array}{c}
    A_-\\
    B_-\end{array}\right],
    \label{M-def}
    \ee
where
    \begin{align}
    &\bM:=\bcG(a_+)\bcG(a_-)^{-1},
    &&\bcG(x):=\bcF_0(x)^{-1}\bcF(x).
    \label{M-G=}
    \end{align}

The matrix $\bM$ that relates the coefficients $A_\pm$ and $B_\pm$ according to (\ref{M-def}) is called the ``transfer matrix'' of the potential $v$. Unlike the fundamental matrix $\bcF(x)$, it is independent of the choice of the solutions $\psi_1$ and $\psi_2$. To see this, let $\breve\psi_1$ and $\breve\psi_2$ be another pair of linearly-independent solutions of (\ref{sch-eq}) in $[a_-,a_+]$. Because these are linear combinations of $\psi_1$ and $\psi_2$, there is an invertible $2\times 2$ matrix $\bL$ such that
    \be
    \big[\!\!\begin{array}{cc}
    \breve\psi_1(x) &   \breve\psi_2(x)\end{array}\!\!\big]=
    \big[\!\!\begin{array}{cc}
    \psi_1(x) & \psi_2(x)\end{array}\!\!\big]\,\bL.
    \label{e10}
    \ee
Now, let $\breve\bcF(x)$, $\breve\bcG(x)$, and $\breve\bM$ be the analogs of $\bcF(x)$, $\bcG(x)$, and $\bM$ that are calculated using the solutions $\breve\psi_1$ and $\breve\psi_2$. In light of (\ref{e5}), (\ref{F-zero}), (\ref{M-G=}) and (\ref{e10}), we see that $\breve\bcF(x)=\bcF(x)\bL$ and $\breve\bcG(x)=\bcG(x)\bL$. These in turn imply
    \[\breve\bM:=\breve\bcG(a_+)\breve\bcG(a_-)^{-1}=\bcG(a_+)\bL[\bcG(a_-)\bL]^{-1}=\bcG(a_+)\bcG(a_-)^{-1}=\bM.\]

Another useful property of the transfer matrix is that it has a unit determinant;
    \be
    \det\bM=1.
    \ee
This follows from (\ref{M-G=}), $\det\bcF_0(a_\pm)=-2ik$, and  $\det\bcF(a_\pm)=W[\psi_1,\psi_2]$.

\begin{figure}
    \begin{center}
    \includegraphics[scale=0.3]{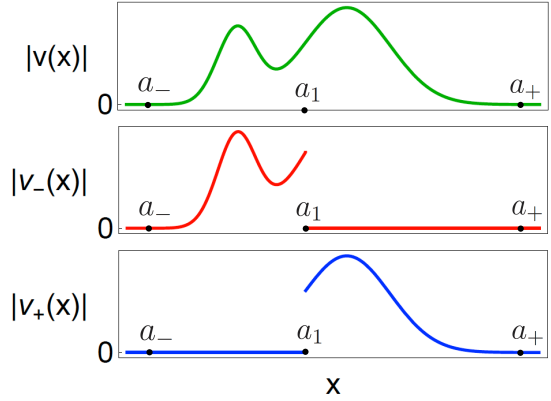}
    \end{center}
    \vspace{-12pt}
    \caption{Plots of $|v(x)|$ and $|v_\pm(x)|$ for a finite range potential $v$ and its truncations $v_\pm$ as defined by (\ref{e11}). The supports of $v$, $v_-$, and $v_+$ are respectively the intervals $[a_-,a_+]$, $I_-:=[a_0,a_1[$, and $I_+:=[a_1,a_+]$.}
    \label{fig2}
    \end{figure}

Next, consider dividing the interval $[a_-,a_+]$ into two pieces, $I_-:=[a_-,a_1[$ and $I_+:=[a_1,a_+]$, and let $v_\pm:\R\to\C$ be the truncations of $v$ that are given by 
    \be
    v_\pm(x):=\left\{\begin{array}{ccc}
    v(x) & \for & x\in I_\pm,\\
    0 & \for & x\notin I_\pm.
    \end{array}\right.
    \label{e11}
    \ee
Fig.~\ref{fig2} shows the graphs of $|v(x)|$ and $|v_\pm(x)|$ for a generic example of a finite-range potential $v$ and intervals $I_\pm$.

Because the support\footnote{The support of a potential $v:\R\to\C$ is the smallest closed interval in $\R$ outside which $v(x)$ vanishes.} of $v_\pm$ lies in $I_\pm$ and $I_\pm\subset [a_-,a_+]$, $v_\pm$ are finite-range potentials, and the support of $v_-$ is to the left of that of $v_+$. Therefore we can associate to each of them a transfer matrix $\bM_\pm$. In view of (\ref{e11}) and the above construction of the transfer matrix, we have
    \begin{align}
    &\bM_-=\bcG(a_1)\bcG(a_-)^{-1},
    &&\bM_+=\bcG(a_+)\bcG(a_1)^{-1}.
    \label{e12}
    \end{align}
A straightforward consequence of (\ref{M-G=}) and (\ref{e12}) is
    \be
    \bM=\bM_+\bM_-.
    \label{compose-1}
    \ee
This equation states that we can determine the transfer matrix of the potential $v$ by dissecting it into two pieces with disjoint supports.\footnote{See Ref.~\cite{bookchapter} for an alternative derivation of Eq.~(\ref{compose-1}).}

It is clear that we can apply the above procedure to $v_+$, i.e., dissect it into potentials with smaller support. Repeating this $n$ times, we obtain an increasing sequence of numbers $a_0,a_1,a_2,\cdots,a_n$, with $a_0:=a_-$ and $a_n:=a_+$, the intervals
   $I_1:=[a_0,a_1[$, $I_2:=[a_1,a_2[$, $\cdots$ $I_{n-1}:=[a_{n-2},a_{n-1}[$, and $I_n:=[a_{n-1},a_n]$, and the potentials,
    \be
    v_j(x):=\left\{\begin{array}{ccc}
    v(x) & \for & x\in I_j,\\
    0 & \for & x\notin I_j,
    \end{array}\right.
    \label{e11-j}
    \ee
with transfer matrices $\bM_j$ and $j\in\{1,2,\cdots,n\}$, such that $v=v_1+v_2+\cdots v_n$ and
    \be
    \bM=\bM_n\bM_{n-1}\cdots\bM_1.
    \label{compose}
    \ee
This relation is the celebrated ``composition property'' of the transfer matrix, which has motivated its introduction in the 1940's \cite{jones-1941} and made it into a useful tool for dealing with a variety of physics problems \cite{abeles,thompson,yeh,abrahams-1980,ardos-1982,pendry-1982,levesque,hosten,sheng-1996,pereyra,griffiths-2001,yeh-book}.

We can generalize the above description of the transfer matrix for general short-range potentials $v:\R\to\C$ simply by letting $a_\pm$ tend to $\pm\infty$. In particular, we can identify the transfer matrix for such a potential with a $2\times 2$ matrix satisfying (\ref{M-def}), where $A_\pm$ and $B_\pm$ are the coefficients entering the asymptotic expression (\ref{asym1}) for the solutions of the Schr\"odinger equation (\ref{sch-eq}). The global existence of the solutions of this equation implies that (\ref{M-def}) determines $\bM$ in a unique manner provided that we demand that it does not depend on $A_-$ and $B_-$, \cite{epjp-2019}.

\subsection{Scattering by a short-range potential in one dimension}
\label{Sec2-3}

A standard scattering setup consists of three ingredients:
    \begin{enumerate}
    \item the source of the incident wave,
    \item the scatterer, which we model using a scattering potential, and
    \item the detector(s) observing the scattered wave.
    \end{enumerate}
We assume that there is a single source for the incident wave located at a pre-determined position far from the origin of the adopted coordinate system and that the wave interacts with the scatterer in a region containing the origin.

In one dimension, the source of the incident wave is placed at either $x=-\infty$ or $x=+\infty$. Suppose that the source is located at $x=-\infty$. Then it emits a wave that after interaction with the potential is partly reflected back towards $x=-\infty$ and partly transmitted through the interaction region and continues its propagation toward $x=+\infty$. This corresponds to a solution of the relevant wave equation that is right-going at $x=+\infty$. We call it a ``left-incident wave.'' Similarly, if the source is located at $x=+\infty$, it will emit a wave that is left-going at $x=-\infty$. Therefore we call it a ``right-incident wave.''

Now, consider a time-harmonic wave (\ref{time-harmonic}) with $\psi$ solving the time-independent Schr\"odinger equation (\ref{sch-eq}) for a short-range potential $v:\R\to\C$. Then, according to (\ref{asym1}), the left- and right-incident waves are respectively given by the solutions, ${\psi^l}$ and ${\psi^r}$, of (\ref{sch-eq}) that satisfy
    \begin{align}
    &{\psi^l}(x)\to \left\{\begin{array}{ccc}
    A^l_-e^{ikx}+B^l_-e^{-ikx} &\for& x\to-\infty,\\
    A^l_+e^{ikx} &\for&x\to+\infty,\end{array}\right.
    \label{psi-left}\\[6pt]
    &{\psi^r}(x)\to \left\{\begin{array}{ccc}
    B^r_-e^{-ikx} &\for& x\to-\infty,\\
    A^r_+e^{ikx}+B^r_+e^{-ikx} &\for&x\to+\infty,\end{array}\right.
    \label{psi-right}
    \end{align}
where we use the subscripts ``$l$'' and ``$r$'' to distinguish between the coefficients $A_\pm$ and $B_\pm$ of (\ref{asym1}) for ${\psi^l}$ and ${\psi^r}$. Notice that $B^l_+=A^r_-=0$. This follows from the fact that ${\psi^l}$ is right-going at $x=+\infty$, and ${\psi^r}$ is left-going at $x=-\infty$. 

\begin{figure}
    \begin{center}
    \includegraphics[scale=.25]{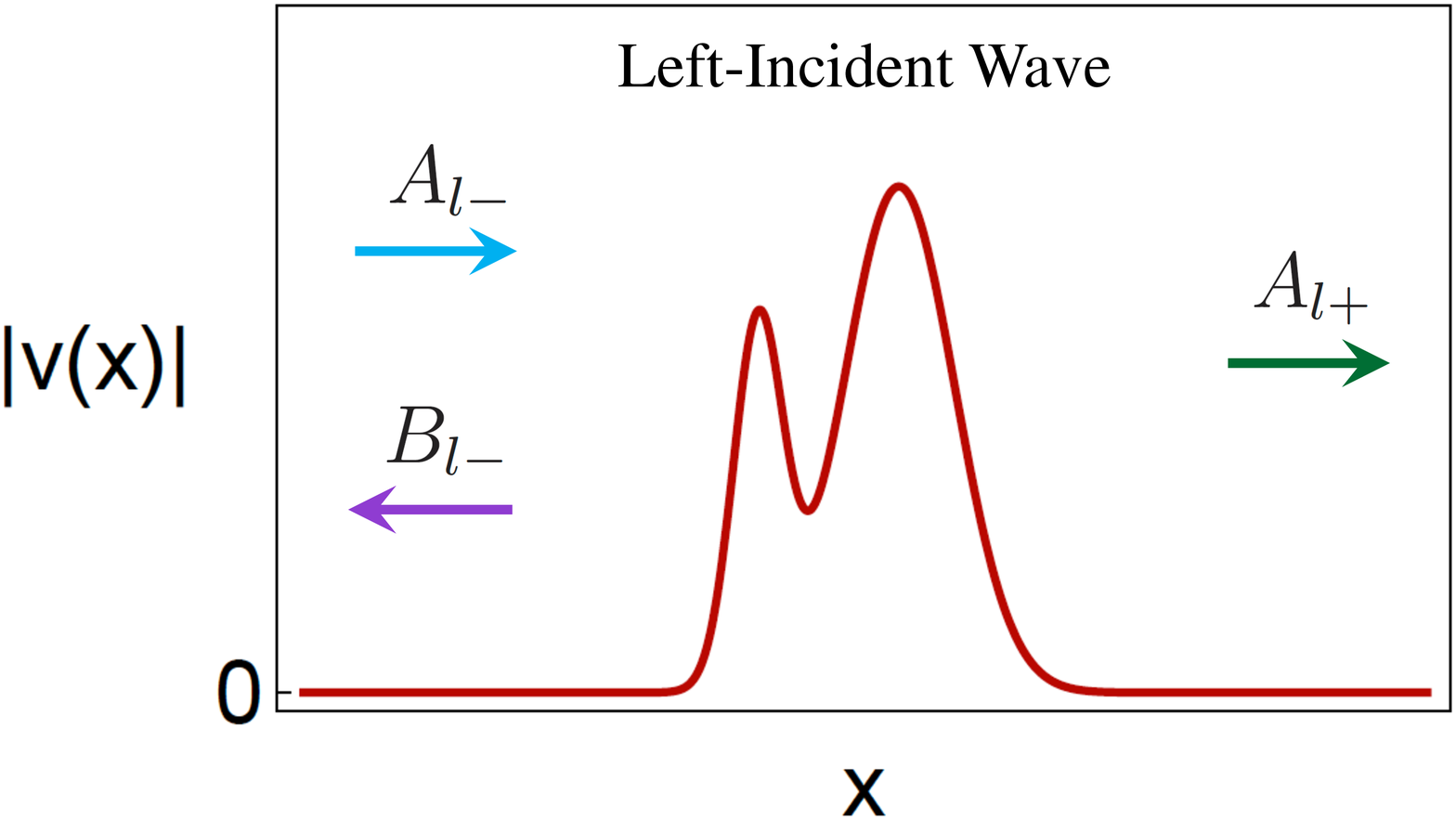}
    \includegraphics[scale=.25]{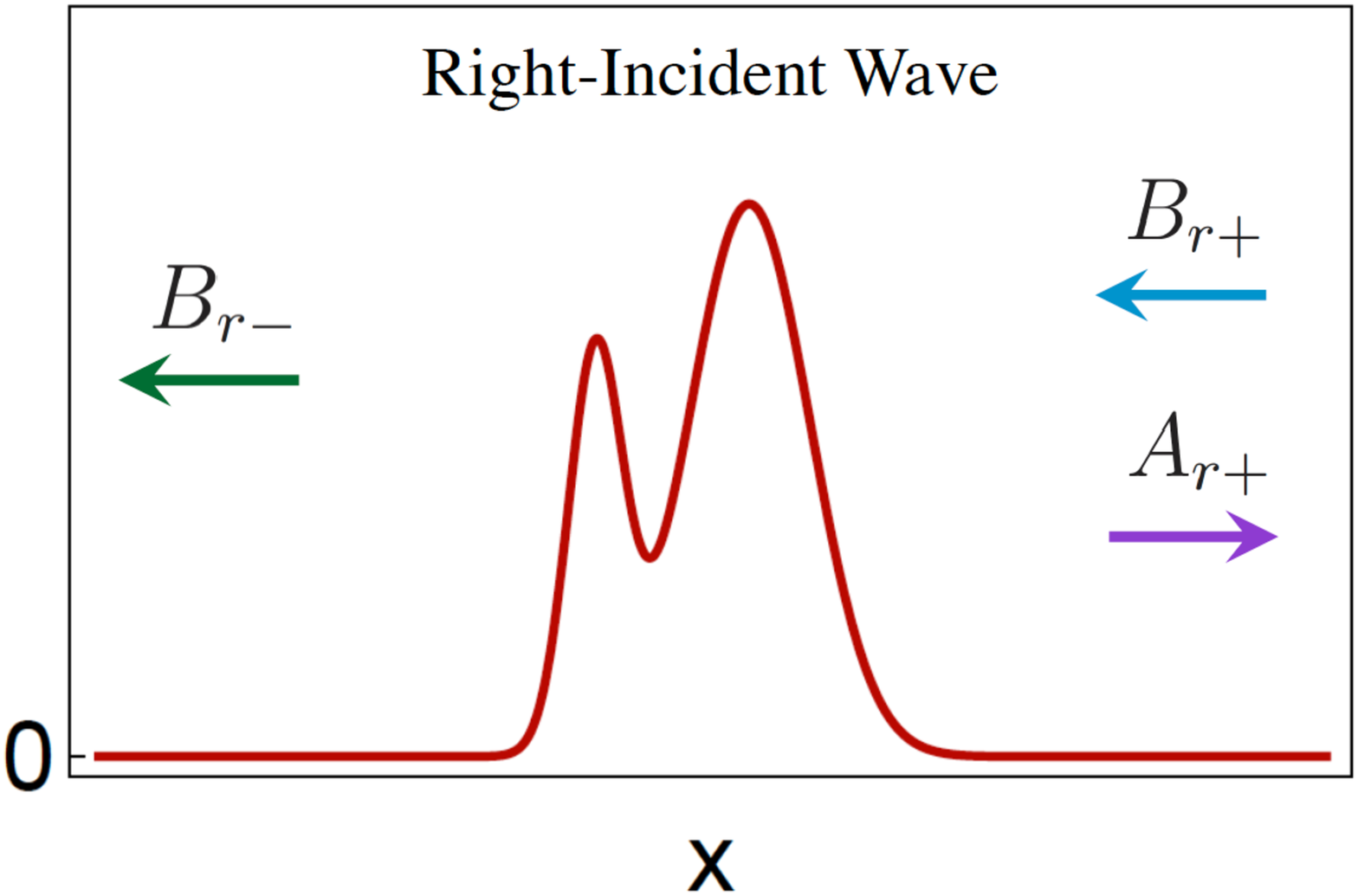}
    \end{center}
    \vspace{-12pt}
    \caption{Schematic view of the graph of $|v(x)|$ for a short-range potential and its scattering effects on   
    left-incident and right-incident waves. Blue, purple, and green arrows represent the incident, reflected, and transmitted waves, respectively.}
    \label{fig3}
    \end{figure}

Now,  let us consider a left-incident wave $\psi^l$. Because $A^l_-$ is the coefficient of the right-going component of ${\psi^l}$ at $x=-\infty$, we identify it with the complex amplitude of the incident wave emitted by the source. This in turn suggests that $B^l_-$ and $A^l_+$ are the amplitudes of the reflected and transmitted waves. See Fig.~\ref{fig3}.
It is customary to call the ratio of the intensity of the reflected (respectively transmitted) wave to the intensity of the incident wave the reflection (respectively transmission) coefficient \cite{Griffiths}. According to this terminology the ``reflection and transmission coefficients'' for a left-incident wave are respectively given by $|R^l|^2$ and $|T^l|^2$, where
    \begin{align}
    &R^l:=\frac{B^l_-}{A^l_-},
    &&T^l:=\frac{A^l_+}{A^l_-}.
    \label{e21}
    \end{align}
We call these the ``left reflection and transmission amplitudes,'' respectively. The reader should be warned that in most standard physics texts the authors use the lower case letters $r$ and $t$ for the reflection and transmission amplitudes and reserve $R$ and $T$ for the reflection and transmission coefficients. Here we employ the mathematicians' convention of using capital letters for the left reflection and transmission amplitudes and not introducing special symbols for the left reflection and transmission coefficients. The is also the convention adopted in the review article \cite{muga-2004}.

Repeating the argument of the preceding paragraph for the right-incident wave $\psi^r$, we identify $B^r_+$, $B^r_-$, and $A^r_-$ respectively with the complex amplitudes of the incident, reflected, and transmitted waves. This in turn justifies the following definitions of the ``right reflection and transmission amplitudes.''
    \begin{align}
    &R^r:=\frac{A^r_+}{B^r_+},
    &&T^r:=\frac{B^r_-}{B^r_+}.
    \label{e22}
    \end{align}

A rather surprising fact regarding transmission amplitudes for the left- and right-incident waves is that they coincide \cite{Faddeev-yakubovski,Ahmed-2001}. This phenomenon, which is known as the ``transmission reciprocity,'' follows from the fact that the Wronskian of solutions of the Schr\"odinger equation (\ref{sch-eq}) does not depend on $x$. If we use (\ref{Wronskian-def}), (\ref{psi-left}), and (\ref{psi-right}) to compute the Wronskian of $\psi^l$ and $\psi^r$ in the limit $x\to\pm\infty$, we find
    \begin{align}
    &\lim_{x\to-\infty}W[{\psi^l}(x),{\psi^r}(x)]=-2ikA^l_-B^r_-=-2ikA^l_-B^r_+T^r,
    \label{e23}\\
    &\lim_{x\to+\infty}W[{\psi^l}(x),{\psi^r}(x)]=-2ikA^l_+ B^r_+=-2ikA^l_-B^r_+T^l,
    \label{e24}
    \end{align}
where we have also benefitted from (\ref{e21}) and (\ref{e22}). Because $A^l_-$ and $B^r_+$ are respectively the amplitudes of the left- and right-incident waves, they do not vanish. This observation together with the $x$-independence of $W[{\psi^l}(x),{\psi^r}(x)]$ and Eqs.~(\ref{e23}) and (\ref{e24}) imply $T^l=T^r$. In the following we will drop the superscript $l$ and $r$ and use $T$ to denote $T^l$ and $T^r$, i.e.,
    \be
    T:=T^l=T^r.
    \label{TTT}
    \ee
    In view of (\ref{e21}), (\ref{e22}), and (\ref{TTT}), we can write (\ref{psi-left}) and (\ref{psi-right}) in the form,
    	 \begin{align}
    &{\psi^l}(x)\to A^l_-\times 
    \left\{\begin{array}{ccc}
    e^{ikx}+R^l\, e^{-ikx} &\for& x\to-\infty,\\
    T\,e^{ikx} &\for&x\to+\infty,\end{array}\right.
    \label{psi-left-RT}\\[3pt]
    &{\psi^r}(x)\to B^r_+\times \left\{\begin{array}{ccc}
    T\,e^{-ikx} &\for& x\to-\infty,\\
    e^{-ikx}+R^r\,e^{ikx} &\for&x\to+\infty.\end{array}\right.
    \label{psi-right-RT}
    \end{align}

Solving the scattering problem for a short-range potential means finding the reflection and transmission amplitudes, $R^{l/r}$ and $T$, as functions of the wavenumber $k$ and the physical parameters entering the expression for the potential. An important property of the transfer matrix $\bM$ is that it carries the complete information about $R^{l/r}$ and $T$. To see this, we examine the implications of (\ref{M-def}) for the left- and right-incident waves $\psi^{l/r}$. 

Setting $\psi={\psi^l}$ and $\psi={\psi^r}$ in (\ref{M-def}) and noting that  $B^l_+=A^r_-=0$, we have   \begin{align}
    &\left[\begin{array}{c}
    A^l_+ \\ 0\end{array}\right]=
    \left[\begin{array}{cc}
    M_{11} & M_{12}\\
    M_{21} & M_{22}\end{array}\right]
    \left[\begin{array}{c}
    A^l_- \\ B^l_-\end{array}\right],
    &&\left[\begin{array}{c}
    A^r_+ \\ B^r_+\end{array}\right]=
    \left[\begin{array}{cc}
    M_{11} & M_{12}\\
    M_{21} & M_{22}\end{array}\right]
    \left[\begin{array}{c}
    0 \\ B^r_-\end{array}\right],
    \label{e25}
    \end{align}
where $M_{ij}$ are the entries of $\bM$. This gives rise to a system of four independent linear equations for $M_{ij}$. Solving this system and using (\ref{e21}), (\ref{e22}), and (\ref{TTT}), we obtain
    \begin{align}
    &M_{11}=T-\frac{R^lR^r}{T},
    &&M_{12}=\frac{R^r}{T},
    &&M_{21}=-\frac{R^l}{T},
    &&M_{22}=\frac{1}{T}.
    \label{Mij=}
    \end{align}
These relations imply
    \be
    \det\bM=1,
    \label{det=1}
    \ee
and
    \begin{align}
    &R^l=-\frac{M_{21}}{M_{22}},
    &&R^r=\frac{M_{12}}{M_{22}},
    &&T=\frac{1}{M_{22}}.
    \label{RRT=}
    \end{align}
This establishes the equivalence of the solution of the scattering problem for a short-range potential and the determination of its transfer matrix.

The fact that we can solve the scattering problem for a finite-range potential $v$ by evaluating its transfer matrix signifies the importance of the composition property (\ref{compose}). This is because with the help of this property, we can reduce the calculation of the transfer matrix for $v$, and hence the treatment of its scattering problem, to that of its truncations $v_1,v_2,\cdots,v_n$ given by (\ref{e11-j}). By increasing $n$, we can shrink the supports $I_j$ of $v_j$ and approximate them by rectangular barrier potentials,
    \[\breve v_j(x)=\left\{\begin{array}{ccc}
    \fz_j &\for& x\in I_j,\\
    0 &\for& x\notin I_j,\end{array}\right.
    \quad\quad\quad\fz_j:=v(\mbox{$\frac{1}{2}(a_{j-1}+a_j)$}),\]
 whose transfer matrix is known.\footnote{We give a derivation of the transfer matrix of rectangular barrier potentials in Subsection~\ref{Sec3-3}.} In this way, we can obtain an approximate expression for the transfer matrix of $v$ as the product of $n$ known $2\times 2$ matrices. We can improve the accuracy of this approximation by selecting a finer slicing of the support of $v$, i.e., increasing $n$. This however leads to the problem of dealing with the numerical errors associated with the multiplication of a large number of matrices.

 \subsection{Multi-delta-function and locally periodic potentials}
 \label{Sec2-4}

An instructive evidence for the effectiveness of the transfer-matrix method is its application in dealing with multi-delta function potentials \cite{reading-1972,Hakke-1981,Besprosvany-2001,pra-2012},
    \be
    v(x)=\sum_{j=1}^n\fz_{j}\,\delta(x-a_j),
    \label{multi-delta}
    \ee
where $n$ is a positive integer, $\fz_j$ and $a_j$ are respectively complex and real parameters, $\fz_j\neq 0$ for all $j\in\{1,2,\cdots,n\}$, and $a_1<a_2<\cdots<a_n$. In view of the composition property of the transfer matrix, we can express the transfer matrix of this potential in the form (\ref{compose}), where $\bM_j$ is the transfer matrix for the delta-function potential \cite{jpa-2006b},
    \be
    v_j(x)=\fz_j\,\delta(x-a_j).
    \label{delta}
    \ee
In other words, the composition property of the transfer matrix reduces the solution of the scattering problem for the multi-delta-function potential to that of a single delta-function potential.

To determine the transfer matrix $\bM_j$ of (\ref{delta}), we need to solve the corresponding Schr\"odinger equation (\ref{sch-eq}). This is done in almost every standard textbook in quantum mechanics for the case that $\fz_j$ is real. The same treatment applies for complex $\fz_j$. It involves identifing the Schr\"odinger equation (\ref{sch-eq}) for (\ref{delta}) with
    \begin{align}
    &-\psi''(x)=k^2\psi(x)~~~\for~~~x\neq a_j,
    \label{delta1}\\
    &\psi(a_j^+)=\psi(a_j^-),\quad\quad\psi'(a_j^+)=\psi'(a_j^-)+\fz\,\psi(a_j^-),
    \label{delta2}
    \end{align}
where, for every function $f:\R\to\C$, $f(a_j^-)$ and $f(a_j^+)$ respectively stand for the left and right limit of $f(x)$ as $x\to a_j$;
	\[f(a_j^\pm):=\lim_{x\to a_j^\pm}f(x).\]
Eq.~(\ref{delta1}) implies
    \be
    \psi(x)=A_\pm e^{ikx}+B_\pm e^{-ikx}~~~\for~~~\pm(x-a_j)>0,
    \label{delta3}
    \ee
where $A_\pm$ and $B_\pm$ are complex coefficients. Using (\ref{delta3}), we can express (\ref{delta2}) as
    \bea
    e^{ika_j} A_++e^{-ika_j}B_+&=&e^{ika_j} A_-+e^{-ika_j}B_-,\nn\\
    ik\left(e^{ika_j} A_+-e^{-ika_j}B_+\right)&=&e^{ika_j} \left(ik+\fz_j\right)A_-
    -e^{-ika_j}\left(ik-\fz_j\right)B_-.\nn
    \eea
These in turn imply
    \be
    \bcF_0(a_j)\left[\begin{array}{c}
    A_+\\
    B_+\end{array}\right]=\bcZ_j\,\bcF_0(a_j)
    \left[\begin{array}{c}
    A_-\\
    B_-\end{array}\right],
    \label{delta4}
    \ee
where $\bcF_0(x)$ is defined in (\ref{F-zero}), and
    \[\bcZ_j:=\left[\begin{array}{cc}
    1&0\\
    \fz_j&1\end{array}\right].\]
Comparing (\ref{delta4}) with (\ref{M-def}), we infer that
    \be
    \bM_j=\bcF_0(a_j)^{-1}\bcZ_j\bcF_0(a_j)=
    \frac{1}{2k}
    \left[\begin{array}{cc}
    2k-i\fz_j & -i \fz_j\, e^{-2i a_j k}\\
    i \fz_j\, e^{2i a_j k} & 2k+i\fz_j\end{array}\right].
    \label{M-delta}
    \ee
It is also useful to introduce,
    \begin{align}
    &\bT(x):=e^{ikx\bsigma_3}=\left[\begin{array}{cc}
    e^{ikx} & 0\\
    0 & e^{-ikx}\end{array}\right],
    &&\mathring\bM_j:=\frac{1}{2k}\left[\begin{array}{cc}
    2k-i\fz_j & -i \fz_j \\
    i \fz_j & 2k+i\fz_j\end{array}\right],
    \label{TM=}
    \end{align}
where $\bsigma_3$ is the third of the Pauli matrices:
	\begin{align*}
	&\bsigma_1:=\left[\begin{array}{cc}
	0 & 1\\
	1 & 0\end{array}\right],
	&&\bsigma_2:=\left[\begin{array}{cc}
	0 & -i\\
	i & 0\end{array}\right],
	&&\bsigma_3:=\left[\begin{array}{cc}
	1 & 0\\
	0 & -1\end{array}\right].
	\end{align*}
Eqs.~(\ref{TM=}) allow use to express the right-hand side of (\ref{M-delta}) as $ \bT(a_j)^{-1}\mathring\bM_j\bT(a_j)$ and lead to
    \be
    \bM_j= \bT(a_j)^{-1}\mathring\bM_j\bT(a_j)=\bT(-a_j)\,\mathring\bM_j\bT(a_j).
    \label{M-delta2}
    \ee
Clearly $\mathring\bM_0$ is the transfer matrix for the potential $\mathring v_j(x):=\fz_j\delta(x)$.

Having obtained the transfer matrix of (\ref{delta}), we can use (\ref{compose}) and (\ref{M-delta2}) to express the transfer matrix of the multi-delta-function potential as
    \be
    \bM=    \bT(a_n)^{-1}\mathring\bM_n \bT(a_n-a_{n-1})\mathring\bM_{n-1}\bT(a_{n-1}-a_{n-2})
    \cdots \bT(a_2-a_1)\mathring\bM_1\bT(a_1).
    \label{M-multi-delta}
    \ee

If the delta-functions $v_j$ contributing to (\ref{multi-delta}) are equally spaced and have the same strength, i.e., there is some length scale $\ell$ such that,
    \begin{align}
    &a_{j+1}-a_j=\ell,
    &&\fz_j=\fz_1,
    \label{delta-locally-periodic}
    \end{align}
then $\bT(a_{j+1})=\bT(a_1+j\ell)$, $\mathring\bM_j=\mathring\bM_1$, and (\ref{M-multi-delta}) becomes
    \be
    \bM=\bT(a_n)^{-1}\left[\mathring\bM_1\bT(\ell)\right]^n\,\bT(\ell)^{-1}\bT(a_1)=
    \bT(-a_1-n\ell+\ell)\,\bL^n\,\bT(a_1-\ell),
    \label{M-multidelta1}
    \ee
where
    \be
    \bL:=\mathring\bM_1\bT(\ell)=\bT(a_1)\bM_1\bT(\ell-a_1).
    \label{L=}
    \ee
Eq.~(\ref{M-multidelta1}) reduces the determination of $\bM$ to the computation of $\bL^{n}$. There is a well-known method based on the Cayley-Hamilton theorem and properties of Chebyshev polynomials  \cite{griffiths-2001} that allows for computing the positive integer powers of $2\times 2$ matrices with unit determinant such as $\bL$. See also \cite{pereyra}. In Appendix, we outline another more direct approach which makes use of the canonical Jordan form of complex matrices and arrives at the same conclusion~\cite{yeh,jones-talk}, namely
    \be
    \bL^n=U_{n+1}(\gamma)\bL-U_n(\gamma)\bI,
    \label{Ln=gen-X}
    \ee
where, for all $n\in\Z^+$ and $z\in\C$,
    \bea
    &&U_n(z):=\left\{\begin{aligned}
    &\displaystyle\frac{\sin (n-1)z}{\sin z} &&\mbox{when $z/\pi$ is not an integer},\\%[6pt]
    &(-1)^{\displaystyle nz/\pi}(n-1) && \mbox{when $z/\pi$ is an integer},
    \end{aligned}\right.\nn\\[6pt]
    &&\gamma:=\cos^{-1}\left(\mbox{\large$\frac{1}{2}$}\,{\rm tr}\,\bL\right)=
    \cos^{-1}\left[\mbox{\large$\frac{1}{2}$}\left(e^{-i\ell k} M_{1_{11}}+e^{i\ell k}M_{1_{22}}\right)\right],\nn
    \eea
and $M_{1_{ij}}$ are the entries of $\bM_1$.

If we substitute (\ref{Ln=gen-X}) in (\ref{M-multidelta1}) and use (\ref{L=}) to express the result in terms of $\bM_1$, we obtain the following expression for the transfer matrix of multi-delta-function potentials satisfying (\ref{delta-locally-periodic}).
    \be
    \bM=U_{n+1}(\gamma)\bT\big((1-n)\ell\big)\bM_1-U_n(\gamma)\bT(-n\ell).
    \label{M-locally-periodic}
    \ee
In the remainder of this section we show that this formula has universal validity for every locally period potential with period $\ell$.

\begin{figure}
    \begin{center}
    \includegraphics[scale=0.60]{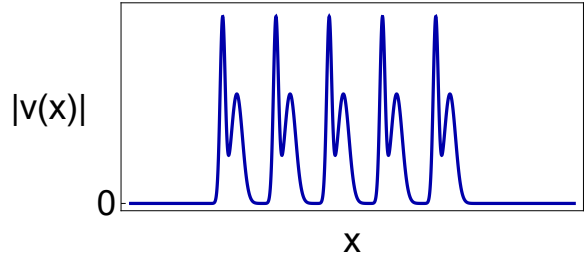}
    \end{center}
    \vspace{-12pt}
    \caption{Graph of $|v(x)|$ for a typical locally periodic potential $v(x)$.}
    \label{fig4}
    \end{figure}

A finite-range potential $v:\R\to\C$ is said to be ``locally period,'' if we can generate it by a period extension of a finite-range potential $v_1:\R\to\C$ with a smaller support. This means that $v$ satisfies,
    \be
    v(x)=\sum_{j=1}^n v_1(x-j\ell+\ell),
    \label{locally-periodic}
    \ee
where $\ell$ is a positive real parameter not smaller than the length
of the support of $v_1$, i.e., if $[a_{1-},a_{1+}]$ is the support
of $v_1$, then $\ell\geq a_{1+}-a_{1-}$. The multi-delta-function
potentials (\ref{multi-delta}) fulfilling
(\ref{delta-locally-periodic}) are particular examples of locally
periodic potentials, where $v_1(x)=\fz_1\delta(x-a_1)$,
$a_{1\pm}=a_1$, and $\ell$ can take any positive real value.
Fig.~\ref{fig4} shows the graph of $|v(x)|$ for a typical locally
periodic potential.%

By universal validity of (\ref{M-locally-periodic}), we mean that it gives the transfer matrix of every locally periodic potential $v$ if it has a period $\ell$ and $\bM_1$ labels the transfer matrix of its generator, namely $v_1$. The proof of this statement relies on the transformation property of the transfer matrix under space translations.

Suppose that $v:\R\to\C$ is a short-range potential. A space translation, $x\to \check x:=x-a$, maps $v$ to the short-range potential $\check v$ that is given by $\check v(x):=v(\check x)=v(x-a)$. This implies that if $\psi(x)$ is a solution of the Schr\"odinger equation~(\ref{sch-eq}) for $v(x)$, then $\check\psi(x):=\psi(x-a)$ is a solution of this equation for $\check v(x)$. Using this relation and (\ref{asym1}), we find that under this translation the coefficients $A_\pm$ and $B_\pm$ of its asymptotic expression for $\psi(x)$ transform according to $A_\pm\to \check A_\pm:=e^{-ika}A_\pm$ and $B_\pm\to\check B_\pm:= e^{ika}B_\pm$. In light of these relations and the fact that the transfer matrix $\check\bM$ of the translated potential $\check v(x):=v(x-a)$ satisfies
    \[\left[\begin{array}{c}
    \check A_+\\ \check B_+\end{array}\right]=\check\bM
    \left[\begin{array}{c}
    \check A_-\\ \check B_-\end{array}\right],\]
we find the following transformation rule for the transfer matix.
    \be
    \bM\to\check\bM=\bT(a)^{-1}\bM\,\bT(a),
    \label{M-translation}
    \ee
where $\bT(x)$ is defined in (\ref{TM=}).

Now, consider a generic locally periodic potential (\ref{locally-periodic}) with generator $v_1$ and period $\ell$, and let $a_j:=(j-1)\ell$. It is easy to see that we can express it in the form, $v=v_1+v_2+\cdots+v_n$, where $v_j:\R\to\C$ with $j\in\{2,3,\cdots,n\}$, are obtained from $v_1$ via the translation $x\to x-a_j$, i.e., $v_j(x):=v_1(x-a_j)$. This relation identifies the support of $v_j$ with the interval $[a_{j-},a_{j+}]$, where $a_{j\pm}:=a_{1\pm}+a_j=a_{1\pm}+(j-1)\ell$. In particular, because $\ell\geq a_{1+}-a_{1-}$, the support of $v_{j-1}$ is to the left of that of $v_j$. We can therefore use the composition property of transfer matrices to express the transfer matrix of $v$ in the form,
    \be
    \bM=\bM_n\bM_{n-1}\cdots\bM_1,
    \label{compose-3}
    \ee
 where $\bM_j$ denotes the transfer matrix of $v_j$.

Because $v_j$ is related to $v_1$ by the translation $x\to x-a_j$, $\bM_j=\bT(a_j)^{-1}\bM_1\,\bT(a_j)$. Substituting this equation in (\ref{compose-3}), we find
    \be
    \bM=\bT(a_n)^{-1}\left[\bM_1\bT(\ell)\right]^n\,\bT(\ell)^{-1}=
    \bT(-n\ell+\ell)\,\bL^n\,\bT(-\ell),
    \label{M-locally-periodic-gen}
    \ee
where $\bL:=\bM_1\bT(\ell)$. Because $a_1=0$, (\ref{M-locally-periodic-gen}) coincides with (\ref{M-multidelta1}). This allows us to conclude that (\ref{M-locally-periodic}) gives the transfer matrix of the locally period potential (\ref{locally-periodic}). The only difference is that now $\bM_1$ is no longer given by (\ref{M-delta}), and we must find a way to compute it. In practice one uses a numerical scheme to perform this computation. This does not, however, overshadow the practical significance of (\ref{M-locally-periodic}). For typical multi-layer systems modeled by locally period potentials, the number $n$ of layers is quite large \cite{yeh-book}. This means that the period of the potential is much smaller than the size of its support; $\ell\ll L$. As a result, it is much more reliable and cost effective to compute $\bM_1$ numerically and use (\ref{M-locally-periodic}) to determine $\bM$ than to perform a direct numerical calculation of $\bM$.

\subsection{Jost solutions and analyticity of the transfer matrix}
\label{Sec2-5}

In Subsec.~\ref{Sec2-3} we show that the solution of the scattering problem for a short-range potential means the determination of its reflection and transmission amplitudes or equivalently its transfer matrix as functions of the incident wavenumber $k$. In this section we elaborate on the analyticity properties of these quantities. Therefore, we make their $k$-dependence explicit.

A straightforward consequence of (\ref{RRT=}) is that $T(k)$ does not vanish for any $k\in\R^+$. This means that perfect absorption of a left- or right-incident wave by a short-range scattering potential is forbidden. Because $T(k)\neq 0$,  (\ref{e21}) and (\ref{e22}) imply that $A^l_+(k)\neq 0$ and $B^r_-(k)\neq 0$. Dividing $\psi^{l/r}$ by these quantities, we obtain a pair of solutions of the Schr\"odinger equation~(\ref{sch-eq}) that are given by
    \begin{align}
    &\psi_+(x,k):=\frac{{\psi^l}(x,k)}{A^l_+(k)},
    &&\psi_-(x,k):=\frac{{\psi^r}(x,k)}{B^r_-(k)}.
    \label{jost}
    \end{align}
These are called the ``Jost solutions." According to (\ref{psi-left}) -- (\ref{e22}) and (\ref{RRT=}), they satisfy
    \begin{align}
    &\psi_+(x,k)\to \left\{\begin{array}{ccc}
    M_{22}(k)\, e^{ikx}-M_{21}(k)\,e^{-ikx} &\for& x\to-\infty,\\
    e^{ikx} &\for&x\to+\infty,\end{array}\right.
    \label{psi+}\\[6pt]
    &\psi_-(x,k)\to \left\{\begin{array}{ccc}
    e^{-ikx} &\for& x\to-\infty,\\
    M_{12}(k)\,e^{ikx}+M_{22}(k)\,e^{-ikx} &\for&x\to+\infty.\end{array}\right.
    \label{psi-}
    \end{align}
It is clear from these equations that we can use the knowledge of the Jost solutions to determine 
the transfer matrix and solve the scattering problem.

The independent variable $k$ entering the above equations represents the wavenumber for the incident wave which is a positive real quantity. However, it proves useful to view $k$ as taking values in the complex plane. A basic mathematical result with direct implications in scattering theory is that if a scattering potential $v:\R\to\C$ satisfies
    \be
    \int_{-\infty}^\infty dx\,\left(1+|x|\right)|v(x)|<\infty,
    \label{Faddeev}
    \ee
then for each $x\in\R$ the corresponding Jost solutions $\psi_\pm(x,k)$ are analytic (holomorphic) functions of $k$ in the upper half-plane, $\{k\in\C~|~\IM(k)>0\}$, and that they are continuous functions of $k$ in $\{k\in\C~|~\IM(k)\geq 0\}\setminus\{0\}$, where ``$\IM$'' stands for the imaginary part of its argument, \cite{kemp}.

The inequality (\ref{Faddeev}) is known as the Faddeev condition. A standard mathematical symbol for potentials satisfying the Faddeev condition is $L^1_1(\R)$. More generally we have the classes of potentials,
    \[ L^1_\sigma(\R):=\left\{ v:\R\to\C~\Big|~\int_{-\infty}^\infty dx\,\left(1+|x|^\sigma\right)|v(x)|<\infty\right\},\]
with $\sigma\in[0,\infty)$. The larger the value of $\sigma$ is, the faster the potentials belonging to $L^1_\sigma(\R)$ decay to zero as $x\to\pm\infty$. This turns out to affect the amount of information one can acquire about the low-energy behavior of the reflection and transmission amplitudes of the potential \cite{newton-1986,aktosun-2001}.

A potential $v:\R\to\C$ is said to be ``exponentially decaying,'' if there are positive real numbers $C,M$, and $\mu$ such that,
    \be
    |v(x)|\leq C\,e^{-\mu|x|}~~~\for~~~|x|\geq M.
    \label{exp-condi}
    \ee
Clearly, such a potential belongs to $L^1_\sigma$ for all $\sigma\geq 0$. If $v$ satisfies (\ref{exp-condi}) for some $C,M,\mu\in\R^+$, its Jost solutions are analytic functions of $k$ in $\{k\in\C~|~\IM(k)>-\mu/2~\&~k\neq 0\}$ and, in particular, in the positive real axis in the complex $k$-plane \cite{blashchak-1968,bolle-1985}. Every finite-range potential $v$ satisfies (\ref{exp-condi}) for $C=1$, $M=|a_+|+|a_-|$, and every $\mu\in\R^+$, where $[a_-,a_+]$ is the support of the potential. This shows that the Jost solutions of finite-range potentials are analytic functions of $k$ in the whole complex $k$-plane except possibly at $k=0$. Because $v(x)=0$ for $x\notin[a_-,a_+]$, (\ref{psi+}) and (\ref{psi-}) imply
	\begin{align*}
	&\psi_+(x,k)= M_{22}(k)\, e^{ikx}-M_{21}(k)\,e^{-ikx}~~~\for~~~x\leq a_-,\\
        &\psi_-(x,k)= M_{12}(k)\,e^{ikx}+M_{22}(k)\,e^{-ikx}~~~\for~~~x\geq a_+.
    	\end{align*}
In view of these relations and the fact that $\psi_\pm(x,k)$ are analytic functions of $k$ in $\C\setminus\!\{0\}$, we conclude that the same applies to $M_{12}(k)$, $M_{21}(k)$, and $M_{22}(k)$.  

Because $\det\bM(k)=1$, 
	\be
	M_{11}(k)=\frac{1+M_{12}(k)M_{21}(k)}{M_{22}(k)}.
	\label{M11=}
	\ee 
This equation seems to suggest that $M_{11}(k)$ may have singularities in $\C\setminus\{0\}$. This is however not true. To see this, we consider the effect of complex-conjugation of the potential, 
	\be
	v(x)\to \overline{v}(x):=v(x)^*,
	\label{time-reversal-v}
	\ee 
which we interpret as its time-reversal transformation \cite{bookchapter}, on its transfer matrix. By complex-conjugating both sides of the Schr\"odinger equation (\ref{sch-eq}) and pursuing a similar approach to the one we adopted in our derivation of the transformation rule for the transfer matrix under space translations, we find that under time-reversal transformation it transforms according to, 
	\be
	\bM(k)\to \overline{\bM}(k):=\bsigma_1\bM(k)^*\bsigma_1,
	\label{time-reversal-M}
	\ee
where $k\in\R^+$. Equivalently, we have the following time-reversal transformation rule for the entries of the transfer matrix  \cite{bookchapter,jpa-2014c}.
%	\be
%	\bM(k)\to \overline{\bM}(k):=\bsigma_1\bM(k)^*\bsigma_1~~~\for~~~k\in\R^+.
%	\label{time-reversal-M}
%	\ee
%This means that
	\be
	\begin{aligned}
	&M_{11}(k)\to \overline{M}_{11}(k):=M_{22}(k)^*,\quad 
	&&M_{12}(k)\to \overline{M}_{12}(k):=M_{21}(k)^*,\\
	&M_{21}(k)\to \overline{M}_{21}(k):=M_{12}(k)^*,\quad
	&&M_{22}(k)\to \overline{M}_{22}(k):=M_{11}(k)^*.
	\end{aligned}
	\label{time-reversal-Mij}
	\ee
Since $\overline{v}$ is also a finite-range potential, $\overline{M}_{22}(k)$ is an analytic function of $k$ in $\C\setminus\{0\}$. Therefore, it cannot have poles in $\C\setminus\{0\}$. In light of (\ref{time-reversal-Mij}), this implies that the same holds to $M_{11}(k)$. This observation together with (\ref{M11=}) show that the zeros\footnote{A complex number $z_0$ is called a zero of a function $f:\C\to\C$, if it belongs to the domain of definition of $f$ and $f(z_0)=0$.} of $M_{22}(k)$ that lie in $\C\setminus\{0\}$ are removable singularities of $M_{11}(k)$. Therefore, we can identify it with its continuous extension to $\C\setminus\{0\}$ which is analytic in $\C\setminus\{0\}$, \cite{Saff}. This completes the proof that all the entries of the transfer matrix of every finite-range potential are analytic functions in $\C\setminus\{0\}$ and in particular in the positive real $k$-axis. 

As we explain in the following subsections, the zeros of $M_{ij}(k)$ that lie on the positive real $k$-axis correspond to certain physical phenomena of particular interest. The analyticity of $M_{ij}(k)$ imply that their real zeros are isolated points\footnote{A zero $z_0$ of a function $f:\C\to\C$ is said to be isolated, if there is some $\delta\in\R^+$ such that $f(z)\neq 0$ for $0<|z-z_0|<\delta$.} \cite{Saff}. This in turn means that the associated physical effects cannot be realized for a finite range of wavenumbers $[k_-,k_+]$; either they occur for all wavenumbers or a discrete set of them with no accumulation point.

\subsection{Spectral singularities and lasing}
\label{Sec2-6}

Consider a short-range potential $v:\R\to\C$, and suppose that the $M_{22}$ entry of its transfer matrix vanishes for a wavenumber $k_\star\in\R^+$, i.e., 
	\be
	M_{22}(k_\star)=0.
	\label{SS-def}
	\ee
Then in view of the analyticity of $M_{11}(x)$ in the positive real $k$-axis and Eq.~(\ref{M11=}), we have $M_{12}(k_\star)M_{21}(k_\star)=-1$ and consequently,
	\begin{align}
	&M_{12}(k_\star)\neq 0, 
	&&M_{21}(k_\star)=-\frac{1}{M_{12}(k_\star)}.
	\label{SS-e1}
	\end{align} 
Eqs.~(\ref{psi+}), (\ref{psi-}), and (\ref{SS-e1}) imply that
	\be
	\psi_-(x,k_\star)=M_{12}(k_\star)\psi_+(x,k_\star)\to
	\left\{	\begin{array}{ccc}
	e^{-ik_\star x} & \for & x\to-\infty,\\[6pt]
	M_{12}(k_\star)\,e^{ik_\star x} & \for &x\to+\infty.
	\end{array}\right.
	\label{Jost-SS}
	\ee
This identifies $\psi_-(x,k_\star)$ with a solution of the Schr\"odinger equation for $k=k_\star$ that is left-going as $x\to-\infty$ and right-going as $x\to+\infty$. In other words, it corresponds to a purely outgoing wave at $x=\pm\infty$. If such a solution exists, the potential acts as a source that emits waves to $x=\pm\infty$. 

The energy $k_\star^2$ corresponding to the Jost solution $\psi_\pm(x,k_\star)$, is a point of the continuous spectrum of the Schr\"odinger operator $-\partial_x^2+v(x)$ at which these solutions become linearly-dependent. This is called a ``spectral singularity'' in mathematical literature \cite{schwartz}. It was discovered by Naimark in the 1950's \cite{Naimark} and has since been a subject of research in operator theory. For a basic mathematical review intended for physicists and further references, see \cite{guseinov}.

Suppose that $k_\star^2$ is a spectral singularity of a potential $v$, then (\ref{SS-def}) and (\ref{SS-e1}) hold, and (\ref{RRT=}) implies that $k_\star$ is a common pole of the reflection and transmission amplitudes of the potential. In particular, $R^{l/r}(k_\star)=\infty$ and $T(k_\star)=\infty$. Poles of the reflection and transmission amplitudes produce the resonances (and anti-resonances) of the potential, if they lie away from the real axis in the complex $k$-plane. Spectral singularities are the poles of $R^{l/r}(k)$ and $T(k)$ that are on the real axis. Because the imaginary part of $k$ for a resonance is a measure of its width, spectral singularities may be interpreted as certain ``zero-width resonances'' \cite{prl-2009}.\footnote{The bound states in the continuum also admit an interpretation as zero-width resonances. As explained in Ref.~\cite{acta-2013}, these are not to be confused with spectral singularities.} It is however important to notice that they correspond to Jost solutions of the Schr\"odinger equation which do not decay in time or space.

Next, consider the case that $v(x)$ is a real-valued potential. Then in view of (\ref{time-reversal-v}), 
$\overline v=v$, $\overline\bM=\bM$, and (\ref{time-reversal-Mij}) implies
	\begin{align}
	&M_{11}(k)=M_{22}(k)^*,
	&&M_{12}(k)=M_{21}(k)^*.
	\label{Mij-real}
	\end{align}
We can use (\ref{RRT=}) and (\ref{Mij-real}) to show,
    \bea
    &&|R^l(k)|=|R^r(k)|,
    \label{Real-RR}\\
    &&|R^{l/r}(k)|^2+|T(k)|^2=1,
    \label{unitarity}
    \eea
where again $k\in\R^+$, \cite{jpa-2014c}. Eqs.~(\ref{Real-RR}) and (\ref{unitarity}) are respectively known as the ``reflection reciprocity'' and ``unitarity" relations. The latter implies that, for a real potential, $R^{l/r}$ and $T$ are bounded functions. In particular, they cannot have poles. This proves that real potentials cannot possess spectral singularities. 

Perhaps the simplest example of a complex potential that can give rise to a spectral singularity is a delta-function potential, 
	\be
	v(x)=\fz\,\delta(x),
	\label{delta-z}
	\ee 
with a complex coupling constant $\fz$, \cite{jpa-2006b}.\footnote{It was indeed the study of this potential \cite{jpa-2006b} and the double-delta-function potential \cite{jpa-2009} that led to our present understanding of the physical aspects of spectral singularities \cite{prl-2009}.} We have already computed the transfer matrix of this potential in Subsec.~\ref{Sec2-4}. It is given by (\ref{M-delta}) with $a_j=0$ and $\fz_j=\fz$. According to (\ref{M-delta}), the condition (\ref{SS-def}) for the existence of a spectral singularity takes the simple form $\fz=2ik_\star$. Because $k_\star$ is real and positive, the delta-function potential (\ref{delta-z}) has a spectral singularity $k_\star^2:=|\fz|^2/4$ provided that $\fz$ is purely imaginary and $\IM(\fz)>0$. 

Now, consider the general case where real and imaginary parts of $\fz$ can take arbitrary values, $\delta:=\RE(\fz)$ and $\zeta:=\IM(\fz)$. We can use (\ref{RRT=}) and (\ref{M-delta}) to express the reflection and transmission amplitudes of the delta-function potential (\ref{delta-z}) in the form,
	\begin{align}
	&R^{l/r}(k)=\frac{-i \fz}{2k+i\fz}=\frac{2k}{2k-\zeta+i\delta}-1,
	&&T(k)=\frac{2k}{2k+i\fz}=\frac{2k}{2k-\zeta+i\delta}.
	\label{RT-delta-1}
	\end{align}
Suppose that $\zeta>0$ and $k_\star:=\zeta/2$. Then (\ref{RT-delta-1}) gives
	\begin{align}
	&R^{l/r}(k_\star)=-1-\frac{i\zeta}{\delta},
	&&T(k)=-\frac{i\zeta}{\delta}.
	\label{RT-delta}
	\end{align}
The fact that for $\delta=0$, $R^{l/r}(k_\star)$ and $T(k_\star)$ diverge is consistent with the fact that $k_\star^2$ is a spectral singularity. This implies that the delta-function potential (\ref{delta-z}) emits out-going waves with wavenumber $k_\star$. For $\delta\neq 0$,  $|R^{l/r}(k_\star)|^2$ and $|T(k_\star)|^2$ take finite values, but we can make them grow indefinitely by reducing the value of $|\delta/\zeta|$. This suggests that this potential acts as a tunable amplifier; as we make $\delta$ approach zero, it amplifies both the left- and right-incident waves with wavenumber $\zeta/2$.

The amplification of waves due to the presence of a spectral singularity is not an exclusive feature of the delta function potential (\ref{delta-z}). It applies to generic complex-valued short-range potentials and serves as a basic principle that most optical amplifiers operate upon. The principal and probably the most important example is a laser which produces intense coherent radiation by amplifying a single Fourier mode of an extremely weak background noise. 

A laser consists of an active dielectric material, which is usually characterized by a complex relative permittivity profile $\widehat\varepsilon$, and a pumping mechanism that injects energy to the active material to adjust the imaginary part of $\widehat\varepsilon$. For a homogeneous slab laser, $\widehat\varepsilon$ takes a constant value $\widehat\varepsilon_{\rm s}$ inside the slab, and  the corresponding optical potential (\ref{optical}) is the complex rectangular barrier potential,
	\be
	v(x)=\left\{\begin{array}{ccc}
	\fz & \for & x\in[0,L],\\
	0 & \for & x\notin[0,L],
	\end{array}\right.
	\label{barrier}
	\ee
where $\fz:=k^2(1-\widehat\varepsilon_{\rm s})$, and $L$ marks the slab's thickness. The relative permittivity of the slab $\widehat\varepsilon_{\rm s}$ and consequently the hight $\fz$ of the barrier potential depend on the  wavenumber $k$ through a dispersion relation. When the laser is turned on, the pump begins boosting the value of $\IM(\fz)$ for certain range of wavenumbers. Once $\IM(\fz)$ attains a value $\IM(\fz_\star)$ at which the potential has a spectral singularity $k_\star^2$, the system starts emitting a laser light with wavenumber $k_\star$. The condition, $\IM(\fz)=\IM(\fz_\star)$, determines the onset of lasing. In optics literature, this is called the ``laser threshold condition.'' A detailed study of the spectral singularities of the barrier potential (\ref{barrier}) shows that this condition follows from the requirement that $k_\star^2$ is a spectral singularity of the potential \cite{pra-2011a,pra-2015a}. It is important to notice that the standard (textbook) derivation of the laser threshold condition relies on accounting for the energy loses of the system and balancing them with energy injected by the pump \cite{silfvast}. The connection to the concept of spectral singularity offers an alternative purely mathematical method of deriving the same condition. This involves the computation of the $M_{22}(k)$ entry of the transfer matrix and finding the values of the physical parameters of the system for which $M_{22}(k)$ has a real and positive zero $k_\star$. This method has the advantage of being applicable to lasers with a variety of geometries and special properties \cite{pla-2011,prsa-2012,jpa-2012,pra-2013b,pra-2013d,ramezani-2014,hang-2016,ap-2016a,ap-2016b,kalozoumis-2017,Jin-2018,Konotop-2019b,Moccia-2020,Zezyulin-2020}. 

The application of spectral singularities in determining the laser threshold condition has also motivated the introduction of a nonlinear generalization of spectral singularity \cite{prl-2013}. This has been used to offer a useful mathematical derivation of the well-known linear relationship between the laser output intensity and the gain coefficient of a slab laser \cite{pra-2013c,sam-2014,ap-2018}. It has also led to certain surprising predictions regarding lasing in the oblique transverse magnetic modes of the slab \cite{jo-2017}. For other related developments, see Refs.~\cite{konotop-review,Zezyulin-2018,Muller-2018,josab-2020}.

\subsection{Time-reversed spectral singularities and anti-lasing}
\label{Sec2-7}

Spectral singularities of a short-range potential $v(x)$ are characterized by the real and positive zeros of the $M_{22}(k)$ entry of its transfer matrix $\bM(k)$. Under time-reversal transformation $M_{11}(k)$ is mapped to $\overline{M}_{22}$. Therefore, spectral singularities of the time-reversed potential $\overline v(x)$ correspond to the real and positive zeros of $M_{11}(k)$. This suggests using the terms ``time-reversed spectral singularity'' of $v(x)$ for energies $\overline{k}_\star^2$ at which 
	\be
	M_{11}(\overline{k}_\star)=0.
	\label{CPA-condi}
	\ee 

According to (\ref{time-reversal-v}), solutions $\overline\psi(x)$ of the Schr\"odinger equation for the time-reversed potential $\overline{v}(x)$ coincide with the complex-conjugate of the solutions $\psi(x)$ of the Schr\"odinger equation for $v(x)$. This implies that under time-reversal transformation the left- and right-going waves are mapped to right- and left-going waves, respectively. Because at a spectral singularity the Jost solutions are out-going at both $x=\pm\infty$, at a time-reversed spectral singularity they satisfy purely incoming boundary conditions at $x=\pm\infty$. 

To arrive at a quantitative assessment of the physical implications of time-reversed spectral singularities, suppose that $\overline{k}_\star^2$ is a spectral singularity of $\overline{v}(x)$, then according to (\ref{Jost-SS}) the Jost solutions $\overline\psi_\pm(x,k)$ of the Schr\"odinger equation~(\ref{sch-eq}) for the time-reversed potential $\overline{v}(x)$ satisfy,
	\be
	\overline{\psi}_-(x,\overline{k}_\star)
	=\overline{M}_{12}(\overline{k}_\star)\:\overline{\psi}_+(x,\overline{k}_\star)\to
	\left\{	\begin{array}{ccc}
	e^{-i\overline{k}_\star x} & \for & x\to-\infty,\\[6pt]
	\overline{M}_{12}(\overline{k}_\star)\,e^{i\overline{k}_\star x} & \for &x\to+\infty.
	\end{array}\right.
	\label{Jost-SS-TR}
	\ee
Now, let us introduce $\psi_\times(x,k):=A_-\,\overline{\psi}_-(x,k)^*$, where $A_-$ is a nonzero complex amplitude. It is clear that $\psi_\times$ solves the Schr\"odinger equation~(\ref{sch-eq}) for the potential ${v}(x)$ and fulfills	
	\be
	\psi_\times(x,\overline{k}_\star)\to
	\left\{	\begin{array}{ccc}
	A_- e^{i\overline{k}_\star x} & \for & x\to-\infty,\\[3pt]
	B_+ e^{-i\overline{k}_\star x} & \for &x\to+\infty,
	\end{array}\right.
	\label{CPA-Sol}
	\ee	
where $B_+:=A_-\,{M}_{21}(\overline{k}_\star)$, and we have made use of (\ref{time-reversal-Mij}). 

The wave function $\psi_\times(x,\overline{k}_\star)$ describes a setup in which there are two sources of the incident waves. One is placed at $x=-\infty$ and emits a right-going wave of amplitude $ {A}_-$, while the other is at $x=+\infty$ and emits a left-going wave with amplitudes $ {B}_+$. Since there is no left-going wave at $x=-\infty$ or right-going wave at $x=+\infty$, the potential absorbs both the incoming incident waves. Therefore, it acts as a ``perfect coherent absorber (CPA),'' \cite{chong-2010,longhi-2010-cpa,wan-2011}. A  dielectric slab whose optical potential realizes the condition (\ref{CPA-condi}) for a wavenumber $\overline{k}_\star$ functions as a time-reversed laser provided that the amplitudes $A_-$ and $B_+$ of the incident waves satisfy
	\be
	\frac{ {B}_+}{ {A}_-}=M_{21}(\overline{k}_\star).
	\label{CPA-condi-2}
	\ee
 Such an optical system is called an ``antilaser.'' 
 
It is important to realize that a potential acts as a CPA, if the following requirements are met.
	\begin{enumerate}
	\item It must have a time-reversed spectral singularity $\overline{k}_\star^2$, i.e., (\ref{CPA-condi}) holds.
	\item The complex amplitudes of the incident waves must fulfill (\ref{CPA-condi-2}).
	\end{enumerate}
The first of these condition cannot be fulfilled by a real potential, because real potentials are time-reversal invariant and cannot possess a spectral singularity. The second condition is a constraint on the incident waves, not the potential. It restricts both the modulus and phase of their amplitudes. 

Recalling our treatment of the delta-function potential~(\ref{delta-z}), we can immediately conclude that this potential has a time-reversed spectral singularity, $\overline{k}_\star^2=|\fz|^2/4$, provided that $\RE(\fz)=0$ and $\IM(\fz)<0$. Other examples of complex potentials and the corresponding permittivity or refractive index profiles that can support a time-reversed spectral singularity and therefore function as a CPA have been studied in Refs.~\cite{ap-2016b,kalozoumis-2017,Konotop-2019b,Zezyulin-2020,chong-2010,longhi-2010-cpa,longhi-2010-CPA-laser,Baranov-2017,sarisaman-2019}.

There is a special class of potentials where a spectral singularity coincides with a time-reversed spectral singularity, i.e., there is $k_\star\in\R^+$ such that 
	\be
	M_{11}(k_\star)=M_{22}(k_\star)=0.
	\label{self-dual-SS}
	\ee
In this case, $k_\star^2$ is called a self-dual spectral singularity \cite{jpa-2012}. In the absence of sources, such a potential serves as a source of left- and right-going waves, and its optical realization corresponds to a laser sufficiently pumped to fulfill the laser threshold condition. If one tries to inject an incident wave from the left or right that has the wavenumber $k_\star$, the optical device will seize to maintain its physical state of satisfying $M_{22}(k_\star)=0$, because the reflection and transmission coefficients at $k=k_\star$ blow up and the device must otherwise emit waves with infinite intensity which would violate energy conservation. This would happen to any laser that is subject to a coherent incident beam with the same wavenumber as the emitted wave by the laser. Much more interesting is the situation where the device is subject to a pair of left- and right-going incident waves with wavenumber $k_\star$ and amplitudes $B_+$ and $A_-$ fulfilling (\ref{CPA-condi-2}). In this case, it stops acting as a laser and functions as an antilaser \cite{longhi-2010-CPA-laser,chong-2011,Wong-2016}.

The simplest examples of potentials capable of supporting a self-dual spectral singularity are $\cP\cT$-symmetric scattering potentials, which satisfy $v(x)^*=v(-x)$, \cite{longhi-2010-CPA-laser,chong-2011}. This is because for a short-range $\cP\cT$ symmetric  potential, $M_{11}(k)=M_{22}(k)^*$, \cite{bookchapter,jpa-2014c,longhi-2010-CPA-laser}. According to this equation, every spectral singularity of a $\cP\cT$-symmetric potential is self-dual. Because the $\cP\cT$-symmetric lasers can display both lasing and antilasing, they are called ``$\cP\cT$-lasers.'' We should however warn that the laser-antilaser action is associated with the concept of self-dual spectral singularity which does not require $\cP\cT$-symmetry \cite{jpa-2012,kalozoumis-2017,konotop-2017}.

\subsection{Reflectionless and invisible potentials}
\label{Sec2-8}

The real and positive zeros of the diagonal entries of the transfer matrix correspond the spectral singularities and their time-reversal. As we argue in the preceding subsection, these provide a mathematical basis for lasing and anti-lasing. In this subsection, we examine the physical meaning of the real and positive zeros of the off-diagonal entries of the transfer matrix.

A simple consequence of  (\ref{RRT=}) is that whenever $M_{12}(\ktl)=0$ for some $\ktl\in\R^+$, $R^r(\ktl)=0$, and the potential is reflectionless for a right-incident wave with wavenumber $\ktl$.  In this case, we say that the potential is ``right-reflectionless'' for $k=\ktl$. Similarly, if $M_{21}(\ktr)=0$ for some $\ktr\in\R^+$, $R^l(\ktr)=0$, and we say that the potential is ``left-reflectionless'' for $k=\ktr$.

Because real-valued potentials satisfy the reflection reciprocity relation (\ref{Real-RR}), their right-reflection\- lessness for a wavenumber $\ktl$ implies their left-reflectionlessness for the same wavenumber and vice versa. In other words, unidirectional reflectionlessness is forbidden for real potentials.

Examples of real potentials that are reflectionless at every wavenumber have been known since the 1930's \cite{Epstein-1930,Eckart-1930}. They are systematically studied and classified in the 1950's \cite{kay-1956}. The principal example is the P\"oschl-Teller potential,
    \be
    v(x)=-\frac{l(l+1)}{a^2\cosh^2(x/a)},\nn
    \ee
where $l$ is a positive integer, and $a$ is a positive real parameter \cite{lekner-2007}.

The term ``reflectionless'' does not necessarily mean ``invisible.'' A  potential is said to be ``left- (respectively right-) invisible'' for some wavenumber $k_{_\Join}$, if it is left- (respectively right-) reflectionless and has unit transmission amplitude for $k=k_{_\Join}$, i.e., $R^l(k_{_\Join})=0$ (respectively $R^r(k_{_\Join})=0$) and $T(k_{_\Join})=1$, \cite{lin-2011}. Unidirectional invisibility means that the potential is invisible from one direction only. This again occurs at particular values of the wavenumber. According to (\ref{Mij=}), the transfer matrix of a potential that is unidirectionally invisible for $k=k_{_\Join}$ satisfies
    \be
    \bM(k_{_\Join})=\left\{
    \begin{aligned}
    &\left[\begin{array}{cc}
    1 & R^r(k_{_\Join})\\
    0 & 1\end{array}\right] && \mbox{for unidirectional left-invisibility},\\[6pt]
    &\left[\begin{array}{cc}
    1 & 0\\
    -R^l(k_{_\Join}) & 1
    \end{array}\right] && \mbox{for unidirectional right-invisibility}.
    \end{aligned}\right.
    \label{Uni-M}
    \ee
where the $R^r(k_{_\Join})$ and $R^l(k_{_\Join})$ do not vanish.

The classical real reflectionless potentials studied in \cite{kay-1956} are not invisible for all wavenumbers, because they do not satisfy $T(k)=1$ for all $k\in\Z^+$. We can however use (\ref{unitarity}) to conclude that $T(k)$ is a phase factor, i.e., there is a function $\tau:\R\to\R$ such that $T(k)=e^{i\tau(k)}$. This implies that if $\tau(k_{_\Join})$ in an integer multiple of $2\pi$ for some $k_{_\Join}\in\R^+$, then the potential is bidirectionally invisible for $k=k_{_\Join}$.

The situation is different for complex potentials. Unlike real potentials, complex potentials can display unidirectional reflectionlessness and invisibility. The principal example is \cite{lin-2011,poladian,greenberg,kulishov}
    \be
    v(x):=\left\{\begin{array}{ccc}
    \fz \, e^{2\pi i x/L} &\for& x\in[0,L],\\
    0 &\for& x\notin[0,L],\end{array}\right.
    \label{exp-pot}
    \ee
where $\fz$ is a real or complex coupling constant, and $L$ is a real and positive parameter. As we demonstrate in Subsec.~\ref{Sec3-4}, under the condition that $|\fz|L^2\ll 1$, this potential is unidirectionally right-invisible for wavenumber $\ktl:=\pi/L$, \cite{lin-2011}. This is however an approximate effect which goes away once $|\fz|$ becomes comparable to $L^{-2}$, \cite{longhi-2011,Graefe-2011,pra-2014a}. The potential~(\ref{exp-pot}) does however display exact unidirectional invisibility for particular values of $\fz$, \cite{jpa-2016}.  Exact unidirectional invisibility can also be realized using piecewise constant complex potentials which model effectively one-dimensional multilayer optical material \cite{Ge-2012}. The simplest examples allowing for an explicit analytic investigation are the complex barrier potentials of the form,
    \be
    v(x)=\left\{\begin{array}{ccc}
    \fz_- &\for& x\in[-\frac{L}{2},0[,\\[3pt]
    \fz_+&\for& x\in [0,\frac{L}{2}],\\[3pt]
    0 &\for& x\notin [-\frac{L}{2},\frac{L}{2}],
    \end{array}\right.
    \label{bilayer}
    \ee
where $L$ is a positive real parameter, and $\fz_\pm$ are complex coupling constants, \cite{pra-2013}.

Finite-range potentials such as (\ref{bilayer}) cannot be reflectionless or invisible for a finite range of wavelengths $[k_-,k_+]$, because the $M_{12}$ and $M_{21}$ entries of their transfer matrix are analytic functions of $k$ in $\C\setminus\{0\}$. The classical reflectionless potentials \cite{kay-1956} are examples of bidirectionally reflectionless potentials that possess this property for all wavenumbers. There is a large class of potentials displaying unidirectional invisibility at all wavelengths. These are potentials $v(x)$ whose Fourier transformation, $\widetilde v(\fK):=\int_{-\infty}^\infty dx\,e^{-i\fK x} v(x)$, vanishes either for $\fK> 0$ or $\fK< 0$, \cite{horsley-2015,longhi-2015,horsley-longhi}. The existence of these potentials do not contradict the analyticity of the transfer matrix of finite-range potentials in $\C\setminus\{0\}$, because the condition, ``$\widetilde v(\fK)=0$ only for $\fK> 0$ or $\fK< 0$,'' cannot be met by a finite-range or exponentially decaying potential, i.e., the potentials considered in Refs.~\cite{horsley-2015,longhi-2015,horsley-longhi} are not exponentially decaying at $x=\pm\infty$. By truncating these potentials, one can construct finite-range potentials with approximate unidirectional invisibility in a rather wide wavenumber spectrum. An example of such a finite-range potential has been investigated experimentally in Ref.~\cite{jiang-2017}.

\subsection{Lippmann-Schwinger equation}
\label{Sec2-9}

An old and useful method of deriving qualitative information about the behavior of the solutions of the Schr\"odinger equation~(\ref{sch-eq}) is to express it in the form of an integral equation. To do this, first we write (\ref{sch-eq}) as 
	\be 
	\psi''(x)+k^2\psi(x)=\xi(x),
	\label{sch-eq-nh}
	\ee
where 
	\be
	\xi(x):=v(x)\psi(x).
	\label{chi=}
	\ee 
Disregarding the fact that $\xi$ is related to $\psi$, we can view (\ref{sch-eq-nh}) as a non-homogeneous linear equation \cite{boyce-diprima} and write its general solution as 
	\be
	\psi(x)=\psi_0(x)+\psi_p(x),
	\label{LS-e1}
	\ee
where $\psi_0$ is a solution of the homogeneous equation, 
	\be
	\psi''(x,k)+k^2\psi(x,k)=0,
	\label{homog}
	\ee
and $\psi_p$ is a particular solution of (\ref{sch-eq-nh}). We can express the latter in the form
	\be
	\psi_p(x)=\int_{x_0}^x dy\:G(x,y)\xi(y),
	\label{LS-e2}
	\ee
where $G(x,y)$ is the Green's function for the Helmholtz operator $\partial_x^2+k^2$, and $x_0\in\R\cup\{-\infty,+\infty\}$ is arbitrary. We can use any pair of linearly-independent solutions, $\psi_{0,\pm}$, of (\ref{homog}) to compute $G(x,y)$ according to,
	\be
	G(x,y)=\frac{\psi_{0-}(x)\psi_{0+}(y)-\psi_{0-}(y)\psi_{0+}(x)}{W[\psi_{0+}(y),\psi_{0-}(y)]},
	\label{Green=}
	\ee
where $W[\psi_{0+},\psi_{0-}]$ is the Wronskian of $\psi_{0\pm}$.	
For example, setting $\psi_{0\pm}(x):=e^{\pm ikx}$ in (\ref{Green=}), we obtain
	\be
	G(x,y)=\frac{\sin [k(x-y)]}{k}.
	\label{Green=1}
	\ee
	
Because solutions of (\ref{homog}) are linear combinations of $\psi_{0\pm}$, there are constant coefficients $c_\pm$ such that
	\be
	\psi_0(x)=c_-\psi_{0-}(x)+c_+\psi_{0+}(x).
	\label{homog-soln=}
	\ee
In view of  (\ref{chi=}), (\ref{LS-e1}), (\ref{LS-e2}), and (\ref{homog-soln=}), every solution $\psi$ of the Schr\"odinger equation (\ref{sch-eq}) satisfies
	\bea
	\psi(x)&=&c_-\psi_{0-}(x)+c_+\psi_{0+}(x)+\int_{x_0}^x dy\:G(x,y)v(y)\psi(y)\nn\\
	&=&e^{-ikx}\left[c_--\frac{1}{2ik}\int_{x_0}^x dy\:e^{iky}v(y)\psi(y)\right]+
	e^{ikx}\left[c_++\frac{1}{2ik}\int_{x_0}^x dy\:e^{-iky}v(y)\psi(y)\right].
	\label{LS-psi=}
	\eea
This is an integral equation involving the constants $c_\pm$ and $x_0$.\footnote{Since changing $x_0$ has the effect of shifting the values of $c_\pm$ by constant amounts, $c_\pm$ and $x_0$ are not independent.}

Next, consider the following normalized scattering solutions of the Schr\"odinger equation~(\ref{sch-eq}).		\begin{align}
	&\widehat\psi^l(x):=\frac{\psi^l(x)}{A_-^{l}},
	&&\widehat\psi^r(x):=\frac{\psi^r(x)}{B_+^{r}},
	\label{hat-psi}
	\end{align}
where $\psi^{l}$ and $\psi^{r}$ are respectively the solutions satisfying (\ref{psi-left-RT}) and (\ref{psi-right-RT}). These equations imply that
	 \begin{align}
    &{\widehat\psi^l}(x)\to \left\{\begin{array}{ccc}
    e^{ikx}+R^l\, e^{-ikx} &\for& x\to-\infty,\\
    T\,e^{ikx} &\for&x\to+\infty,\end{array}\right.
    \label{hat-psi-left-RT}\\[3pt]
    &{\widehat\psi^r}(x)\to \left\{\begin{array}{ccc}
    T\,e^{-ikx} &\for& x\to-\infty,\\
    e^{-ikx}+R^r\,e^{ikx} &\for&x\to+\infty.\end{array}\right.
    \label{hat-psi-right-RT}
    \end{align}
Being solutions of the Schr\"odinger equation (\ref{sch-eq}), ${\widehat\psi^{l/r}}$ satisfy the integral equation (\ref{LS-psi=}) for appropriate values of $c_\pm$ and $x_0$. These are the values for which the right-hand side of (\ref{LS-psi=}) fulfill the asymptotic boundary conditions (\ref{hat-psi-left-RT}) and (\ref{hat-psi-right-RT}). Imposing this condition, we discover that ${\widehat\psi^l}(x)$ and ${\widehat\psi^r}(x)$ are respectively the solutions of (\ref{LS-psi=}) for 
	\begin{align}
	&c_-=0, &&c_+=T, &&x_0=+\infty,
	\label{ccx-left}
	\end{align}
and 
	\begin{align}
	&c_-=T, &&c_+=0, &&x_0=-\infty.
	\label{ccx-right}
	\end{align}
In particular, we can write $\widehat\psi^{l/r}(x)$ in the form,
	\begin{align}
	&\widehat\psi^{l}(x)=e^{ikx}+\widehat\psi^{l}_s(x),
	&&\widehat\psi^{r}(x)=e^{-ikx}+\widehat\psi^{r}_s(x),
	\label{hat-psi-scatter}
         \end{align}
where $e^{\pm ikx}$ and $\widehat\psi^{l/r}_s(x)$ are respectively the solutions of the homogeneous equation (\ref{homog}) and particular solutions of the non-homogenous equation (\ref{sch-eq-nh}). They represent the incident and scattered waves. According to (\ref{hat-psi-right-RT}) and (\ref{hat-psi-scatter}), 
	 \begin{align}
    	&{\widehat\psi_s^l}(x)\to \left\{\begin{array}{ccc}
    	R^l\, e^{-ikx} &\for& x\to-\infty,\\
    	(T-1)\,e^{ikx} &\for&x\to+\infty,\end{array}\right.
    	\label{hat-psi-left-RT-s}\\[3pt]
    	&{\widehat\psi^r_s}(x)\to \left\{\begin{array}{ccc}
    	(T-1)\,e^{-ikx} &\for& x\to-\infty,\\
    	R^r\,e^{ikx} &\for&x\to+\infty.\end{array}\right.
    	\label{hat-psi-right-RT-s}
    	\end{align}

Using (\ref{hat-psi-scatter}) and the fact that ${\widehat\psi^{l/r}}$ satisfy (\ref{LS-psi=}) with $c_\pm$ and $x_0$ given by (\ref{ccx-left}) and (\ref{ccx-right}), we can show that
	\bea
	\widehat\psi^{l}_s(x)&=&\frac{e^{-ikx}}{2ik}\int_x^\infty dy\:e^{iky}v(y)\widehat\psi^{l}(y)+
	\frac{e^{ikx}}{2ik}\left[2ik(T-1)-\int_x^\infty dy\:e^{-iky}v(y)\widehat\psi^{l}(y)\right],
	\label{SL-e11}\\
	\widehat\psi^{r}_s(x)&=&
	\frac{e^{-ikx}}{2ik}\left[2ik(T-1)-\int_{-\infty}^x dy\:e^{iky}v(y)\widehat\psi^{r}(y)\right]+
	\frac{e^{ikx}}{2ik}\int_{-\infty}^x dy\:e^{-iky}v(y)\widehat\psi^{r}(y).
	\label{SL-e12}
	\eea
If we compare the $x\to-\infty$ limit of (\ref{SL-e11}) and the $x\to+\infty$ limit of (\ref{SL-e12}) with the ones given by (\ref{hat-psi-left-RT-s}) and (\ref{hat-psi-right-RT-s}), we obtain
	\begin{align}
	&R^l=\frac{1}{2ik}\int_{-\infty}^\infty dy\:e^{iky}v(y)\widehat\psi^{l}(y),
	\quad\quad R^r=\frac{1}{2ik}\int_{-\infty}^\infty dy\:e^{-iky}v(y)\widehat\psi^{r}(y),
	\label{LS-RR=}\\
	&T=1+\frac{1}{2ik}\int_{-\infty}^\infty dy\:e^{-iky}v(y)\widehat\psi^{l}(y)=
	1+\frac{1}{2ik}\int_{-\infty}^\infty dy\:e^{iky}v(y)\widehat\psi^{r}(y).
	\label{LS-T=}
	\end{align}
With the help of the latter equation, we can write (\ref{SL-e11}) and (\ref{SL-e12}) as
	\bea
	\widehat\psi^{l}_s(x)&=&\frac{e^{-ikx}}{2ik}\int_x^\infty dy\:e^{iky}v(y)\widehat\psi^{l}(y)+
	\frac{e^{ikx}}{2ik} \int_{-\infty}^x dy\:e^{-iky}v(y)\widehat\psi^{l}(y) ,
	\label{SL-e13}\\
	\widehat\psi^{r}_s(x)&=&
	\frac{e^{-ikx}}{2ik} \int_x^{\infty} dy\:e^{iky}v(y)\widehat\psi^{r}(y)+
	\frac{e^{ikx}}{2ik}\int_{-\infty}^x dy\:e^{-iky}v(y)\widehat\psi^{r}(y),
	\label{SL-e14}
	\eea
or alternatively
	\be
	\widehat\psi^{l/r}_s(x)=\int_{-\infty}^\infty dy\: G^+(x,y)v(y)\,\widehat\psi^{l/r}(y),
	\label{LS-eq1}
	\ee
where
	\be
	G^+(x,y):=\frac{e^{ik|x-y|}}{2ik}.
	\label{G+=}
	\ee
Substituting (\ref{LS-eq1}) in (\ref{hat-psi-scatter}), we see that $\widehat\psi^{l/r}$ solve the Lippmann-Schwinger equation \cite{Lippmann-Schwinger},
	\be
	\psi(x)=\psi_0(x)+\int_{-\infty}^\infty dy\: G^+(x,y)v(y)\,\psi(y),
	\label{LS-eq2}
	\ee
where $\psi_0(x)$ represents the incident wave; $\psi_0(x):=\psi^l_{0}(x):=e^{ikx}$ for $\psi=\widehat\psi^{l}$, and $\psi_0(x):=\psi^r_{0}(x):=e^{-ikx}$ for $\psi=\widehat\psi^{r}$.

It turns out that the Green function $G^+(x,y)$ coincides with the $\epsilon\to 0^+$ limit of the integral kernel of the operator $(-\hat H_0+k^2+i\epsilon)^{-1}$, where $\hat H_0:=\hat p^2$ is the Hamiltonian operator for a free particle of mass $m=1/2$, and $\hat p$ is the standard momentum operator\footnote{Because we use units in which $\hbar=1$, $\hat p\,\psi(x):=-i\psi'(x)$.}, i.e.,
	\be
	G^+(x,y):=\lim_{\epsilon\to 0^+}\br x|(-\hat H_0+k^2+i\epsilon)^{-1}|y\kt.
	\label{SL-G+}
	\ee
To see this, we use the spectral resolution of $(k^2-\hat H_0+i\epsilon)^{-1}$, namely
	\[(-\hat H_0+k^2+i\epsilon)^{-1}=\int_{-\infty}^\infty dp\:
	\frac{|p\kt\br p|}{k^2-p^2+i\epsilon},\]
together with the well-known relation, $\br x|p\kt=e^{-ipx}/\sqrt{2\pi}$, to show that
	\[\br x|(-\hat H_0+k^2+i\epsilon)^{-1}|y\kt=-\frac{1}{2\pi}\int_{-\infty}^\infty dp\:
	\frac{e^{ip(x-y)}}{p^2-k^2-i\epsilon}.\]
We can easily turn the integral on the right-hand side of this equation into a contour integral and evaluate it using the residue theorem \cite{Saff}.\footnote{The term $i\epsilon$ entering the integrand determines the location of the poles and affects the outcome of this calculation.}  The result is (\ref{G+=}). 	

It is important to observe that we can start from the Lippmann-Schwinger equation~(\ref{LS-eq2}) and derive the formulas (\ref{LS-RR=}) and  (\ref{LS-T=}) for the reflection and transmission amplitudes by examining its $x\to\pm\infty$ limits. These formulas seem to be of limited practical value, because they involve the unknown scattering solutions $\widehat\psi^{l/r}$. They become useful, only if we can compute these functions. In the following subsection we outline a perturbative solution of the Lippmann-Schwinger equation~(\ref{LS-eq2}) that produces a series expansion for $\widehat\psi^{l/r}$. Substituting this in (\ref{LS-RR=}) and  (\ref{LS-T=}), we find series expansions for the reflection and transmission amplitudes.

\subsection{Born series and Born approximation}
\label{Sec2-10}

To determine a perturbative series solution of the Lippmann-Schwinger equation (\ref{LS-eq2}) , we 
introduce an auxiliary perturbation parameter $\zeta$, let $u(x):=v(x)/\zeta$, so that $v(x):=\zeta\,u(x)$, and write this equation in the Dirac bra-ket notation. For $\psi=\widehat\psi^{l/r}$, this gives
	\be
	|\widehat\psi^{l/r}\kt=|\widehat\psi^{l/r}_0\kt+
	\zeta\,\hat G^+\hat u \,|\widehat\psi^{l/r}\kt,
	\label{LS-eq3}
	\ee
where 
	\begin{align*}
	&\br x|\widehat\psi^{l}_0\kt:=\psi_{0+}(x)=e^{ikx},
	&&\br x|\widehat\psi^{r}_0\kt:=\psi_{0-}(x)=e^{-ikx},  
	&&\hat G^+:=\lim_{\epsilon\to 0^+}(-\hat H_0+k^2+i\epsilon)^{-1},
	\end{align*} 
$\hat u$ is the operator defined by $\hat u\psi(x):=u(x)\psi(x)$, and we have used the resolution of the identity operator $I$ in the position representation, i.e., $\int_{-\infty}^\infty dy\:|y\kt\br y|=I$.

Next, we write (\ref{LS-eq3}) in the form, 
	\be
	|\widehat\psi^{l/r}\kt=(1-\zeta\,\hat G^+\hat u)^{-1}|\widehat\psi^{l/r}_0\kt.
	\label{LS-eq-5}
	\ee 
If we expand $(1-\zeta\,\hat G^+\hat u)^{-1}$ as a formal geometric series in $\zeta$, i.e., employ
	\[(1-\zeta\,\hat G^+\hat u)^{-1}=\sum_{n=0}^\infty(\zeta\,\hat G^+\hat u)^n,\]
(\ref{LS-eq-5}) reads
	\be
	\widehat\psi^{l/r}(x)=\sum_{n=0}^\infty \widehat\psi^{l/r}_{n}(x),
	\label{Born-series-1}
	\ee
where $\widehat\psi^{l/r}_{n}(x):=\br x|(\hat G^+\hat v)^n|\widehat\psi^{l/r}_0\kt$ and $\hat v:=\zeta\,\hat u$. 
	
To derive similar power series expansions for the reflection and transmission amplitudes, first we introduce 
	\be
	f^{l/r}(\pm):=\frac{1}{2ik}\int_{-\infty}^\infty dy\: e^{\mp iky}v(y)\widehat\psi^{l/r}(y)=
	\frac{1}{2ik}\,\br\psi_{0\pm}|\hat v|\widehat\psi^{l/r}\kt.
	\label{f+-}
	\ee
We call these the ``left/right scattering amplitude'' of the potential $v$. In light of (\ref{Born-series-1}), they admit the series expansions,
	\be
	f^{l/r}(\pm)=\sum_{n=1}^\infty f_n^{l/r}(\pm),
	\label{Born-series-2}
	\ee
where
	\be
	f^{l/r}_n(\pm):=\frac{1}{2ik}
	\,\br\psi_{0\pm}|\hat v(\hat G^+\hat v)^{n-1}|\widehat\psi^{l/r}_0\kt.
	\label{f-n=}
	\ee
We can use (\ref{Born-series-2}) to construct series expansions for the reflection and transmission amplitudes, because according to (\ref{LS-RR=}), (\ref{LS-T=}), and (\ref{f+-}),
	\begin{align}
	& R^l= f^l(-), && R^r= f^r(+), && T=1+f^l(+)=1+f^r(-).
	\label{LS-RRT=}
	\end{align}
	
Eqs.~(\ref{Born-series-1}) and (\ref{Born-series-2}) are known as the ``Born series for the scattering solutions'' and the ``Born series for the scattering amplitudes.'' If we terminate these series, we find the following approximate expressions for the normalized scattering solutions and the scattering amplitudes.
	\begin{align}
	&\widehat\psi^{l/r}(x)\approx\sum_{n=0}^N \widehat\psi_n^{l/r}(x),
	&&f^{l/r}(\pm)\approx\sum_{n=1}^N f_n^{l/r}(\pm).
	\end{align}
This procedure is known as the ``$N$-th order Born approximation.''  The term ``Born approximation'' is sometimes used to mean the first-order Born approximation, also called the ``first Born approximation.'' For the scattering amplitudes it means 
	\be
	f^{l/r}(\pm)\approx f^{l/r}_1(\pm).
	\label{1st-BA}
	\ee 

It is instructive to examine the implications of the first Born approximation for the scattering, reflection, and transmission amplitudes. According to (\ref{f-n=}), 
	\[f^{l/r}_1(\pm)=\frac{\br\psi_{0\pm}|\hat v|\widehat\psi^{l/r}_0\kt}{2ik}.\]
It is not difficult to show that this relation implies,
	\begin{align}
	&f^{l}_1(+)=f^{r}_1(-)=\frac{\widetilde v(0)}{2ik},
	&&f^{l}_1(-)=\frac{\widetilde v(-2k)}{2ik},
	&&f^{r}_1(+)=\frac{\widetilde v(2k)}{2ik},
	\label{fff=}
	\end{align}
where $\widetilde v(k)$ stands for the Fourier transform of $v(x)$, i.e.,
	\be
	\widetilde v(k):=\sF_k\{v(x)\}:=\int_{-\infty}^\infty dx\:e^{-ikx}v(x).
	\label{FT}
	\ee
Substituting (\ref{fff=}) in (\ref{1st-BA}) and using the result in (\ref{LS-RRT=}), we find
	\begin{align}
	&R^l(k)\approx \frac{\widetilde v(-2k)}{2ik},
	&&R^r(k)\approx \frac{\widetilde v(2k)}{2ik},
	&&&T(k)\approx 1+\frac{\widetilde v(0)}{2ik}.
	\label{RRT-1st-BA}
	\end{align}

Eqs.~(\ref{RRT-1st-BA}) reveal an interesting connection between the reflection amplitudes and the Fourier transform of the potential. In particular, they show that whenever the first Born approximation provides a reliable description of the scattering properties of the potential, we can recover its expression using the formula for either of the reflection amplitudes. More specifically, we have
	\bea
	v(x)&\approx&i\sF^{-1}_x\left\{kR^r(\mbox{$\frac{k}{2}$})\right\}=
	%\partial_x\sF^{-1}_x\left\{R^r(k/2)\right\}=
	2\partial_x\sF^{-1}_{2x}\left\{R^r(k)\right\},
	\label{IS-Born-1}\\
	v(x)&\approx&-i\sF^{-1}_x\left\{kR^l(\mbox{$-\frac{k}{2}$})\right\}=
	%-\partial_x\sF^{-1}_x\left\{R^l(-k/2)\right\}=
	-2\partial_x\sF^{-1}_{-2x}\left\{R^l(k)\right\},
	\label{IS-Born-2}
	\eea
where $\sF^{-1}_x\{g(k)\}$ stands for the inverse Fourier transform of $g(k)$, i.e.,
	\[\sF^{-1}_x\{g(k)\}:=\frac{1}{2\pi}\int_{-\infty}^\infty dk\:
	e^{ikx}g(k).\]
Eqs.~(\ref{IS-Born-1}) and (\ref{IS-Born-2}) provide a simple approximate inverse scattering prescription. Notice that in order to implement this scheme we need to know either of $R^l(k)$ or $R^r(k)$ for both $k>0$ and $k<0$, and make sure that they have differentiable inverse Fourier transforms. For the potentials belonging to the Faddeev class $L^1_1(\R)$, $M_{ij}(k)$ are analytic functions in the upper-half complex $k$-plane and continuous in $\R\setminus\{0\}$. Therefore, one can attempt to define $M_{ij}(-k)$ for $k\in\R^+$ by analytically continuing $M_{ij}(k)$ through the upper half-plane, and use (\ref{RRT=}) to determine $R^{l/r}(k)$ for $k\in\R^-$. For a discussion of the transformation rule for $M_{ij}(k)$, $R^{l/r}(k)$, and $T(k)$ under the transformation $k\to-k$, see Ref.~\cite{jpa-2014c}.

As a simple example, let us examine the application of the first Born approximation to the delta-function potential, $v(x)=\fz\,\delta(x)$. Because for this potential $\widetilde v(k)=\fz$, (\ref{RRT-1st-BA}) gives
	\begin{align}
	&R^{l/r}(k)\approx -\frac{i\fz}{2k},
	&&T(k)\approx 1-\frac{i\fz}{2k}.\nn
	\end{align}
Comparing these relations with the exact expressions for the reflection and transmission amplitudes given in (\ref{RT-delta}), i.e.,
	\begin{align}
	&R^{l/r}(k)= -\frac{i\fz}{2k+i\fz}=-\frac{i\fz}{2k}+O(\fz^2),
	&&T(k)=\frac{2k}{2k+i\fz}= 1-\frac{i\fz}{2k}+O(\fz^2),\nn
	\end{align}
we see that they agree up to the linear terms in the coupling constant $\fz$.\footnote{$O(\fz^d)$ stands for terms of order $d$ and higher in powers of $\fz$.} This is indeed expected, because the first Born approximation is a first-order approximation in the strength of the potential. Similarly, if we try to use the above approximate inverse scattering scheme to recover the delta-function potential from the exact expression for its reflection amplitudes, we find that the result differs from $\fz\,\delta(x)$ by a term proportional to $\fz^2$.  

We have derived the Born series (\ref{Born-series-1}) and (\ref{Born-series-2}) as power series in the auxiliary perturbation parameter $\zeta$ which drops out of the calculations once we express the final result in terms of potential $v$. For the example of the delta-function potential, $v(x)=\fz\,\delta(x)$, the coupling constant $\fz$ plays the same role as $\zeta$, and (\ref{Born-series-1}) and (\ref{Born-series-2}) are indeed power series in $\fz$. This is generally true for any potential of the form, $v(x)=\fz\, u(x)$, where $\fz$ signifies the strength of the potential $v$ while $u$ is a function determining its shape. For such potentials, the $N$-th order Born approximation provides expressions for the scattering solutions $\widehat\psi^{l/r}(x)$ and the scattering, refection, and transmission amplitudes, $f^{l/r}(\pm)$, $R^{l/r}$, and $T$, that are polynomials of degree not larger than $N$. 

Because the entries of the transfer matrix involve ratios of $R^{l/r}$ and $T$, the $N$-th order Born approximation produces infinite power series expressions for $M_{ij}$. The terms of order $N+1$ and higher  in these series are however unreliable and should be discarded, because they receive contributions from the neglected terms of the Born series (\ref{Born-series-2}). 

In principle, we can obtain a Born series for $M_{ij}$ by substituting those for $R^{l/r}$ and $T$ in (\ref{Mij=}). The explicit form of the coefficients of this series will naturally be too complicated to be useful. There are however a special class of potentials for which this series terminates. Because the dependence of the Jost solutions on $M_{ij}$ is linear, this happy situation corresponds to cases where the $n$-th order time-independent perturbation theory produces the exact expression for the Jost solutions for some $n\in\Z^+$. As shown in Ref.~\cite{pra-2012} by explicit calculations, this happens for the multi-delta-function potentials (\ref{multi-delta}). A much easier way of proving this claim is to show that the entries of the transfer matrix of every multi-delta-function potential have polynomial dependence on its coupling constants. This follows from Eqs.~(\ref{M-delta}) and (\ref{M-multi-delta}), because the first of these equation shows that the entries of the transfer matrix of the single delta-function potential (\ref{delta}) is a polynomial of degree one in $\fz_j$, while the second expresses the transfer matrix of the multi-delta-function potential (\ref{multi-delta}) as the product of the transfer matrices of single delta-function potentials.

\section{Dynamical formulation of time-independent scattering}
\label{Sec3}

\subsection{Dynamics of non-stationary two-level quantum systems}
\label{Sec3-1}

Consider a quantum mechanical system with a two-dimensional Hilbert space $\sH$ and a time-dependent Hamiltonian operator $\cH(t)$ acting in $\sH$. As a vector space $\sH$ is isomorphic to $\C^2$ and we can use standard basis of $\C^2$, i.e., $\{(1,0),(0,1)\}$, to represent the elements of $\sH$ and the linear operators acting in $\sH$ by column vectors and $2\times 2$ matrices; if $\Psi\in\sH$ and $L:\sH\to\sH$ is a linear operator with domain $\sH$, the corresponding column vector $\bPsi$ and $2\times 2$ matrix $\bL$ have the form,
	\begin{align*}
	&\bPsi:=\left[\begin{array}{c}
	\br e_1|\Psi\kt\\
	\br e_2|\Psi\kt\end{array}\right],
	&&\bL:=\left[\begin{array}{cc}
	\br e_1|L\,e_1\kt & \br e_1|L\,e_2\kt\\\
	\br e_2|L\,e_1\kt & \br e_2|L\,e_2\kt\end{array}\right],
	\end{align*}
where $e_1:=(1,0)$ are $e_2:=(0,1)$ are the standard basis vectors, and $\br\cdot|\cdot\kt$ is the Euclidean inner product on $\C^2$; $\br(w_1,w_2)|(z_1,z_2)\kt:=\sum_{j=1}^2w_j^*z_j$ for all $(w_1,w_2),(z_1,z_2)\in\C^2$.

The dynamics of this system is determined by the time-independent Schr\"odinger equation,
	\be
	i\partial_t\bPsi(t)=\bcH(t)\bPsi.
	\label{t-sch-eq-1}
	\ee
We can express the evolving state vector in the form,
	\be
	\bPsi(t)=\bcU(t,t_0)\bPsi(t_0),
	\label{U-def}
	\ee 
where $\bcU(t,t_0)$  is the evolution operator corresponding to an initial time $t_0$. By definition, it satisfies
	\be
	i\partial_t\bcU(t,t_0)=\bcH(t)\bcU(t,t_0),\quad\quad \bcU(t_0,t_0)=\bI,
	\label{t-sch-eq-U}
	\ee
where $\bI$ is the $2\times 2$ identity matrix. If the entries of $\bcH(t)$ are piecewise continuous functions of $t$, (\ref{t-sch-eq-U}) has a unique solution in $\R$.

An important property of the evolution operator is the following useful identity,
	\be
	\bcU(t_2,t_1)\,\bcU(t_1,t_0)=\bcU(t_2,t_0),
	\label{compose-U}
	\ee
where $t_0,t_1$, and $t_2$ are arbitrary real parameters. It states that in order to evolve a state vector from the initial time $t_0$ to a final time $t_2$, we can first evolve it from $t_0$ to $t_1$ and then from $t_1$ to $t_2$. This agrees with our intuitions when $t_0\leq t_1\leq t_2$. To establish (\ref{compose-U}) for arbitrary $t_1$, we first use (\ref{U-def}) to infer $\bcU(t_0,t)=\bcU(t,t_0)^{-1}$. Setting $t=t_1$, this implies
	\be
	\bcU(t_0,t_1)\bcU(t_1,t_0)=\bI.
	\label{U-inv}
	\ee 
It is also easy to see that 
	\begin{align}
	&i\partial_{t_2}[\bcU(t_2,t_1)\bcU(t_1,t_0)]=
	[i\partial_{t_2}\bcU(t_2,t_1)]\bcU(t_1,t_0)=\bcH(t_2)
	\bcU(t_2,t_1)\bcU(t_1,t_0).
	\label{UU=U-proof}
	\end{align}
Eqs.~(\ref{U-inv}) and (\ref{UU=U-proof}) show that $\bcU(t_2,t_1)\bcU(t_1,t_0)$ satisfies (\ref{t-sch-eq-U}) for $t=t_2$. This together with the uniqueness of the solution of (\ref{t-sch-eq-U}) imply (\ref{compose-U}).

Because $\bcH(t)$ is a generic time-dependent matrix Hamiltonian, we cannot obtain a closed-form expression for the solution of (\ref{t-sch-eq-U}). The best we can do is to turn (\ref{t-sch-eq-U}) into the integral equation,
	\be
	\bcU(t,t_0)=\bI-i\int_{t_0}^t dt_1\:\bcH(t_1)\bcU(t_1,t_0),
	\label{integral-sch-eq}
	\ee
and use the latter repeatedly to construct the following series expansion for its solution.
	\be
	\bcU(t,t_0)=\bI+\sum_{{n}=1}^\infty(-i)^{n}
        \int_{t_0}^t\!\! dt_{n} \int_{t_0}^{t_{n}} \!\! dt_{{n}-1}\cdots
        \int_{t_0}^{t_{2}} \!\! dt_{1}\bcH(t_{n})\bcH(t_{{n}-1})\cdots\bcH(t_1).
        \label{dyson-1}
	\ee 
This is known as the ``Dyson series'' for $\bcU(t,t_0)$. 

The term $\bcH(t_{n})\bcH(t_{{n}-1})\cdots\bcH(t_1)$ appearing in (\ref{dyson-1}) is the product of ${n}$ copies of the Hamiltonian matrix $\bcH(t_j)$ whose arguments $t_j$ are arranged in the descending order from the right to the left, i.e., $t_n\geq t_{n-1}\geq\cdots\geq t_1$. For this reason, we say that $\bcH(t_{n})\bcH(t_{{n}-1})\cdots\bcH(t_1)$ is a ``time-ordered product'' of $\bcH(t_j)$'s, \cite{weinberg}. Bringing a randomly ordered product of $\bcH(t_j)$'s to their time-ordered product is called ``chronological ordering'' or ``time ordering.'' We can view this as the action of a linear operator $\sT$ in the space of products of linear operators (in our case $2\times 2$ matrices) $\bL(t_j)$ that are generally time-dependent. $\sT$ is called the ``chronological ordering'' or ``time ordering operator'' \cite{dewitt}. Because $\sT$ is linear we can specify it by demanding that the following conditions hold.
	\begin{itemize}
	\item[-] $\sT\{\bL(t)\}=\bL(t)$;
	\item[-] For every positive integer $n$, every permutation $\sigma$ of $\{1,2,\cdots,n\}$, and time labels $t_1,t_2,\cdots,t_n$ satisfying $t_n\geq t_{n-1}\geq\cdots\geq t_1$, 
	\be
	\sT\{\bL(t_{\sigma(1)})\bL(t_{\sigma(2)})\cdots\bL(t_{\sigma(n)})\}=
	\bL(t_n)\bL(t_{n-1})\cdots\bL(t_1).
	\label{Time-ordering}
	\ee
	\end{itemize}
Eq.~(\ref{Time-ordering}) states that we can rearrange the order of factors appearing in the argument of $\sT$ without affecting their time-ordered product. This is the main utility of the time ordering operation. It allows for treating $t$-dependent non-commuting operators as if they are commuting.

Because all the terms appearing in the right-hand side of (\ref{dyson-1}) are time-ordered, inserting $\sT$ after the equality sign in this equation does not change it. Once we do so, $\sT$ affects all the terms in the series, and we can freely commute the $\bcH(t_j)$'s as if they are scalars. Using this trick we can change the upper boundary of the integrals in (\ref{dyson-1}) to $t$, if we insert a factor of $1/n!$ before each of the multiple integrals to avoid double counting. In this way we can express (\ref{dyson-1}) as
 	\be
	\bcU(t,t_0)=\sT\left\{\bI+\sum_{{n}=1}^\infty\frac{(-i)^{n}}{n!}
        \int_{t_0}^t\!\! dt  \int_{t_0}^{t} \!\! dt_{{n}-1}\cdots
        \int_{t_0}^{t} \!\! dt_{1}\bcH(t_{n})\bcH(t_{{n}-1})\cdots\bcH(t_1)\right\}.
        \label{dyson-2}
	\ee 
Because of the resemblance of the right-hand side of this equation with the Maclorean series for the exponential function, $e^z=1+\sum_{n=0}^\infty z^n/n!$, it is customary to call it a``time-ordered exponential,'' and write (\ref{dyson-2}) in the following abbreviated form.
	\be
	\bcU(t,t_0)=\sT\left\{\exp\left[-i\int_{t_0}^t ds\: \bcH(s)\right]\right\}.
	\label{dyson-3}
	\ee

\subsection{Potential scattering as a dynamical phenomenon}
\label{Sec3-2}

One of the elementary facts about ordinary differential equations is that every linear equation of order $m$ is equivalent to a system of $m$ linear first order equations. For the time-independent Schr\"odinger equation~(\ref{sch-eq}), which is second order, this leads to an equivalent system of two linear equations. The relationship is however not one-to-one; there are infinitely many systems of linear equations whose solution will give the solutions of the Schr\"odinger equation~(\ref{sch-eq}). 

Because (\ref{sch-eq}) is homogeneous, the equivalent first order systems take the form of a time-dependent Schr\"odinger equation,
	\be
	i\partial_x\bPsi(x)=\bcH(x)\bPsi(x),
	\label{x-sch-eq}
	\ee
where $x$ plays the role of time. However, as we just noted, the choice of $\bPsi$ and $\bcH(x)$ is not unique. The only constraint is that the components of $\bPsi(x)$ are linear combinations of the solutions $\psi(x)$ of (\ref{sch-eq}) and its first derivative $\psi'(x)$ with generally $x$-dependent coefficients, i.e., there are functions $g_{ij}(x)$ with $i,j\in\{1,2\}$ such that
	\be
	\bPsi(x)=\left[\begin{array}{c} 
	g_{11}(x)\,\psi(x)+g_{12}(x)\,\psi'(x)\\ 
	g_{21}(x)\,\psi(x)+g_{22}(x)\,\psi'(x)\end{array}\right].
	\label{Psi-gen}
	\ee
The non-uniqueness of $\bPsi(x)$ and $\bcH(x)$ is related to the freedom in the choice of $g_{ij}(x)$.\footnote{For specific choices considered in literature, see \cite{feshbach-1958,ap-2004}.} There is however a unique choice for $g_{ij}(x)$ and $\Psi(x)$ that proves to be of great importance for scattering theory. Its discovery was motivated by the simple observation that the composition property (\ref{compose-1}) of the transfer matrix has the same structure as the identity (\ref{compose-U}) satisfied by the evolution operators \cite{ap-2014,pra-2014a}. This naturally led to a search for a concrete connection between the transfer matrix of a given short-range potential $v$ and the evolution operator for an associated quantum system. 

If the transfer matrix $\bM$ of a potential $v$ is to be identified with the evolution operator of a quantum system, then this evolution operator and the corresponding Hamiltonian should be represented by $2\times 2$ matrices. This in turn means that we seek for the matrix Hamiltonian $\bcH(x)$ of a two-level quantum system. According to (\ref{M-def}), $\bM$ maps $\left[\begin{array}{c} A_-\\ B_-\end{array}\right]$ to $\left[\begin{array}{c} A_+\\ B_+\end{array}\right]$. Therefore, if we are to set 
	\be
	\bM=\bcU(x_+,x_-),
	\label{M=U}
	\ee
for some $x_\pm$, then the two-component state vector $\bPsi(x)$ must satisfy
	\begin{align}
	&\bPsi(x_\pm)=\left[\begin{array}{c} A_\pm\\ B_\pm\end{array}\right].
	\label{Psi-Psi}
	\end{align}
The fact that $\left[\begin{array}{c} A_\pm\\ B_\pm\end{array}\right]$ are determined through (\ref{asym1}) by the asymptotic expressions for $\psi(x)$ at $x=\pm\infty$ suggests that we identity $x_\pm$ with $\pm\infty$. Doing this in (\ref{M=U}) and (\ref{Psi-Psi}), we find
	\bea
	\bM&=&\bcU(+\infty,-\infty)=\sT\left\{\exp\left[-i\int_{-\infty}^\infty dx\: \bcH(x)\right]\right\}\nn\\
	&=&\bI+\sum_{{n}=1}^\infty(-i)^{n}
        \int_{-\infty}^\infty\!\! dx_{n} \int_{-\infty}^{x_{n}} \!\! dx_{{n}-1}\cdots
        \int_{-\infty}^{x_{2}} \!\! dx_{1}\bcH(x_{n})\bcH(x_{{n}-1})\cdots\bcH(x_1),
        \label{M=exp}
        \eea
and
	\be
	\lim_{x\to\pm\infty}\bPsi(x)=\left[\begin{array}{c} A_\pm\\ B_\pm\end{array}\right],
	\label{Psi-IC}
	\ee
where we have also benefitted from (\ref{dyson-1}) and (\ref{dyson-3}).

If we substitute (\ref{Psi-gen}) in (\ref{Psi-IC}) and use (\ref{asym1}) to simplify the resulting equation, we obtain
	\bea
	\lim_{x\to\pm\infty}\left\{A_\pm[g_{11}(x)+ikg_{12}(x)]e^{ikx}+
	B_\pm[g_{11}(x)-ikg_{12}(x)]e^{-ikx}\right\}&=&A_\pm,\nn\\
	\lim_{x\to\pm\infty}\left\{A_\pm[g_{21}(x)+ikg_{22}(x)]e^{ikx}+
	B_\pm[g_{21}(x)-ikg_{22}(x)]e^{-ikx}\right\}&=&B_\pm.\nn
	\eea
Demanding that the content of the braces on the left-hand sides of these equations be equal to their right-hand side and using the fact that $A_\pm$ and $B_\pm$ are arbitrary complex numbers, we arrive at the following system of linear equations for $g_{ij}(x)$.
	\begin{align}
	&g_{11}(x)+ikg_{12}(x)=e^{-ikx}, && g_{11}(x)-ikg_{12}(x)=0,\nn\\
	&g_{21}(x)+ikg_{22}(x)=0, &&g_{21}(x)-ikg_{22}(x)=e^{ikx}.\nn
	\end{align}
They have the following unique solution.
	\begin{align*}
	&g_{11}(x)=g_{21}(x)^*=\frac{1}{2}\:e^{-ikx},
	&&g_{12}(x)=g_{22}(x)^*=-\frac{i}{2k}\:e^{-ikx}.
	\end{align*} 
Substituting these in (\ref{Psi-gen}) gives
	\be
	\bPsi(x)=\frac{1}{2}\left[\begin{array}{c} 
	e^{-ikx}[\psi(x)-ik^{-1}\psi'(x)]\\ 
	e^{ikx}[\psi(x)+ik^{-1}\psi'(x)]\end{array}\right]=
	\frac{1}{2}\: e^{-ikx\bsigma_3}\left[\begin{array}{c} 
	\psi(x)-ik^{-1}\psi'(x)\\ 
	\psi(x)+ik^{-1}\psi'(x)\end{array}\right].
	\label{Psi-gen-1}
	\ee
	
Next, we differentiate $\bPsi(x)$ and use (\ref{sch-eq}) to show that
	\be
	i\bPsi'(x)=\frac{v(x)\psi(x)}{2k}\left[\begin{array}{c} 
	e^{-ikx}\\ 
	-e^{ikx}\end{array}\right]=\frac{-iv(x)\psi(x)}{2k}\:
	e^{-ikx\bsigma_3}\left[\begin{array}{c} 
	1\\ 
	-1\end{array}\right].
	\label{dPsi-gen-1}
	\ee 
If we plug (\ref{Psi-gen-1}) and  (\ref{dPsi-gen-1}) in (\ref{x-sch-eq}) and demand that the resulting equation holds independently of the choice of $\psi(x)$, we obtain a linear system of four algebraic equations for the entries of the matrix Hamiltonian $\bcH(x)$. Solving this system, we find \cite{ap-2004}
	\be
	\bcH(x)=\frac{v(x)}{2k}\left[\begin{array}{cc} 
	1 &e^{-2ikx}\\ 
	-e^{2ikx}&-1\end{array}\right]
	=\frac{v(x)}{2k}\:e^{-ikx\bsigma_3}\bcK\,e^{ikx\bsigma_3},
	\label{sH=}
	\ee
where 
	\[\bcK:=\bsigma_3+i\bsigma_2=\left[\begin{array}{cc} 1 &1\\ -1&-1\end{array}\right].\]

The matrix Hamiltonian (\ref{sH=}) is manifestly non-Hermitian. Therefore, the evolution operator it defines is non-unitary. It is also easy to verify that it has a single eigenvalue, namely $0$. Because $\bcH(x)$ is not the null matrix, this shows that it is not even diagonalizable. We can however check that whenever $v$ is a real-valued potential, it satisfies
	\be
	\bcH(x)^\dagger=\bsigma_3\bcH(x)\bsigma_3^{-1},
	\label{ph}
	\ee
i.e., it is $\bsigma_3$-pseudo-Hermitian \cite{p1}. This observation underlines the importance of pseudo-Hermitian operators in the standard quantum mechanics, where the potential $v$ is real-valued and the Hamiltonian operator, $-\partial_x^2+v(x)$, acts as a self-adjoint operator in the Hilbert space of square-integrable functions $L^2(\R)$. The solution of the scattering problem for such a potential is equivalent to finding $\bcU(+\infty,-\infty)$, where $\bcU(x,x_0)$ is the evolution operator for an effective two-level quantum system with a pseudo-Hermitian matrix Hamiltonian.\footnote{This implies that the evolution operator is pseudo-unitary \cite{jmp-2004}.}
	
For a potential $v(x)$ that is not real-valued, (\ref{ph}) may not hold. But it is easy to show that $\bcH(x)$ commutes with its $\bsigma_3$-pseudo-adjoint, $\bcH(x)^\sharp:=\bsigma_3^{-1}\bcH(x)^\dagger\bsigma_3$. This implies that, for both real and complex potentials, $\bcH(x)$ is $\bsigma_3$-pseudo-normal. 

If there is an interval $[b_-,b_+]$ in which $v(x)=0$ for all $x\in[b_-,b_+]$, then according to (\ref{sH=}),  $\bcH(x)$ also vanishes, and $\bcU(b_+,b_-)=\bI$. We can use this simple observation to show that if $v$ is a finite-range potential with support $[a_-,a_+]$, then its transfer matrix satisfies
	\bea
	\bM&=&\bcU(a_+,a_-)=\sT\left\{\exp\left[-i\int_{a_-}^{a_+} dx\: \bcH(x)\right]\right\}.
	\label{M-finite-range}
	\eea
This follows from (\ref{M=exp}) and
	\begin{align*}
	&\bcU(+\infty,a_+)=\bcU(a_-,-\infty)=\bI,  
	&&\bcU(+\infty,-\infty)=\bcU(+\infty,a_+)\bcU(a_+,a_-)\bcU(a_-,-\infty).
	\end{align*} 

In the remainder of this article, we explore some of the implications and applications of Eqs.~(\ref{M=exp}) and (\ref{M-finite-range}). These are the basic relations underlying a ``dynamical formulation of time-independent scattering theory.''

\subsection{Transfer matrix as a non-unitary S-matrix}
\label{Sec3-3}

The quantity $\bcU(+\infty,-\infty)$ that gives the transfer matrix of the potential $v$ also appears in the standard description of time-dependent scattering theory. It gives the celebrated  ``scattering operator,'' also known as the ``S-matrix,'' of a quantum system \cite{weinberg}, if we identify $\bcU(x,x_0)$ with the evolution operator for the system in the interaction picture. In our case though the quantum system is an effective non-unitary system which we have introduced for the purpose of devising an alternative formulation of time-independent potential scattering. We nevertheless wish to determine the Hamiltonian operator $\bH(x)$ in the standard Schr\"odinger picture of this system.

We begin our analysis by recalling the definition of the interaction-picture state vectors and the Hamiltonian for a generic quantum system with a Schr\"odinger-picture Hamiltonian $\hat H(t)=\hat H_0+\hat V(t)$, where $\hat H_0$ and $\hat V(t)$ respectively represent the free Hamiltonian operator and the operator for the interaction potential. In the Schr\"odinger picture of dynamics, the state vectors $\Phi(t)$ evolve according to $\Phi(t)=\hat U(t,t_0)\Phi(t_0)$, where $\hat U(t,t_0)$ is the evolution operator associated with the Hamiltonian $\hat H(t)$. By definition, it satisfies $i\partial_t \hat U(t,t_0)=\hat H(t)\hat U(t,t_0)$ and $\hat U(t_0,t_0)=I$,
where $I$ is the identity operator. The interaction-picture state vectors $\Psi(t)$, evolution operator $\hat \sU(t,t_0)$, and the Hamiltonian $\hat\cH(t)$ are related to $\Phi(t)$, $\hat U(t,t_0)$ and $\hat H(t)$ via
	\begin{align}
	&\Psi(t):=e^{i(t-t_0)\hat H_0}\Phi(t),
	\label{psi-int-pic}\\
	&\hat\sU(t,t_0):=e^{i(t-t_0)\hat H_0}\hat U(t,t_0),
	\label{U-int-pic}\\
	&\hat\cH(t):=e^{i(t-t_0)\hat H_0}\hat H(t)e^{-i(t-t_0)\hat H_0}-\hat H_0,
	\label{H-int-pic}
	\end{align}
where $e^{-i(t-t_0)\hat H_0}$ is the evolution operator for the free Hamiltonians $\hat H_0$, \cite{sakurai}.

For the quantum system we introduced in the preceding subsection, $x$ plays the role of the time label $t$, and $\bPsi(x)$ and $\bcH(x)$ are respectively the interaction-picture state vectors and the Hamiltonian, $\Psi(t)$ and $\hat\cH(t)$. Comparing (\ref{Psi-gen-1}) with (\ref{psi-int-pic}), we see that by setting $x_0=0$ we can identify the evolving state vectors in the Schr\"odinger picture and the free Hamiltonian respectively with 
	\begin{align}	
	&\bPhi(x)=\frac{1}{2}\left[\begin{array}{c} 
	\psi(x)-ik^{-1}\psi'(x)\\ 
	\psi(x)+ik^{-1}\psi'(x)\end{array}\right],
	&&\bH_0:=-k\bsigma_3.
	\label{H-zero}
	\end{align}
The second of these relations together with (\ref{H-int-pic}) lead to the following expression for the Schr\"odinger-picture Hamiltonian.
 	\bea
	\bH(x)&=&e^{ikx\bsigma_3}\bcH(x)e^{-ikx\bsigma_3}-k\bsigma_3
	=\frac{v(x)}{2k}\,\bcK-k\bsigma_3\nn\\
	&=&\frac{1}{2k}\left[\begin{array}{cc}
	v(x)-2k^2 & v(x)\\
	-v(x) & -v(x)+2k^2\end{array}\right].
	\label{bH=}
	\eea
Note that for $v=0$, $\bH(x)$ becomes $\bH_0$. Therefore, $\bV(x):=[v(x)/2k]\,\bcK$ plays the role of the interaction term in $\bH(x)$.	
	
    Again, we can check that $\bH(x)$ is a $\bsigma_3$-pseudo-Hermitian matrix whenever $v$ is real-valued. Otherwise it is $\bsigma_3$-pseudo-normal. An important distinction between $\bcH(x)$ and $\bH(x)$ is that the latter is diagonalizable. This allows for the construction of a biorthonormal system \cite{p1,review} consisting of the eigenvectors of $\bH(x)$ and $\bH(x)^\dagger$. This is a useful tool for the study of the adiabatic geometric phases associated with non-Hermitian Hamiltonian operators \cite{garrison,jmp-2008}. The problem of exploring the adiabatic geometric phases for the Hamiltonian $\bH(x)$ has been addressed in Ref.~\cite{jpa-2014a}. The results reveal the curious fact that the adiabatically evolving state vectors of the Hamiltonian $\bH(x)$ correspond to the semiclassical solutions of the Schr\"odinger equation~(\ref{sch-eq}). In particular, complex geometric phases acquired by these states give the pre-exponential factor in the WKB wave functions! These observations have also motivated the development of a systematic method of computing corrections to the WKB approximation \cite{jpa-2014b}.

Next, we use (\ref{U-int-pic}) and (\ref{H-zero}) to show that the interaction-picture evolution operator $\bcU(x,0)$ is related to evolution operator $\bU(x,0)$ for the Hamiltonian $\bH(x)$ according to $\bcU(x,0)=e^{-ikx\bsigma_3}\bU(x,0)$. This relation together with (\ref{compose-U}) and (\ref{U-inv}) imply that for every $x_\pm\in\R$,
	\bea
	\bcU(x_+,x_-)&=&\bcU(x_+,0)\bcU(0,x_-)=\bcU(x_+,0)\bcU(x_-,0)^{-1}\nn\\
	&=&	e^{-ikx_+\bsigma_3}\bU(x_+,0)\bU(x_-,0)^{-1}e^{ikx_-\bsigma_3}\nn\\
	&=&	e^{-ikx_+\bsigma_3}\bU(x_+,x_-)e^{ikx_-\bsigma_3}.
	\label{sU-bU}
	\eea
In particular,
	\be
	\bM=\bcU(+\infty,-\infty)=\lim_{x_\pm\to\pm\infty}
	e^{-ikx_+\bsigma_3}\bU(x_+,x_-)e^{ikx_-\bsigma_3}.
	\label{M=exp-bH}
	\ee
If $v$ is a finite-range potential with support $[a_-,a_+]$, we can use (\ref{M-finite-range}) and (\ref{sU-bU}) to express the transfer matrix of $v$ in the form,
	\be
	\bM=\bcU(a_+,a_-)=e^{-ika_+\bsigma_3}\bU(a_+,a_-)e^{ika_-\bsigma_3}.
	\label{M-finite-range-bH}
	\ee
This relation is particularly useful for computing the transfer matrix of the barrier potentials (\ref{barrier}) which play an important role in the numerical scattering calculations based on the slicing of the potential and making use of the composition property of the transfer matrix. 

Consider a rectangular barrier potential,
	\be
	v(x):=\left\{\begin{array}{ccc}
	\fz & \for & x\in[a_-,a_+],\\
	0 & \for & x\notin[a_-,a_+],
	\end{array}\right.
	\label{barrier-gen}
	\ee
with a real or complex hight $\fz$ and support $[a_-,a_+]$. Then (\ref{bH=}) gives
	\be
	\bH(x)=\left\{\begin{array}{ccc}
	(\fz/2k)\:\bcK-k\bsigma_3 &\for& x\in[a_-,a_+],\\
	-k\bsigma_3 &\for& x\notin[a_-,a_+].\end{array}\right.
	\label{bH-barrier}
	\ee
The fact that $\bH(x)$ is piecewise constant simplifies the calculation of its evolution operator. In particular, we have
	\begin{align}
	&\bU(a_+,a_-)=e^{-ikL\left(\,\widehat\fz\,\bcK-\bsigma_3\right)},
	\label{bU-barrier}
	\end{align}
where $L:=a_+-a_-$, and 
	\[\widehat\fz:=\frac{\fz}{2k^2}.\] 
Because $\widehat\fz\,\bcK-\bsigma_3$ is a diagonalizable $2\times 2$ matrix, we can easily compute the right-hand side of (\ref{bU-barrier}). The result is
	\be
	\bU(a_+,a_-)=\left[\begin{array}{cc}
	\fc-i(\widehat\fz-1)\fs & -i\widehat\fz\,\fs\\
	i\widehat\fz\,\fs &\fc+i(\widehat\fz-1)\fs\end{array}\right],
	\label{bU-barrier-2}
	\ee
where $\fc:=\cos(kL\fn)$, $\fs:=\sin(kL\fn)/\fn$, and $\fn:=\sqrt{1-2\,\widehat\fz}=\sqrt{1-\fz/k^2}$.
Substituting (\ref{bU-barrier}) in (\ref{M-finite-range-bH}) and making use of (\ref{bU-barrier-2}), we obtain
	\bea
	\bM&=&e^{-ika_+\bsigma_3}e^{-ikL\left(\widehat\fz\,\bcK-\bsigma_3\right)}e^{ika_-\bsigma_3}
	\nn\\[6pt]
	&=&\left[\begin{array}{cc}
	e^{-ikL}[\fc-i(\widehat\fz-1)\fs] & -i e^{ik(a_++a_-)}\,\widehat\fz\,\fs\\
	i e^{-ik(a_++a_-)}\,\widehat\fz\,\fs &e^{ikL}[\fc+i(\widehat\fz-1)\fs]\end{array}\right].
	\label{M-barrier=}
	\eea
	
Having derived an explicit formula for the transfer matrix of the barrier potential (\ref{barrier-gen}), we can use (\ref{RRT=}) to determine its reflection and transmission amplitudes. Furthermore, by investigating the real zeros of the entries of the transfer matrix (\ref{M-barrier=}), we can find the values of $k$, $L$, and $\fz$ for which the potential develops a spectral singularity \cite{pra-2011a} or its time reversal, or becomes reflectionless or invisible. 
	
We would like to remind the reader that the standard treatment of the transfer matrix of the barrier potential (\ref{barrier-gen}) involves solving the Schr\"odinger equation in the intervals $]\!-\infty,-a[$, $[a_-,a_+]$, and $]a_+,+\infty[$ and patching them together in such a way that the resulting global solution is continuous and differentiable at $x=\pm a$. This is in sharp contrast to the above calculation of the transfer matrix of this potential which only involves algebraic operations. Although reducing the solution of physics and mathematics problems to algebraic manipulations is always desirable, we cannot consider the algebraic solution of the scattering problem for the barrier potential (\ref{barrier-gen}) as a solid evidence for the practical superiority of the dynamical formulation of time-independent scattering theory. In the next subsection, we describe a much more significant application of this formulation.

\subsection{An alternative to the Born series and Born approximation}
\label{Sec3-4}

The interaction-picture Hamiltonian (\ref{sH=}) whose time-ordered exponential yields the transfer matrix of the potential $v(x)$ has an extremely simple dependence on $v(x)$; it is the product of $v(x)$ and a matrix that does not involve $v(x)$. In particular, if we write $v(x)=\zeta\, u(x)$, as we did in the derivation of the Born series (\ref{Born-series-2}), the Dyson series in (\ref{M=exp}) becomes a power series in $\zeta$. This is similar to the Born series for the scattering amplitudes $f^{l/r}(\pm)$, but it is a power series for the transfer matrix $\bM$. Because the entries of $\bM$ are rational functions of $f^{l/r}(\pm)$, the relationship between the Born series (\ref{Born-series-2}) and the Dyson series for the transfer matrix (\ref{M=exp}) is quite complicated. Nevertheless, similarly to our derivation of the Born approximation, we can devise an approximation scheme by truncating the series in (\ref{M=exp}), \cite{pra-2014a}. This leads to an alternative to the $N$-th order Born approximation, if we neglect all but the first $N+1$ terms on the right-hand side of  (\ref{M=exp}), i.e.,
	\be
	\bM\approx\bM^{(N)}:=\bI+\sum_{{n}=1}^N(-i)^{n}
        \int_{-\infty}^\infty\!\! dx_{n} \int_{-\infty}^{x_{n}} \!\! dx_{{n}-1}\cdots
        \int_{-\infty}^{x_{2}} \!\! dx_{1}\bcH(x_{n})\bcH(x_{{n}-1})\cdots\bcH(x_1).
        \label{M=approx}
        \ee
        
Let us examine the consequences of setting $N=1$ in (\ref{M=approx}). In view of (\ref{sH=}), this gives
	\bea
	\bM\approx\bM^{(1)}=\bI-\frac{i}{2k}\int_{-\infty}^\infty dx_1\: 
	v(x) \left[\begin{array}{cc}
	1 & e^{-2ikx}\\
	-e^{2ikx} & -1\end{array}\right]=\bI-\frac{i}{2k}
	\left[\begin{array}{cc} 
	 \widetilde v(0) & \widetilde v(2k)\\
	- \widetilde v(-2k) & - \widetilde v(0)\end{array}\right],
	\label{M=approx-1}
	\eea
where $\widetilde v(k)$ is the Fourier transform of $v(x)$ that is given by (\ref{FT}). According to (\ref{M=approx-1}),
	\begin{align}
	&M_{11}\approx1-\frac{i\,\widetilde v(0)}{2k},
	&&M_{12}\approx-\frac{i\,\widetilde v(2k)}{2k},
	&&M_{21}\approx\frac{i\,\widetilde v(-2k)}{2k},
	&&M_{22}\approx1+\frac{i\,\widetilde v(0)}{2k}.
	\nn
	\end{align}
Substituting these in (\ref{RRT=}), we obtain 
	\begin{align}
	&R^l\approx \frac{\widetilde v(-2k)}{2ik-\widetilde v(0)},
	&&R^r\approx \frac{\widetilde v(2k)}{2ik-\widetilde v(0)},
	&&T\approx \frac{2ik}{2ik-\widetilde v(0)}.
	\label{RRT=approx-1}
	\end{align}
These relations are different from those obtained using the first Born approximation, i.e., (\ref{RRT-1st-BA}). But the difference disappears, if we introduce the auxiliary perturbation parameter $\zeta$, write $\widetilde v(k)=\zeta\,\widetilde u(k)$, expand the right-hand sides of the relations in (\ref{RRT=approx-1}) in power series in $\zeta$, and neglect the quadratic and higher order terms in powers of $\zeta$. 

The implementation of the first order approximation (\ref{RRT=approx-1}) is particularly straightforward for the delta-function potential $v(x)=\fz\,\delta(x)$. Because for this potential, $\widetilde v(x)=\fz$, (\ref{RRT=approx-1}) gives
	\begin{align}
	&R^{l/r}\approx \frac{\fz}{2ik-\fz}=\frac{-i\fz}{2k+i\fz},
	&&T\approx \frac{2ik}{2ik-\fz}=\frac{2k}{2k+i\fz}.
	\nn
	\end{align}
Comparing these relations with the exact expressions for the reflection and transmission amplitudes of this potential, namely (\ref{RT-delta}), we see that they coincide; i.e., our first-order approximation is exact for this potential. As we discussed in Subsec.~\ref{Sec2-10}, we can infer this from the fact that the entries of the transfer matrix for the delta function potential are polynomials of degree one in $\fz$. This in turn implies that the Dyson series (\ref{M=exp}) for the delta-function potential must terminate. It is instructive to verify this assertion by direct calculations. 

For $v(x)=\fz\,\delta(x)$, (\ref{sH=}) gives 
	\[\bcH(x)=\frac{\fz}{2k}\,\delta(x)\,e^{-ikx\bsigma_3}\bcK
	e^{-ikx\bsigma_3}=\frac{\fz}{2k}\,\delta(x)\,\bcK.\]
Because $\bcK^2=\bzero$, this implies that $\bcH(x_1)\bcH(x_2)=\bzero$ for all $x_1,x_2\in\R$. Therefore, all the terms on Dyson series (\ref{M=exp}) vanish except the first two, and the first order approximation (\ref{M=approx-1}) is exact, i.e., $\bM=\bM^{(1)}$.

It is important to realize that today we do not know of any potential in one dimension for which the first Born approximation is exact. The situation is different in two dimensions. There are complex potentials in two-dimensions for which the first Born approximation is exact  \cite{pra-2019}. Note however that it took 93 years since the introduction of the Born approximation \cite{Born-1926} to construct such potentials.\footnote{This was achieved as a by-product of the dynamical formulation of time-independent scattering theory in two dimensions \cite{pra-2016}. }

Next, we examine the implications of the second order approximation,
	\be
	\bM\approx\bM^{(2)}:=\bI-i\int_{-\infty}^\infty dx_1\:\bcH(x_1)-
        \int_{-\infty}^\infty\!\! dx_{2} \int_{-\infty}^{x_{2}} \!\! dx_{1}
        \bcH(x_{2})\bcH(x_1).
        \nn
        \ee
Inserting (\ref{sH=}) in this relation and carrying our the necessary calculations, we find
	\begin{align}
	&M_{11}\approx1-\frac{i\,\widetilde v(0)}{2k}
	+\frac{\widetilde v(-2k,2k)-\widetilde v(0,0)}{4k^2},
	&&M_{12}\approx-\frac{i\,\widetilde v(2k)}{2k}-
	\frac{\widetilde v(2k,0)-\widetilde v(0,2k)}{4k^2},
	\label{M=approx-2a}\\
	&M_{21}\approx\frac{i\,\widetilde v(-2k)}{2k}-
	\frac{\widetilde v(-2k,0)-\widetilde v(0,-2k)}{4k^2},
	&&M_{22}\approx1+\frac{i\,\widetilde v(0)}{2k}
	+\frac{\widetilde v(2k,-2k)-\widetilde v(0,0)}{4k^2},
	\label{M=approx-2b}
	\end{align}
where $\widetilde v(k_1,k_2)$ denotes the two-dimensional Fourier transform of the function, 
	\[v(x_1,x_2):=v(x_2)\theta(x_2-x_1)v(x_1),\]
i.e.,
	\bea
	\widetilde v(k_1,k_2)&:=&%\sF_{k_1,k_2}\{w(x_1,x_2)\}:=
	\int_{-\infty}^\infty\!\! dx_1\int_{-\infty}^\infty\!\! dx_2\: e^{-i(k_1x_1+k_2x_2)}
	v(x_1,x_2)\nn\\
	&=&\int_{-\infty}^\infty\!\! dx_2\int_{-\infty}^{x_2}\!\! dx_1\: e^{-i(k_1x_1+k_2x_2)}
	v(x_1) v(x_2),\nn
	\eea
and $\theta(x)$ stands for the Heaviside step function, \cite{pra-2014a}. Substituting (\ref{M=approx-2a}) and (\ref{M=approx-2b}) in (\ref{RRT=}), we obtain the second order approximate expressions for the reflection and transmission amplitudes of the potential, namely
	\bea
	R^l&\approx&\frac{\widetilde v(2k,0)-\widetilde v(0,2k)+2ik\,\widetilde v(2k)}{
	\widetilde v(2k,-2k)+4k^2+2i\,\widetilde v(0)k-\widetilde v(0,0)},
	\label{R1-approx-2}\\[3pt]
	R^r&\approx&\frac{\widetilde v(0,-2k)-\widetilde v(-2k,0)+2ik\,\widetilde v(-2k)}{
	\widetilde v(2k,-2k)+4k^2+2i\,\widetilde v(0)k-\widetilde v(0,0)},
	\label{Rr-approx-2}\\
	T&\approx&\frac{4k^2}{
	\widetilde v(2k,-2k)+4k^2+2i\,\widetilde v(0)k-\widetilde v(0,0)}.
	\label{T-approx-2}
	\eea
These relations produce the exact expressions for the reflection and transmission amplitudes of the double-delta function potentials, $v(x)=\fz_1\,\delta(x-a_1)+\fz_2\,\delta(x-a_2)$.

\subsection{Perturbative unidirectional invisibility}
\label{Subsec-pert}

An interesting application of (\ref{R1-approx-2}) -- (\ref{T-approx-2}) is in the study of the following complex potential which was the focus of attention in connection to unidirectional invisibility \cite{lin-2011,poladian,greenberg,kulishov}.
    \be
    v(x):=\left\{\begin{array}{ccc}
    \fz \, e^{2\pi i n x/L} &\for& x\in[0,L],\\
    0 &\for& x\notin[0,L],\end{array}\right.
    \label{exp-pot-n}
    \ee
where $\fz$ is a real or complex coupling constant, $n$ is a positive integer, and $L$ is a positive real parameter. It is easy to see that this is a locally periodic potential with period $\ell:=L/n$.

In view of (\ref{sH=}) and (\ref{exp-pot-n}), the Dyson series (\ref{M=exp}) for this potential is a power series in $\fz$. In particular, we can derive the outcome of the first (respectively second) Born approximation by expanding the right-hand side of (\ref{R1-approx-2}) -- (\ref{T-approx-2}) in powers of $\fz$ and neglecting the quadratic (respectively cubic) and higher order terms.

The calculation of $\widetilde v(k)$ and $\widetilde v(k_1,k_2)$ for the potential (\ref{exp-pot-n}) is straightforward. It gives
	\begin{align}
	&\widetilde v(k)=i\fz\, \sE(k-n\fK),
	&&\widetilde v(k_1,k_2)=\fz^2\,\sF(k_1-n\fK,k_2-n\fK),
	\label{exp-pot-v-v2}
	\end{align}
where we have introduced $\fK:=2\pi/L$ and 
	\begin{align}
	&\sE(k):=\left\{\begin{array}{ccc}
	\displaystyle\frac{e^{-i kL}-1}{k} &\for & k\neq 0,\\[6pt]
	-iL &\for&k=0,\end{array}\right. 
	&&\sF(k_1,k_2):=\left\{\begin{array}{ccc}
	\displaystyle\frac{\sE(k_2)-\sE(k_1+k_2)}{k_1} &\for & k_1\neq 0,\\[6pt]
	-\sE'(k_2) &\for&k_1=0.\end{array}\right. \nn
	\end{align} 
Because $n$ is a positive integer, (\ref{exp-pot-v-v2}) implies 
	\be
	\widetilde v(0)=\widetilde v(0,0)=0.
	\nn%\label{tvs-0}
	\ee	
Furthermore, if $k=m\fK/2=m\pi/L$ for a positive integer $m$, 
	\begin{align}
	&\widetilde v(-2k)=\widetilde v(-2k,0)=\widetilde v(0,-2k)=0,
	&&\widetilde v(2k)=\fz L\,\delta_{mn},
	\nn\\%\label{tvs=1}\\
	&\widetilde v(2k,0)=-\widetilde v(0,2k)=\frac{i(\fz L)^2(\delta_{m\,2n}-\delta_{mn})}{2\pi n}, 
	&&\widetilde v(2k,-2k)= -\frac{i(\fz L)^2\delta_{mn}}{2\pi(m+n)},
	\nn%\label{tvs-2}
	\end{align}
and (\ref{R1-approx-2}) -- (\ref{T-approx-2}) give
		\bea
	R^l&=&\frac{in}{m}\left[
	\delta_{mn}\,\hat\fz+\frac{n}{\pi m}(\delta_{m\,2n}-\delta_{mn})\,\hat\fz^2
	\right]+O(\hat\fz^3)\label{R1-approx-2-uni}
	\\[3pt]
	R^r&=&O(\hat\fz^3),
	\label{Rr-approx-2-uni}\\
	T&=&1+ \frac{i n^2  \delta_{mn}\,\hat\fz^2}{2\pi m^2(m+n)}+O(\hat\fz^3),
	\label{T-approx-2-uni}
	\eea
where 
	\[\hat\fz:=\frac{\fz L^2}{2\pi n}=\frac{ n\, \fz\, \ell^2}{2\pi}.\]

We can use Eqs.~(\ref{R1-approx-2-uni}) -- (\ref{T-approx-2-uni}) to verify the following assertions.
	\begin{enumerate}
	\item Suppose that the dimensionless parameter $\hat\fz$ takes a value such that $O(\hat\fz^3)$ in (\ref{R1-approx-2-uni}) -- (\ref{T-approx-2-uni}) is negligibly small but the linear and quadratic terms in these equations are not. This is the case where the second Born approximation is valid while the first Born approximation fails. Then the potential~(\ref{exp-pot-n}) displays unidirectional right-reflectionless for incident waves with wavenumber $k=\pi n/L=\pi/\ell$. 
	\item Under the same conditions on $\hat\fz$, the potential displays bidirectional invisibility for all wavenumbers except $k=\pi/\ell$ and $2\pi/\ell$. Again this is valid whenever the second Born approximation is reliable.
	\item If $\hat\fz$ takes smaller values so that we can also neglect the quadratic terms in (\ref{R1-approx-2-uni}) -- (\ref{T-approx-2-uni}), but need to keep the linear terms, i.e., when the first Born approximation is valid, this potential displays unidirectional right invisibility for $k=\pi/\ell$.
	\end{enumerate}
Indeed, $R^l$ turns out to receive nonzero contributions of order $\hat\fz^3$, therefore the unidirectional reflectionlessness, the bidirectional invisibility, and the unidirectional invisibility of the potential~(\ref{exp-pot-n}) that we discussed above are approximate effects \cite{longhi-2011,Graefe-2011,pra-2014a}. They disappear for sufficiently large values of $\hat\fz$.

The approximate unidirectional invisibility that applies to (\ref{exp-pot-n}) is an example of a more general situation, where the potential possesses this property only within the domain of the validity of the first Born approximation. We call this ``perturbative unidirectional invisibility,'' \cite{pra-2014a,pra-2015c}. We can produce a variety of examples of potentials that possess this property at a finite or infinite discrete set of wavenumbers. To do this, we consider a finite-range potential $v$ with support $[0,L]$ that is given by a Fourier series in $[0,L]$ according to
	\be
	v(x):=\left\{\begin{array}{ccc}
	\displaystyle
	\sum_{n=-\infty}^\infty \fz_n e^{2\pi i n x/L} & \for & x\in [0,L],\\
	0 &\for & x\notin[0,L].\end{array}\right.
	\label{F-series}
	\ee
Here $\fz_n$ are real or complex numbers such that the series $\sum_{n=-\infty}^\infty|\fz_n|$ converges. This in turn implies that the right-hand side of (\ref{F-series}) converges and $\fz_n:=L^{-1}\int_0^L dx\, e^{-2\pi i n/L}v(x)$. The potential (\ref{exp-pot-n}) is a special case of (\ref{F-series}) where only one of the Fourier coefficients $\fz_n$ is nonzero. If there is a finite or infinite set $\sS$ of nonzero integers such that $\fz_{n}\neq 0$ if and only if $n\in\sS$, then we can infer from the results we obtained for 
(\ref{exp-pot-n}) that the potential (\ref{F-series}) displays perturbative unidirectional invisibility for wavenumbers $k=\pi |n|/L$. 

An extreme situation is when $\sS$ is the set of positive integers, i.e., $\fz_n\neq 0$ if and only if $n\geq 1$. Under this condition (\ref{F-series}) displays perturbative unidirectional right-invisibility for the wavenumbers $k=\pi n/L$, where $n$ is an arbitrary positive integer. As we argue below, the same applies for the following potentials which are obtained from (\ref{F-series}) by the translation, $x\to x-L/2$.
	\be
	v(x)=\left\{\begin{array}{ccc}
	\displaystyle
	\sum_{n=1}^\infty \fc_{n} e^{2\pi i n x/L} & \for & x\in [-\frac{L}{2},\frac{L}{2}],\\
	0 &\for & x\notin[-\frac{L}{2},\frac{L}{2}],\end{array}\right.
	\label{F-series2}
	\ee
where $\fc_{n}:=(-1)^n\fz_n\neq 0$ for all $n\in\Z^+$. 

First, we recall from Subsec.~\ref{Sec2-4} that under a translation $x\to x-a$, which maps a potential $v(x)$ to $\check v(x):=v(x-a)$, the transfer matrix of $v(x)$ transforms according to (\ref{M-translation}). Expressing this equation in terms of the entries of the transfer matrix and using (\ref{RRT=}), we arrive at the following translation rule for the reflection and transmission amplitudes.
	\begin{align}
	&R^l(k)\to\check R^l(k)=e^{2ika}R^l(k),
	&&R^r(k)\to\check R^r(k)=e^{-2ika}R^r(k),
	&&T(k)\to\check T(k)=T(k).
	\label{RRT-translation}
	\end{align}
These relations show that unidirectional invisibility is invariant under space translations. In particular, (\ref{F-series2}) gives a class of potentials displaying perturbative unidirectional invisibility for the wavenumbers $k=\pi n/L$ provided that $\fc_{n}\neq 0$ for all $n\in\Z^+$. Because 
	\be
	\fc_n=\frac{1}{L}\int_{-L/2}^{L/2} dx\: e^{-2\pi i n x/L}v(x),\nn
	\ee
Eq.~(\ref{F-series2}) identifies $v(x)$ with a potential supported in $[-\frac{L}{2},\frac{L}{2}]$ that satisfies
	\be
	\int_{-L/2}^{L/2} dx\: e^{-2\pi i n x/L}v(x)=0~~~{\for}~~~n\leq 0.
	\label{full-band}
	\ee

Next, take a positive integer $N$, and let $\ell:=L/N$ and $\fK:=\pi/\ell$. Then the wavenumbers at which (\ref{F-series2}) possesses perturbative unidirectional right-invisibility are $k=n\fK/N$ where $n\in\Z^+$. If we fix $\ell$ and increase $N$, the support of (\ref{F-series2}) expands, $\fK/N$ shrinks, and this property of the potential holds for at least one value of the wavenumber in every closed interval of size $\fK/N$ on the positive real $k$-axis. This shows that in the limit $N\to+\infty$, it practically holds for all wavenumbers, i.e., all $k\in\R^+$.  On the other hand, $N\to+\infty$ implies $L\to\infty$, and in view of (\ref{FT}) we can express (\ref{full-band}) as the following  condition on the Fourier transform of $v(x)$.
	\be
	\lim_{N\to\infty} \widetilde v(\mbox{$\frac{2\pi n}{N\ell}$})=0~~~\mbox{for all}~~~n\in\Z^+. \nn
	\ee
It is clear that this holds true, if 
	\be
	\widetilde v(k)=0~~~\mbox{for all}~~~k\in\R^+.
	\label{full-band-main}
	\ee
This argument suggests that if the Fourier transform of a potential vanishes in the negative $k$-axis, then it possesses perturbative unidirectional right-invisibility for all wavenumbers.  Ref.~\cite{horsley-2015} shows that this property holds to all orders of perturbation theory, i.e., potentials fulfilling (\ref{full-band-main}) have exact unidirectional right-invisibility for all wavenumbers. For further discussion of these potentials, see Refs.~\cite{longhi-2015,horsley-longhi}.

\subsection{Dynamical equations for reflection and transmission amplitudes}
\label{Sec3-5}

The dynamical formulation of time-independent scattering theory relies on the identification of the transfer matrix $\bM$ of a short-range scattering potential $v(x)$ with $\bcU(+\infty,-\infty)$, where $\bcU(x,x_0)$ is the evolution operator for the effective Hamiltonian $\bcH(x)$. Let $[a_-,a_+]$ be the support of $v(x)$, where we set $a_\pm=\pm\infty$ whenever $v(x)$ has an infinite range. Then, $\bcU(x,-\infty)=\bcU(x,a_-)$, and 
	\begin{align}
	&i\partial_x\bcU(x,a_-)=\bcH(x)\bcU(x,a_-),
	&&\bcU(a_-,a_-)=\bI,
	&&\bM=\bcU(a_+,a_-).
	\label{sch-eq-sU-1}
	\end{align}
	
Now, consider the one-parameter family of the truncations of the potential $v$,
	\be
	v_s(x):=%v(x)\theta(a-x)=
	\left\{\begin{array}{ccc}
	v(x) &\for& x< s,\\
	0 &\for& x\geq s,\end{array}\right.
	\label{v-truncated}
	\ee
where $s\in[a_-,a_+]$. Let $\bM_s$ and $\bcH_s(x)$ be respectively the transfer matrix and the effective Hamiltonian (\ref{sH=}) for $v_a$, and $\bcU_s(x,x_0)$ be the evolution operator for $\bcH_s(x)$. It is clear from (\ref{v-truncated}) that
	\begin{align}
	&\bcH_s(x)=\left\{\begin{array}{ccc}
	\bcH(x) &\for& x< s,\\
	\bzero &\for& x\geq s,\end{array}\right.
	&&\bcU_s(x,a_-)=\left\{\begin{array}{ccc}
	\bcU(x,a_-) &\for& x< s,\\
	\bcU(s,a_-) &\for& x\geq s,\end{array}\right.
	\end{align}
where in the second equation we have used (\ref{compose-U}) and the fact that $\bcU_s(x,s)=\bI$ for $x\geq s$. Because the support of $v_s$ lies in $[a_-,s]$,
	\be
	\bM_s=\bcU_s(s,a_-)=\bcU(s,a_-).
	\label{Ms=}
	\ee
In view of (\ref{sch-eq-sU-1}) and (\ref{Ms=}), we have
	\begin{align}
	&i\partial_s\bM_s=\bcH(s)\bM_s,
	&&\bM_{a_-}=\bI, 
	&&\bM_{a_+}=\bM.
	\label{sch-eq-M}
	\end{align}
According to these relations, the determination of the transfer matrix of the potential $v$ is equivalent to solving the initial-value problem for a matrix Schr\"odinger equation in $[a_-,a_+]$, namely the one given by the first two equations in (\ref{sch-eq-M}). 

\begin{figure}
    \begin{center}
    \includegraphics[scale=0.25]{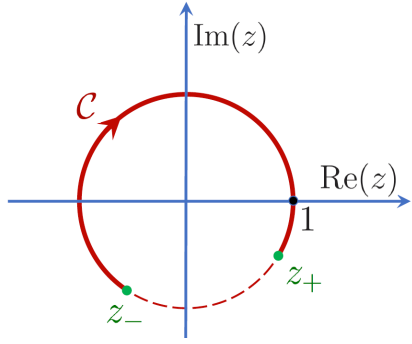}
    \end{center}
    \vspace{-12pt}
    \caption{Graph of the clockwise oriented curve $\cC$ with endpoints $z_\pm$ for the case that $(a_+-a_-)k<\pi$.}
    \label{fig5}
    \end{figure}

Using the analog of (\ref{Mij=}) for the potential $v_s$ to express $\bM_s$ in terms of the reflection and transmission amplitudes of the potential $v_s(x)$, which we respectively denote by $R_s^{l/r}$ and $T_s$, and substituting the result in the matrix Schr\"odinger equation for $\bM_s$, we obtain  a system of first-order differential equations for $R_s^{l/r}$ and $T_s$. These equations are highly nonlinear, but it is possible to reduce them to a single second order linear differential equation defined in a unit circle in the complex plane, \cite{ap-2014}. To describe this equation, we introduce the clockwise-oriented curve,
	\be
	\cC:=\left\{ e^{-2ik x}~|~x\in[a_-,a_+]~\right\},
	\label{C=}
	\ee
in the complex plane, as depicted in Fig.~\ref{fig5}, 
and let $z_\pm:=e^{-2ika_\pm}$ be its end points. Then a lengthy calculation shows that we can partially integrate the differential equations for $R_s^{l/r}$ and $T_s$, and  express their solution in terms of the solution, $S:\cC\to\C$, of the following initial-value problem.
 	\bea
	&&z^2 S''(z)+\frac{\sV_k(z)}{4k^2}S(z)=0~~~\for~~~z\in\cC,
	\label{S-eqn-1}\\
	&&\quad\quad S(z_-)=z_-,\quad\quad\quad S'(z_-)=1,
	\label{S-IC}
	\eea
where $\sV_k:\cC\to\C$ is the function defined by $\sV_k(z):=v(i\ln z/2k)$, so that for all $x\in[a_-,a_+]$,
	\be
	\sV_k(e^{-2ikx})=v(x).
	\label{sV}
	\ee
In this way, we derive a set of equations that express $R_s^{l/r}$ and $T_s$ in terms of $S(z)$. For $s=a_+$, these give  
the following formulas for the reflection and transmission amplitudes of the original potential.
	\begin{align}
	&R^l=-\int_{\cC}dz\:\frac{S''(z)}{S(z)S'(z)},
	&&R^r=\frac{S(z_+)}{S'(z_+)}-z_+,
	&&T=\frac{1}{S'(z_+)}.
	\label{RRT-S}
	\end{align}
	
Observe that both the curve $\cC$ and the function $S$ depend on the wavenumber $k$. Therefore, if we wish to use (\ref{RRT-S}) for the purpose of calculating $R^{l/r}$ and $T$, we need to solve the initial-value problem~(\ref{S-eqn-1}) and (\ref{S-IC}) for each $k\in\R^+$ and  substitute the result in (\ref{RRT-S}). It should be clear that this method is much easier to implement for finite-range potentials. 

Another point worth paying attention to is that whenever $L:=a_+-a_-\geq\pi/k$, $\cC$ runs over the whole unit circle, $|z|=1$, at least once, and $S(z)$ becomes multivalued. This does not however cause any difficulties if we view $\cC$ as a clockwise oriented parameterized curve and solve the initial-value problem accordingly. For example, suppose that $k$ is an integer multiple of $\pi/L$, i.e., there is a positive integer $m$ such that $k=\pi m/L$. Then $\cC$ is a closed curve traversing the unit circle $m$ times.  In this case, we first solve the initial-value problem~(\ref{S-eqn-1}) and (\ref{S-IC}) in $\cC_\times:=\left\{ e^{-2ik x}~|~x\in[a_-,a_+[~\right\}$, label the solution by $S_1$, and use it to introduce a new set of initial conditions, namely
	\begin{align}
	&S(z_-)=\lim_{z\to z_+}S_1(z),
	&&S'(z_-)=\lim_{z\to z_+}S'_1(z).
	\label{IC-S1}
	\end{align}
We then solve the initial-value problem given by (\ref{S-eqn-1}) and (\ref{IC-S1}) in $\cC_\times$, denote the solution by $S_2$, and iterate this procedure by letting $S_{j+1}$ be the solution of (\ref{S-eqn-1}) and  
	\begin{align*}
	&S(z_-)=\lim_{z\to z_+}S_j(z),
	&&S'(z_-)=\lim_{z\to z_+}S'_j(z),
	\end{align*}
in $\cC_\times$ for $j\in\{2,3,\cdots,m-1\}$. The reflection and transmission amplitudes of the potential are given by 
	\begin{align}
	&R^l=-\sum_{j=1}^{m}\oint_{\cC}dz\:\frac{S_j''(z)}{S_j(z)S_j'(z)},
	&&R^r=\frac{S_{m}(z_+)}{S_{m}'(z_+)}-z_+,
	&&T=\frac{1}{S_{m}'(z_+)},
	\label{RRT-S-lpp}
	\end{align}
where  $S_{m}(z_+):=\lim_{z\to z_+}S_{m}(z)$ and $S_{m}'(z_+)=\lim_{z\to z_+}S'_{m}(z)$. Note also that in the $z\to z_+$ limits appearing in the above relations, $z$ tends to $z_+$ along $\cC_\times$ in the clockwise direction.
	
Eqs.~(\ref{RRT-S}) are actually more convenient for constructing finite-range potentials with desirable properties rather than solving the scattering problem for a given potential. For example, suppose that we are interested in finding a finite-range potential $v(x)$ such that for a right-incident wave with wavenumber $k_0$ its reflection and transmission amplitudes, $R^r$ and $T$, take certain values, say $R_0^r$ and $T_0$. To construct $v$, we can proceed as follows.
	\begin{enumerate}
	\item For simplicity we choose $a_-=0$ and $a_+=L\leq\pi/k_0$, so that $z_-=1$ and $z_+=e^{-2ik_0L}$.
	\item We pick a twice differentiable function $S(z)$ satisfying
		\begin{align}
		&S(1)=1, &&S'(1)=1, && \lim_{z\to z_+}S(z)=\frac{R^r_0+z_+}{T_0},
		&& \lim_{z\to z_+}S'(z)=\frac{1}{T_0}.
		\label{SSSS}
		\end{align}
	\item We identify the potential with 
		\be 
		v(x):=\left\{\begin{array}{ccc}
		\displaystyle
		-\frac{4k_0^2 e^{-4ik_0x} S''(e^{-2ik_0x})}{S(e^{-2ik_0x})} & \for & x\in[0,L],\\[9pt]
		0 & \for & x\notin[0,L].\end{array}\right.
		\label{v(x)=}
		\ee
	\end{enumerate}
Eq.~(\ref{v(x)=}) together with (\ref{sV}) and the first two equations in (\ref{SSSS}) ensure that $S(z)$ is a solution of the initial-value problem (\ref{S-eqn-1}) and (\ref{S-IC}) for $k=k_0$. The second and third equations in (\ref{RRT-S}) imply 
	\begin{align}
	&R^r(k_0)=R^r_0, &&T(k_0)=T_0.
	\label{RR-TT}
	\end{align}

The strategy we pursued in constructing the potential (\ref{v(x)=}) is an example of a partial single-mode inverse scattering prescription. Our approach allows for finding a finite-range potential satisfying (\ref{RR-TT}) without much effort. Since we could not control $R^l(k_0)$, our approach does not solve a general single-mode inverse scattering problem. Nevertheless, we can use it to produce a variety of potentials with desirable properties such as spectral singularities and exact unidirectional invisibility \cite{ap-2014,pra-2014a}.

\subsection{Achieving exact and tunable unidirectional invisibility}
\label{Sec3-6}

The complex potentials of the form (\ref{exp-pot-n}) were the first examples of potentials displaying unidirectional invisibility \cite{lin-2011}, but as we discuss in Subsec.~\ref{Sec3-4}, they possess this property approximately. The simplest examples of complex potentials that are capable of possessing exact unidirectional invisibility are piecewise constant potentials obtained by adding two or more complex rectangular barrier potentials (\ref{barrier-gen}), \cite{pra-2013,Ge-2012}. The transfer matrix of such a potential is a product of that of its constituent rectangular barrier potentials. Therefore, we can derive explicit formulas for its refection and transmission amplitudes. Demanding that $T(k_0)=1$ and $R^l(k_0)=0\neq R^r(k_0)$ (or $R^l(k_0)=0\neq R^r(k_0)$), we can use these formulas to obtain conditions among the wavenumber $k_0$ and parameters of the potential that yield exact unidirectional invisibility. It is however not easy to satisfy this condition and at the same time control the value of the non-vanishing reflection amplitude, i.e., $R^r(k_0)$ for the unidirectionally left-invisibility and $R^l(k_0)$ for the unidirectionally right-invisibility. In this subsection, we construct exact unidirectionally right-invisible (respectively left-invisible) potentials with tunable $R^l(k_0)$ (respectively $R^r(k_0)$.) In the next subsection, we outline the application of these potentials in devising an extremely simple single-mode inverse scattering scheme. 
	
First, we construct a finite-range potential with support $[a_-,a_+]$ that is unidirectionally right-invisible for the incident waves with a given wavenumber $k_0$. Following the prescription we proposed in the preceding section, we choose $a_0=0$ and $a_+=L:=\pi n/k_0$, so that $\cC$ is an $n$-fold covering of the unit circle, and set 
	\be
	S(z):=z[\alpha(z-1)2+1],
	\label{S=}
	\ee
where $n$ is a positive integer, and $\alpha$ is a real parameter \cite{pra-2014b}. Inserting (\ref{S=}) in (\ref{v(x)=}) gives
		\be 
		v(x):=
		\left\{\begin{array}{ccc}
		\displaystyle
		\frac{-8\alpha k_0^2 (2 e^{2ik_0x}-3)}{e^{4ik_0x}+\alpha(e^{2ik_0x}-1)^2}&\for&x\in[0,L],\\[9pt]
		0 &\for& x\notin[0,L].\end{array}\right.
		\label{v(x)=SMIS}
		\ee
Because $L=\pi n/k_0$, this is a locally period potential with period $\ell=\pi/k_0$. It is also easy to check that $z_\pm=1$ and $S(1)=S'(1)=1$. Therefore, plugging (\ref{S=}) in the second and third equations in (\ref{RRT-S-lpp}) gives $R^r(k_0)=0$ and $T(k_0)=1$. This shows that the potential (\ref{v(x)=SMIS}) is right-invisible for $k=k_0$. 

For the function $S$ given in (\ref{S=}), the integral in the first equation in (\ref{RRT-S-lpp}) is an $n$-fold contour integral along the unit circle. If we demand that $\alpha>-1/4$, its integrand has a single simple pole in the unit circle, $|z|=1$, and we can easily evaluate this integral using the residue theorem. This gives
	\be
	R^l(k_0)=-\frac{8\pi i n\alpha}{(\alpha+1)^2}.
	\label{RL=SMIS}
	\ee
The fact that $n$ is an arbitrary positive integer and $|\alpha|$ can be an arbitrarily small real number shows that by properly adjusting the values of $n$ and $\alpha$, we can adjust the value of $|R^l(k_0)|$. To arrive at a completely adjustable $R^l(k_0)$, we only need to control its phase. %We can easily do this by translating the support of the potential.

From Subsec.~\ref{Subsec-pert}, we recall that under a translation, $x\to x-a$, the reflection and transmission amplitudes of a potential $v(x)$ transform according to (\ref{RRT-translation}). The first of these equations implies that we can adjust the phase of $R^l(k_0)$ by relocating the support of the potential. 

For example, suppose that we wish to set $R^l(k_0)=R^l_0$ for a given nonzero complex number $R^l_0$ while maintaining the right-invisibility of the potential~(\ref{v(x)=SMIS}). The translation of the potential, $v(x)\to\check v(x):=v(x-a)$, shifts its support to $[a,L+a]$ and, in view of (\ref{RL=SMIS}) and (\ref{RRT-translation}), the reflection and transmission amplitudes of $\check v(x)$ satisfy,
	\begin{align}
	&\check R^l(k_0)=-\frac{8\pi i n\alpha\, e^{2ik_0a}}{(\alpha+1)^2},
	&&\check R^r(k_0)=0, &&T(k_0)=1. \nn
	\end{align}
Let $\varphi_0\in[0,2\pi\,[$ be the phase angle (principal argument) of $R^l_0$, so that $R^l_0/|R^l_0|=e^{i\varphi_0}$. Our aim is to select $a$, $n$, and $\alpha$ such that $\check R^l(k_0)=R^l_0$. It is not difficult to show that we can do this by setting
	\begin{align}
	&\alpha=c_n\left[1-\sqrt{1-\frac{2}{c_n}}\right]-1, 
	&&a=\frac{\pi(m+\frac{1}{4})}{k_0}=(m+\mbox{$\frac{1}{4}$})\ell,
	\label{alpha-a=}
	\end{align}
where 
	\be
	c_n:=\frac{4\pi n}{|R^l_0|},
	\label{cn=}
	\ee 
and $m$ is an arbitrary integer. 

In view of (\ref{alpha-a=}), $0<\alpha<c_n-1$. Therefore, for this choice of $\alpha$, we have $\alpha>-1/4$ and (\ref{RL=SMIS}) holds. We also notice that $n$ is still an arbitrary positive integer, and that increasing its value makes $\alpha$ approach zero. For applications in optics, the imaginary part of the relative permittivity $\widehat\varepsilon(x)$ is a typically small number. In view of (\ref{optical}), an optical realization of the potential (\ref{v(x)=SMIS}) or its translated copy, $\check v(x)$, demands that $\alpha$ be at most of the order or $10^{-3}$. This can be easily arranged by choosing a sufficiently large value for $n$. Another useful information is the freedom in the choice of $m$. Because $m$ can take any positive or negative integer values, we can place the support of the potential $\check v(x)$ at arbitrarily large distances from the support of any other short-range potential. We will use this observation in the next subsection.  

Making the choices (\ref{alpha-a=}) for $a$ and $\alpha$, the potential 
	\be
	\check v(x):=v(x-a)=\left\{\begin{array}{ccc}
	\displaystyle
	\frac{8i\,\alpha k_0^2 (2 e^{2ik_0x}-3i)}{e^{4ik_0x}+\alpha(e^{2ik_0x}+i)^2}&\for&x\in[a,a+L],\\[9pt]
	0 &\for& x\notin[a,a+L],\end{array}\right.
	\label{tv(x)=SMIS}
	\ee	
is unidirectionally invisible for $k=k_0:=\pi n/L=\pi/\ell$, and its left reflection amplitude at this wave number coincides with $R^l_0$. Therefore, it realizes exact tunable unidirectional right invisibility.

Next, consider the transformation rule (\ref{time-reversal-Mij}) for the entries of the transfer matrix under the time-reversal transformation, $v(x)\to\overline v(x):=v(x)^*$. Substituting (\ref{RRT=}) in these relations, we find the following relations for the reflection and transmission amplitudes of the time-reversed potential $\overline v$, \cite{bookchapter}.
	\begin{align}
	&\overline R^l(k)=-\frac{R^{r}(k)^*}{D(k)^*},
	&&\overline R^r(k)=-\frac{R^{l}(k)^*}{D(k)^*},
	&&\overline T(k)=\frac{T(k)^*}{D(k)^*},
	\label{RRT-time-reversed}
	\end{align}
where $D:=T^2-R^lR^r$. Because for the potential (\ref{tv(x)=SMIS}), $R^l(k_0)=R^l_0$, $R^r(k_0)=0$, and $T(k_0)=1$, according to (\ref{RRT-time-reversed}), the reflection and transmission amplitudes of the time-reversal of the potential (\ref{tv(x)=SMIS}), i.e., $\overline{\check v}$, satisfy $\overline{\check R^l}(k_0)=0$, $\overline{\check R^r}=-R_0^{l*}$, and $\overline{\check T}=1$. This shows that for any given nonzero complex number $R^r_0$, the potential $\overline{\check v}$ is unidirectionally invisible from the left at $k=k_0$ and its right reflection amplitude equals $R^r_0$ for this wavenumber provided that in (\ref{alpha-a=}) and (\ref{cn=}) we take $R^r_0:=-R^{l*}_0$. Therefore, $\overline{\check v}$  realizes exact tunable unidirectional left invisibility. Note also that because $\overline{\check v}$ has the same support as ${\check v}$, we can place the support of $\overline{\check v}$ at any distance to the left or right of the support of any other finite-range potential.

\subsection{Unidirectional invisibility and single-mode inverse scattering}
\label{Sec3-7}

By a single-mode inverse scattering problem we mean the problem of constructing a potential $v(x)$ with prescribed values for its left/right reflection and transmission amplitudes, $R^{r/l}_0$ and $T_0$, at a given wavenumber $k_0$, i.e.,
	\begin{align}
	&R^{l/r}(k_0)=R^{l/r}_0,
	&&T(k_0)=T_0.
	\label{SMIN-condi}
	\end{align}
This problem is of direct practical importance provided that we can find a solution for it that is a short-range potential. For example, suppose that we wish to design an optical transmission amplifier operating at a wavelength $\lambda_0=2\pi/k_0$ that is reflectionless for right-incident waves, doubles the intensity of the transmitted wave, and shifts its phase angle by $90^\circ$. Let us further demand that it operates also as a reflection amplifier at the same wavelength, tripling the intensity of a left-incident wave and inducing a $-45^\circ$ phase shift upon reflection. This can be achieved by producing a relative permittivity profile $\widehat\varepsilon(x)$ or equivalently the corresponding short-range potential (\ref{optical}) whose reflection and transmission amplitudes satisfy (\ref{SMIN-condi}) for
	\begin{align*}
	&R^l_0=\sqrt 3\, e^{-i\pi/4}=\sqrt{\frac{3}{2}}(1-i), &&R^r_0=0, && T_0=\sqrt 2\, e^{i\pi/2}=\sqrt 2\, i.
	\end{align*}
Notice also that the knowledge of such a permittivity profile will not be sufficient for its practical realization unless the corresponding potential has a finite range. 
	
It should be clear that the single-mode inverse scattering problem we have described has infinitely many solutions, for we know from the extensive work on inverse scattering that the scattering data can determine the potential if they are available for all wavenumbers. Yet, to the best of the author's knowledge, none of the known inverse scattering prescriptions is capable of producing a closed-form expression for a finite-range potential that solves the single-mode inverse scattering problem. In the following we give an extremely simple and exact solution for this problem. 

We begin by recalling that according to (\ref{Mij=}), the condition (\ref{SMIN-condi}) is equivalent to demanding that the transfer matrix $\bM$ of the desired potential $v$ satisfies 
	\be
	\bM(k_0)=\frac{1}{T_0}\left[\begin{array}{cc}
	T_0^2-R^l_0R^r_0 & R_0^r\\
	-R_0^l & 1\end{array}\right].
	\label{M-zero}
	\ee
The basic idea of our solution of the single-model inverse scattering problem is that we can express the matrix $\bM(k_0)$ as the product of at most four matrices that have the form of transfer matrices of unidirectionally invisible potentials (\ref{Uni-M}). If we associate the latter with tunable unidirectionally invisible potentials we have constructed in Subsec.~\ref{Sec3-6} and make sure their support do not overlap and are arranged in the correct order along the $x$-axis, then the transfer matrix of the sum of these potentials will be equal to $\bM(k_0)$ by virtue of the composition property \cite{pra-2014b}.

To give the details of this construction, first we introduce the matrix-valued functions,
	\begin{align}
	&\bfM_1(\rho):=\left[\begin{array}{cc}
	1 & 0 \\
	\rho\,T_0-R_0^l & 1\end{array}\right],
	&&\bfM_2(\rho):=\left[\begin{array}{cc}
	1 & (T_0-1)/\rho T_0 \\
	0 & 1\end{array}\right],
	\label{M1-M2}\\
	&\bfM_3(\rho):=\left[\begin{array}{cc}
	1 & 0 \\
	-\rho & 1\end{array}\right],
	&&\bfM_2(\rho):=\left[\begin{array}{cc}
	1 & (1-T_0)/\rho \\
	0 & 1\end{array}\right],
	\label{M3-M4}
	\end{align}
where $\rho$ is an arbitrary nonzero complex number. Because $\bfM_j(\rho)$ have the form of the transfer matrix of a unidirectionally invisible potential, we can use the results of Subsec.~\ref{Sec3-6} to construct finite-range potentials $v_j$ whose transfer matrices coincide with $\bfM_j(\rho)$ for $k=k_0$. Furthermore, we can tune the parameters of these potential so that they have non-overlapping supports  placed along the $x$-axis in any order we wish. Now, suppose that $\bfM_0$ is any $2\times 2$ matrix that is obtained by multiplying $\bfM_j(\rho)$'s. Then we can arrange the positions of the supports of $v_j$'s so that the transfer matrix of the sum of these potentials equals $\bfM_0$. This argument reduces the solution of the single-mode inverse scattering problem to the decomposition of $\bM(k_0)$ into a product of matrices of the form $\bfM_j(\rho)$. We use this strategy to address
the single-mode inverse scattering problem by considering the following cases separately. 
	\begin{enumerate}
	\item $R_0^r\neq 0$: In this case, we let $\rho_\star:=(T_0-1)/R^r_0$ and use (\ref{M-zero}) -- (\ref{M3-M4}) to check that indeed
		\be
		\bfM_3(\rho_\star)\,\bfM_2(\rho_\star)\,\bfM_1(\rho_\star)=\bM(k_0).
		\label{Rr-not-zero}
		\ee			
	Because we can construct the unidirectionally invisible potentials $v_1$, $v_2$, and $v_3$ with transfer matrices given respectively by $\bfM_1(\rho_\star)$, $\bfM_2(\rho_\star)$, and $\bfM_3(\rho_\star)$ for $k=k_0$ and supports $I_1$, $I_2$, and $I_3$ such that $I_j$ is to the left of $I_{j+1}$ for $j\in\{1,2\}$, Eq.~(\ref{Rr-not-zero}) identifies $v_1+v_2+v_3$ with the solution $v$ of the single-mode inverse scattering problem for this case. Notice that this construction is valid for the special case where $R_0^l=0$, i.e., $v$ is unidirectionally left-reflectionless. In other words, we have a solution of the single-mode inverse scattering problem for a general unidirectionally left-reflectionless potential. 
		
	\item $R_0^r=0$ and $R_0^l\neq 0$: In this case, we wish to construct a unidirectionally right-reflectionless potential with given left reflection and transmission amplitudes. Because, in view of (\ref{RRT-time-reversed}), the right reflection amplitude of the time-reversed potential $\overline v$ satisfies $\overline{R^r}(k_0)=-R_0^{l*}/T_0^{*2}\neq 0$. We can follow the approach we pursed in dealing with the unidirectionally left-reflectionless potentials (in case 1) to construct $\overline v$ and then obtain $v(x)$ by complex conjugation; $v(x)=\overline v(x)^*$. 
	
	\item $R_0^l=R_0^r=0$: In this case, for every nonzero complex number $\rho$, (\ref{M-zero}) -- (\ref{M3-M4}) imply 
		\be
		\bfM_4(\rho)\,\bfM_3(\rho)\,\bfM_2(\rho)\,\bfM_1(\rho)=\left[\begin{array}{cc}
		T_0 & 0 \\
		0 & T_0^{-1}\end{array}\right]=\bM(k_0).
		\label{Rr-Rl-zero}
		\ee
For simplicity we can set $\rho=1/T_0$. Again we can construct unidirectional potentials $v_j$ with transfer matrices $\bfM_j(\rho)$ for $k=k_0$ and supports $I_j$ such that $I_j$ is to the left of $I_{j+1}$ for $j\in\{1,2,3\}$. Eq.~(\ref{Rr-Rl-zero}) then shows that we can identify $v$ with $v_1+v_2+v_3+v_4$.
	\end{enumerate}
For a discussion of concrete optical applications of the above single-mode inverse scattering prescription, see Ref.~\cite{pra-2014b,pra-2015b}.

\section{Concluding remarks}

The quantum scattering theory, which was funded by Born in 1926 \cite{Born-1926}, has been a subject of study by at least three generations of physicists and mathematicians. This makes one doubt if there is anything left to be discovered in its basic structures and methods. The progress made during the past decade has proved otherwise. 

We can trace back the origin of the developments we have reported in this article to the attempts made in the period 2002-2004 towards devising a consistent unitary quantum theory using a given non-Hermitian Hamiltonian operator with a real spectrum. This turned out to be possible provided that we modify the inner product of the Hilbert space in such a way that the Hamiltonian acts as a self-adjoint operator in the modified Hilbert space \cite{p2,p3,bbj-2002,jpa-2013,jpa-2004,ps-2010}. The fact that this is not possible for the Hamiltonian operator $-\partial_x^2+\fz\,\delta(x)$ when $\fz$ is purely imaginary suggested the presence of a spectral singularity \cite{jpa-2006b}. A detailed examination of double-delta-function potentials with complex coupling constants revealed the characterization of spectral singularities in terms of real zeros of the $M_{22}$ entry of the transfer matrix \cite{jpa-2009}. This in turn paved the way towards uncovering the physical meaning and implications of spectral singularities \cite{prl-2009} and led to the introduction and study of its nonlinear generalization \cite{prl-2013,konotop-review}. This followed by the work on coherent perfect absorbers \cite{chong-2010} and unidirectional invisibility \cite{lin-2011}, which could also be related to the real zeros of entries of the transfer matrix. These developments provided the motivation for a fresh look at the transfer matrix. Among the outcomes are a detailed study of the geometric aspects of the transfer matrix \cite{sanchecz} and the discovery of an alternative dynamical formulation of time-independent scattering theory. The latter was motivated by a purely theoretical curiosity regarding the similarity between the composition property of the transfer matrix and a basic identity satisfied by the evolution operators of quantum mechanics. 

In the present article, we have provided the necessary background on the general aspects of potential scattering as well as an accessible survey of the ideas and methods developed within the framework of the dynamical formulation of time-independent scattering theory. Our main intention for writing this article was to bring the reader to forefront of research on the subject. But to keep the size of this article reasonable we decided to omit the more recent progress made in the context of the two- and higher-dimensional generalizations of the dynamical formulation of time-independent scattering theory \cite{pra-2016}. We suffice to mention that it is possible to define an operator-valued transfer matrix in these dimensions that shares the basic properties of the transfer matrix in one dimension. In particular, we can express it as the time-ordered exponential of a pseudo-normal Hamiltonian operator and use it to address a number of previously unsolved basic problems of scattering theory. Among these are a singularity-free treatment of the single- and multi-delta-function potentials in two and three dimensions \cite{pra-2016,jpa-2018}, the discovery of a new class of exactly-solvable scattering potentials with potential applications in laser optics and quantum computation \cite{pra-2017}, the first theoretical realization of exact omnidirectional \cite{ol-2017} and unidirectional invisibility \cite{pra-2019} in a finite spectral band $[k_-,k_+]$, the characterization of scattering potential with identical scattering properties below a prescribed wavenumber \cite{jmp-2019}, the introduction of the quasi-exactly solvable scattering potentials, and the discovery of the scattering potentials for which the first Born approximation is exact \cite{pra-2019}. 

Another major outcome of this line of research, which we could not cover in this article, is the development of a fundamental notion of transfer matrix for electromagnetic scattering by isotropic media and the ensuing dynamical formulation of electromagnetic scattering theory \cite{jpa-2020}. An important application of this formulation is a remarkable construction of isotropic permittivity profiles in three dimensions which are perfectly invisible for arbitrary electromagnetic waves below a prescribed critical wavenumber.

\subsection*{Acknowledgments}
\addcontentsline{toc}{section}{Acknowledgments}

In June 2018, I had delivered three pedagogical lectures \cite{ICTS-Lectures} at the International Center for Theoretical Sciences (ICTS), Bangalore, India during the 18th International Workshop on Pseudo-Hermitian Hamiltonians in Quantum Physics. The work on this article was initiated as an attempt to improve and expand a set of notes I prepared for these lectures at ICTS. I am indebted to the organizers of this workshop and the administration of ICTS for their hospitality during my visit. I would also like to thank Hugh Jones for sending me the PDF of Ref.~\cite{jones-talk} and Hamed Ghaemi-Dizicheh for reading the first draft of this article and helping me find and correct several typos. This work has been supported by the Scientific and Technological Research Council of Turkey (T\"UB\.{I}TAK) in the framework of the project 120F061 and by Turkish Academy of Sciences (T\"UBA).

\section*{Appendix~A: Positive integer powers of $2\times 2$ matrices with unit determinant}
\addcontentsline{toc}{section}{Appendix}

Let $\bL$ be a complex $2\times 2$ matrix such that $\det\bL=1$. We wish to compute $\bL^n$ for integers $n\geq 2$. Our approach is based on separate examinations of the cases where $\bL$ is diagonalizable and non-diagonalizable.

Suppose that $\bL$ is diagonalizable. Then, because $\det\bL=1$, there is an invertible matrix $\bA$ and a nonzero complex number $\lambda$ such that
    \be
    \bL=\bA\,\bL_d\bA^{-1},
    \label{L=ALA}
    \ee
where $\bL_d:=\left[\begin{array}{cc}
    \lambda & 0\\
    0 & \lambda^{-1}\end{array}\right]$. Because $\bL_d$ is diagonal, we can express it as the following linear combination of the identity matrix $\bI$ and the diagonal Pauli matrix $\bsigma_3$.
    \be
    \bL_d=\alpha_+\bI+\alpha_-\bsigma_3,
    \label{Ld=}
    \ee
where $\alpha_\pm:=(\lambda\pm\lambda^{-1})/2$. It is easy to see that
    \begin{align}
    &\alpha_+=\mbox{\large$\frac{1}{2}$}\,{\rm tr}\,\bL,
    &&\alpha_-=\sqrt{\left(\mbox{\large$\frac{1}{2}$}\,{\rm tr}\,\bL\right)^2-1}.
    \label{alpha-pm}
    \end{align}

Next, recall the identity
    \be
    e^{i\gamma\bsigma_3}=\cos\gamma\,\bI+i\sin\gamma\,\bsigma_3,
    \label{id-app}
    \ee
which holds for every complex number $\gamma$. Because $\alpha_+^2-\alpha_-^2=1$, we can respectively identify $\alpha_+$ and $\alpha_-$ with $\cos\gamma$ and $i\sin\gamma$ for some $\gamma\in\C$. In other words, there is complex number $\gamma$ such that
    \begin{align}
    &\cos\gamma=\mbox{\large$\frac{1}{2}$}\,{\rm tr}\,\bL,
    &&\sin\gamma=\sqrt{1-\left(\mbox{\large$\frac{1}{2}$}{\rm tr}\,\bL\right)^2},
    \label{gamma=}
    \end{align}
where we have made use of (\ref{alpha-pm}). These equations determine $\gamma$ in a unique manner, if we demand that its phase angle lies in $[0,2\pi[$. Furthermore, together with (\ref{Ld=}) -- (\ref{id-app}), they imply $\bL_d=e^{i\gamma\bsigma_3}$. Substituting this relation in (\ref{L=ALA}) and using it to evaluate $\bL^n$, we find
    \be
    \bL^n=\bA\, e^{in\gamma\bsigma_3}\bA^{-1}=\cos n\gamma\,\bI+i\sin n\gamma\,\bA\,\bsigma_3
    \bA^{-1},
    \label{Ln-app-2}
    \ee
where we have also benefitted from the identity (\ref{id-app}) with $n\gamma$ playing the role of $\gamma$.

If ${\rm tr}\,\bL\neq\pm 2$, $\sin\gamma\neq 0$, and we can use (\ref{Ln-app-2}) with $n=1$ to show that %$i\bA\,\bsigma_3\bA^{-1}=\left(\bL-\cos\gamma \bI\right)/ \sin\gamma$
$i\bA\,\bsigma_3\bA^{-1}=\csc\gamma\,\bL-\cot\gamma\,\bI$. Inserting this equation in (\ref{Ln-app-2}), we arrive at
    \be
    \bL^n=\frac{\sin n\gamma}{\sin\gamma}\,\bL-\frac{\sin[(n-1)\gamma]}{\sin\gamma}\,\bI.
    \label{Ln-di-1}
    \ee
If ${\rm tr}\,\bL=\pm 2$, $\alpha_+=\pm$, $\alpha_-=0$, $\gamma\to \pi(1\mp 1)/2$, and (\ref{Ld=}) gives $\bL=\pm\bI$. Therefore,
    \be
    \bL^n=(\pm 1)^{n}\bI.
    \label{Mn-id-1b}
    \ee
Taking the limit of the right-hand side of (\ref{Ln-di-1}) as $\gamma\to \pi(1\mp 1)/2$, we also recover (\ref{Mn-id-1b}). This shows that if we define the functions $U_n:\C\to\C$ according to
    \be
    U_n(z):=\left\{\begin{aligned}
    &\displaystyle\frac{\sin (n-1)z}{\sin z} &&\mbox{when $z/\pi$ is not an integer},\\%[6pt]
    &(-1)^{\displaystyle nz/\pi}(n-1) && \mbox{when $z/\pi$ is an integer},
    \end{aligned}\right.
    \nn
    \ee
then the positive integer powers of every diagonalizable $2\times 2$ matrix $\bL$ that has a unit determinant are given by
    \be
    \bL^n=U_{n+1}(\gamma)\bL-U_n(\gamma)\bI.
    \label{Ln=gen}
    \ee

Next, we consider the case that $\bL$ is not diagonalizable. Then we can express it in its canonical Jordan form \cite{Axler}. In view of the fact that $\det\bL=1$, this gives    \be
    \bL=\bA\, \bJ_\pm\, \bA^{-1},
    \label{J-pm}
    \ee
where $\bA$ is an invertible $2\times 2$ matrix, and $\bJ_\pm:= \left[\begin{array}{cc}
\pm1&1\\
0 & \pm1\end{array}\right]$. Because $\sigma_3\bJ_-\sigma_3^{-1}=-\bJ_+$ and $\bsigma_3^{-1}=\bsigma_3$, we can use (\ref{J-pm}) to show that
    \be
    \bL=\pm \bA_\pm\,\bJ_+\bA_\pm^{-1}.
    \label{A-pm}
    \ee
where $\bA_+:=\bA$ and $\bA_-:=\bA\bsigma_3$. It is easy to see that
    \[\bJ_+^n=\left[\begin{array}{cc}
    1 &n\\
    0 & 1\end{array}\right]=n\bJ_+-(n-1)\bI.\]
In view of this relation and (\ref{A-pm}),
    \bea
    \bL^n&=&(\pm)^n \bA_\pm\bJ_+^n\bA_\pm^{-1}\nn\\
    &=&(\pm)^n\left[n\, \bA_\pm\bJ_+\bA_\pm^{-1}-(n-1)\bI\right]\nn\\
    &=&(\pm)^n[\pm n\,\bL-(n-1)\bI].
    \label{Ln=4}
    \eea
Notice that according to (\ref{J-pm}), ${\rm tr}(\bL)={\rm tr}(\bJ_\pm)=\pm 2$. Therefore, if we again define $\gamma$ using (\ref{gamma=}), so that $\gamma=\pi(1\mp 1)/2$, we find that in the limit $\gamma\to \pi(1\pm 1)/2$, (\ref{Ln-di-1}) reproduces (\ref{Ln=4}). This completes the proof that (\ref{Ln=gen}) also holds for the cases where $\bL$ is non-diagonalizable.

\np

\ed

\bibitem{sanchez} 
\bibitem{bender-1998} C.~M.~Bender and S.~Boettcher, Real spectra in non-Hermitian Hamiltonians having $\cP\cT$ symmetry, Phys.\ Rev.\ Lett.~{\bf 80}, 5243-5246 (1998).

\bibitem{born-wolf} M.~Born and E.~Wolf, {\em Principles of Optics,} Cambridge University Press, Cambridge, 1999.

\bibitem{am-boyce-diPrima} W.~E.~Boyce and R.~C.~DiPrima, {\em Elementary Differential
Equations and Boundary Value Problems}, 10th Edition, Wiley, Hoboken, N.~J., 2012.

\bibitem{am-flugge} S.~Fl\"ugge, {\em Practical Quantum Mechanics}, Springer, Berlin, 1999.

%\bibitem{am-ge-2011} L.~Ge, Y.~D.~Chong, S.~Rotter, H.~E.~T\"ureci, and A.~D.~Stone, Phys.\ Rev.\ A~\textbf{84}, 023820 (2011).

\bibitem{stone-2012} L.~Ge, Y.~D.~Chong, and A,~D.~Stone, Conservation relations and anisotropic transmission resonances in one-dimensional $\cP\cT$-symmetric photonic heterostructures, Phys.\ Rev.~A {\bf 85} 023802 (2012).

\bibitem{am-jones-2012} H.~F.~Jones, Analytic results for a PT-symmetric optical structure,
J.~Phys.~A: Math.\ Theor.\ {\bf 45}, 135306 (2012).

\bibitem{am-gupta-2014} X.~Liu, S.~Dutta Gupta, and G.~S.~Agarwal, Regularization of the spectral singularity in $\cP\cT$-symmetric systems by all-order nonlinearities: Nonreciprocity and optical isolation, Phys. Rev. A {\bf 89}, 013824 (2014).

\bibitem{am-messiah} A.~Messiah, {\em Quantum Mechanics}, Dover, New York, 1999.

\bibitem{am-SS-review} A.~Mostafazadeh, Physics of Spectral Singularities, in Proceedings of XXXIII Workshop on Geometric Methods in Physics, held in Bialowieza, Poland, June 29-July 5, 2014, Trends in Mathematics, pp.~145-165, Springer International Publishing, Switzerland, 2015; preprint arXiv:1412.0454.

\bibitem{am-muga} J.~G.~Muga, J.~P.~Palao, B.~Navarro, and I.~L.~Egusquiza, Complex absorbing potentials, Phys.\ Rep.~{\bf 395}, 357-426 (2004).

\bibitem{am-sanchez-soto} L.~ L.~S\'{a}nchez-Soto, J.~J.~Monz\'{o}na, A.~G.~ Barriuso, and J.~ F.~Cari\~{n}ena, The transfer matrix: A geometrical perspective, Phys.\ Rep.\ {\bf 513}, 191-227 (2012).

\bibitem{am-seigert-1939} A.~J.~F.~Seigert, On derivation of the dispersion formula for nuclear reactions, Phys.\ Rev.\ {\bf 56}, 750-752 (1939).

\bibitem{am-tureci-2006} H.~E.~T\"ureci,  A.~D.~Stone, and B.~Collier,
Self-consistent multimode lasing theory for complex or random lasing media,
Phys.~Rev.~A {\bf 74}, 043822 (2006).